%% file: main.tex
\documentclass[conference]{IEEEtran}
\usepackage{tabularx}
\usepackage{mathrsfs}
\usepackage{tikz}
\usepackage{amsmath}
\usepackage{comment}
\let\labelindent\relax
\usepackage{enumitem}
\usepackage{algorithm}
\usepackage{algpseudocode}
\usepackage{float}
\usepackage{caption}
\usepackage{subcaption}

\usepackage{filecontents}
\usepackage{siunitx}
\usepackage{varwidth}
\usepackage{amsmath}
\usepackage{multirow}
\usepackage{url}
\usepackage{framed}
\usepackage{mdframed}
\usepackage{mathtools, nccmath}
\usepackage{balance}
\usepackage{threeparttable}
\usepackage{amsfonts}
\usepackage{mathrsfs}
\usepackage{tikz}
\usepackage{float}
\usepackage{amsmath}
\usepackage{comment} 
\usepackage{algorithm}
\usepackage{algpseudocode}
\usepackage{float}
\usepackage{filecontents}
\usepackage{amsmath} 
\usepackage{multirow}
\usepackage{url}
\usepackage{framed}
\usepackage{mathtools, nccmath}
\usepackage{xspace}
\usepackage{booktabs,siunitx,caption}
\usepackage{varwidth}
\usepackage{placeins}
\usepackage{algpseudocode}
\DeclareMathOperator*{\mins}{min} % thin space, limits underneath in displays
\long\def\authornote#1{%
  \leavevmode\unskip\raisebox{-3.5pt}{\rlap{$\scriptstyle\diamond$}}%
  \marginpar{\raggedright\hbadness=10000
    \def\baselinestretch{0.8}\tiny
    \it #1\par}}

\newcommand{\sysname}{{HAMP}\xspace}

\usepackage{pifont}
\usepackage{picture,xcolor}

\usepackage{tikz}
\newcommand*{\circled}[2][]{\tikz[baseline=(C.base)]{
    \node[inner sep=0pt] (C) {\vphantom{1g}#2};
    \node[draw, circle, inner sep=2pt, yshift=1pt] 
        at (C.center) {\vphantom{1g}};}}
\begin{document}
%\raggedbottom
%-------------------------------------------------------------------------------

%don't want date printed
%\date{}

\title{%Don't be Too Smug: 
	Overconfidence is a Dangerous Thing: 
	Mitigating Membership Inference Attacks by Enforcing Less Confident Prediction}

\author{\IEEEauthorblockN{Zitao Chen}
\IEEEauthorblockA{University of British Columbia\\
zitaoc@ece.ubc.ca}
\and
\IEEEauthorblockN{Karthik Pattabiraman}
\IEEEauthorblockA{University of British Columbia\\
karthikp@ece.ubc.ca}}

\IEEEoverridecommandlockouts
\makeatletter\def\@IEEEpubidpullup{6.5\baselineskip}\makeatother
\IEEEpubid{\parbox{\columnwidth}{
    Network and Distributed System Security (NDSS) Symposium 2024\\
    26 February - 1 March 2024, San Diego, CA, USA\\
    ISBN 1-891562-93-2\\
    https://dx.doi.org/10.14722/ndss.2024.23014\\
    www.ndss-symposium.org
}
\hspace{\columnsep}\makebox[\columnwidth]{}}

\maketitle
%-------------------------------------------------------------------------------
\begin{abstract} 
\input{abstract}

\end{abstract}

\section{Introduction}
\input{intro}

\section{Background}
\input{background}

\input{attack_defense}

\section{Methodology}
\input{methodology}

\section{Evaluation}
\input{evaluation}

\section{Discussion}
\input{limitation}

\section{Related Work}
\input{related_work}

\section{Conclusion}
\input{conclusion}

\section*{Acknowledgment}
This work was funded in part by the Natural Sciences and Engineering Research Council of Canada (NSERC), and a Four Year Fellowship and a Public Scholar Award from the University of British Columbia (UBC).

%-------------------------------------------------------------------------------
\bibliographystyle{IEEEtranS}
\bibliography{\jobname}

\appendix
\section{Appendix} 
\input{appendix}

%%%%%%%%%%%%%%%%%%%%%%%%%%%%%%%%%%%%%%%%%%%%%%%%%%%%%%%%%%%%%%%%%%%%%%%%%%%%%%%%
\end{document}

%% file: abstract.tex
Machine learning (ML) models are vulnerable to \emph{membership inference attacks} (MIAs), which determine whether a given input is used for training the target model. 
While there have been many efforts to mitigate MIAs, they often suffer from limited privacy protection, large accuracy drop, and/or requiring additional data that may be difficult to acquire.  

This work proposes a defense technique, \sysname that can achieve both strong membership privacy and high accuracy, {without} requiring extra data. 
To mitigate MIAs in different forms, we observe that they can be unified as they all exploit the ML model's \emph{overconfidence in predicting training samples} through different proxies. 
This motivates our design to \emph{enforce less confident prediction by the model}, hence forcing the model to behave similarly on the training and testing samples. 
\sysname consists of  {a novel training framework} with {high-entropy soft labels} and {an entropy-based regularizer} to constrain the model's prediction while still achieving high accuracy. 
To further reduce privacy risk, \sysname uniformly modifies {all} the prediction outputs to become low-confidence outputs while preserving the accuracy, which effectively obscures the differences between the prediction on members and non-members.

We conduct extensive evaluation on five benchmark datasets, and show that \sysname provides consistently high accuracy and strong membership privacy.
Our comparison with seven state-of-the-art defenses shows that \sysname achieves a superior  privacy-utility trade off than those techniques.

%% file: intro.tex
\label{sec:intro}
Machine learning (ML) models are often trained with the sensitive or private user data like clinical records~\cite{kourou2015machine}, financial information~\cite{ngai2011application} and personal photos~\cite{kemelmacher2016megaface}. 
Unfortunately, ML models can also unwittingly leak private information~\cite{shokri2017membership,fredrikson2015model,tramer2016stealing,ganju2018property,carlini2021extracting}. 
One prominent example is \emph{Membership inference attacks} (MIAs)~\cite{shokri2017membership,nasr2018comprehensive,yeom2018privacy,song2021systematic,li2020membership,ye2021enhanced,carlini2022membership}, which determine whether an input is used for training the target model,  
Hence, MIAs constitute a fundamental threat to data privacy. 
For instance, by knowing that an individual's clinical  record was used to train a hospital's diagnostic model, the adversary can directly infer his/her health status. 

MIAs exploit the ML model's differential behaviors on members and non-members%, commonly due to the model's overfitting on member samples
~\cite{shokri2017membership,nasr2018comprehensive,yeom2018privacy,li2020membership,song2021systematic,choquette2021label,carlini2022membership}.
\emph{Members} are the samples used to train the model (i.e., training samples) and \emph{non-members} are the samples not used for training (e.g., testing samples). 
Existing MIAs can be divided into score-based~\cite{shokri2017membership,nasr2018comprehensive,hui2021practical,yeom2018privacy,song2021systematic,carlini2022membership} and label-only attacks~\cite{choquette2021label,li2020membership}, where the former requires access to the model's \emph{output score} indicating the class probability, while the latter needs only the prediction label. 
These attacks all seek to learn distinctive statistical features from the model's predictions in different ways, such as training an attack inference model~\cite{nasr2018comprehensive,shokri2017membership}, computing metrics like prediction loss~\cite{yeom2018privacy} and  entropy~\cite{shokri2017membership,song2021systematic}, or using Gaussian likelihood estimate~\cite{carlini2022membership}.

Defenses against MIAs can be categorized into provable and practical defenses.
\emph{Provable} defenses provide provable %privacy 
guarantees through differential privacy (DP)~\cite{abadi2016deep}, but they often incur severe accuracy degradation.
\emph{Practical} defenses, instead, offer empirical membership privacy with the goal of maintaining high model accuracy~\cite{nasr2018machine,tang2021mitigating,shejwalkar2019membership,jia2019memguard}.  
However,  existing defenses still suffer from the following limitations:
(1) limited privacy protection~\cite{jia2019memguard,nasr2018machine};
(2) large accuracy drop~\cite{abadi2016deep,nasr2018machine,tang2021mitigating};
(3) requiring additional public datasets that may not always be available in practice~\cite{papernot2016semi,shejwalkar2019membership}. 
To the best of our knowledge, no technique satisfies all these constraints, though they may address individual issues, e.g., high model accuracy but with limited privacy protection~\cite{jia2019memguard}; or strong privacy but with significant accuracy loss~\cite{abadi2016deep}.

\textbf{Our Approach.}
This paper proposes a practical defense called \sysname that can achieve both \textbf{H}igh \textbf{A}ccuracy and \textbf{M}embership \textbf{P}rivacy {without} requiring additional data.
Existing MIAs employ diverse approaches in inferring membership, e.g., score-based MIAs may exploit prediction loss or entropy~\cite{yeom2018privacy,song2021systematic,nasr2018comprehensive} while label-only MIAs~\cite{choquette2021label,li2020membership} can leverage adversarial robustness. 
Despite the different manifestations of these attacks, we identify {a common exploitation} thread among them - {they are all learning to distinguish whether the model is \emph{overly confident} in predicting the training samples} via different proxies. 
Our defense is therefore to \emph{reduce the model's overconfident prediction on training samples while preserving the model's prediction performance}, which can simultaneously reduce membership leakage (from different MIAs) and maintain model accuracy.

\sysname consists of a training- and testing-time defense.

\emph{Training-time defense}. 
Our key idea is to explicitly enforce the model to be less confident in predicting training samples during training. 
We first identify that the prevailing use of \emph{hard labels} in common training algorithms is one of the main factors that lead to the model's excessive confidence in predicting training samples. 
Hard labels assign 1 to the ground-truth label class and 0 elsewhere. The model is trained to produce outputs that match the labels, i.e., near 100\% probability for the ground-truth class and 0\% otherwise.
On the other hand, a non-member sample that is not seen during training, is usually predicted with lower confidence, and can hence be distinguished by the adversary from member samples. 

We therefore propose a new training framework that gets rid of hard labels and instead uses 
(1) \emph{High-entropy soft labels}, which are soft labels with high entropy that assign a much lower probability to the ground-truth class and non-zero probability for other classes. This explicitly enforces the model to make less confident prediction on training samples. 
(2) \sysname also consists of an \emph{entropy-based regularizer}, which is to penalize the model for predicting any high-confidence outputs via regularizing the prediction entropy during training.   

The proposed training framework is able to significantly reduce the model's overconfident prediction and improve membership privacy, without (severely) degrading the model accuracy. Section~\ref{sec:overview} explains how it prevents privacy leakage from different sources (output scores and prediction labels).  On the other hand, stronger membership privacy can also be achieved (e.g., by increasing the strength of regularization), but it would be at the cost of accuracy, which is undesirable as both privacy and accuracy are important considerations.
This motivates our testing-time defense, whose goal is to gain higher membership privacy without degrading accuracy.

\emph{Testing-time defense}. 
We propose to uniformly modify \emph{all} the outputs (from members and non-members) into low-confidence outputs, without changing the prediction labels. 
Our idea is to leverage the output scores from the \emph{randomly-generated samples}, which are often predicted with low confidence due to the high dimensionality of the input space.

In our defense, all the values in each output score are replaced by those from random samples, and we keep the relative ordering of different classes unchanged to maintain the same prediction labels (e.g., a dog image is still predicted as a dog but with different output scores). 
Both the high-confidence outputs (on training samples) and low-confidence outputs (on testing samples) are uniformly replaced by such low-confidence outputs from random samples.
This further reduces the membership leakage from the output scores.

\textbf{Evaluation.}
We evaluate \sysname on five benchmark datasets (Purchase100, Texas100, Location30, CIFAR100 and CIFAR10), and perform comprehensive evaluation on 
 a total of nine diverse MIAs (including the state-of-art LiRA attack~\cite{carlini2022membership}). 

We compare \sysname with seven leading defenses: AdvReg~\cite{nasr2018machine}, MemGuard~\cite{jia2019memguard}, SELENA~\cite{tang2021mitigating}, DMP~\cite{shejwalkar2019membership}, Label Smoothing (LS)~\cite{szegedy2016rethinking}, Early-stopping~\cite{song2021systematic}, and DP-SGD~\cite{abadi2016deep}. 

An ideal privacy defense should offer strong protection for both members and non-members. Therefore, we follow  Carlini et al.~\cite{carlini2022membership} to use attack true positive rate (TPR) controlled at low false positive rate (FPR), and attack true negative rate (TNR) at low false negative rate (FNR) to evaluate membership privacy. 
The former metric evaluates the privacy protection for members, and the latter for non-members.

\begin{figure}[t]
    \centering
    \includegraphics[width=\columnwidth, height=1.4in]{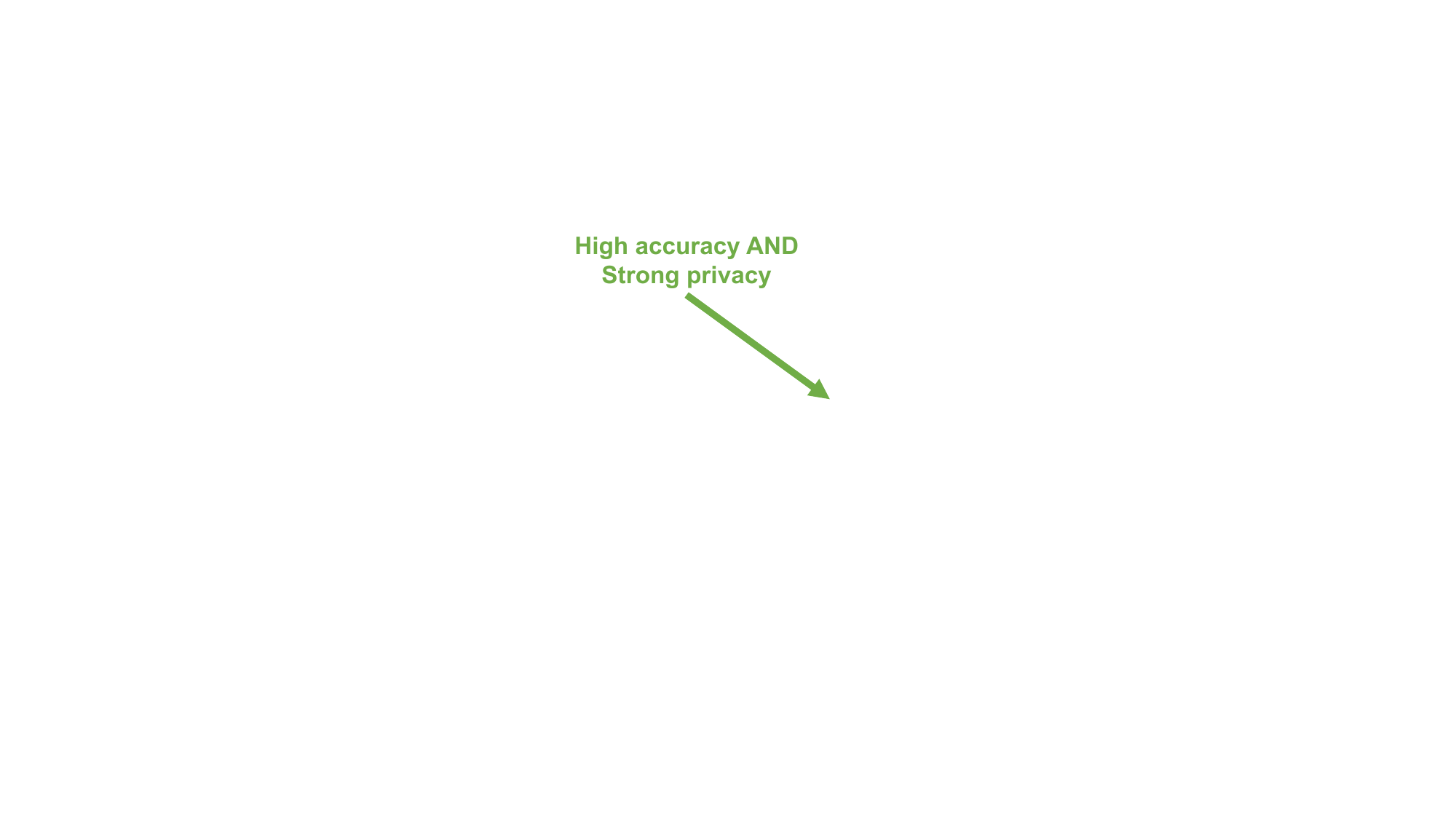}
    \caption{Privacy and utility evaluation on each defense (results averaged across datasets). 
    Negative accuracy delta means accuracy drop compared with the undefended models. 
    DP-SGD is reported at  $\epsilon=4$. 
    \sysname \emph{simultaneously} achieve strong membership privacy (for both members and non-members) and high prediction accuracy, hence providing a better privacy-utility trade off than existing defenses.} 
    \label{fig:defense-comparison} 
  \vspace{-2mm}
\end{figure}

\textbf{Contributions. } We summarize our contributions below. 

\begin{itemize}[leftmargin=*]
\itemsep0em  
\item Develop a novel training framework with high-entropy soft labels and an entropy-based regularizer to enforce less confident prediction by the model, which can significantly mitigate diverse MIAs and incur minimal accuracy drop. 
\item Propose a novel testing time defense technique to modify all the output scores into low-confidence outputs, which further improves membership privacy without degrading accuracy.
\item Integrate the training and testing framework as \sysname, and conduct rigorous evaluation under a wide range of attacks on five different datasets.  We compare \sysname against seven leading defenses and show that \sysname outperforms existing defenses by achieving a superior privacy-utility trade off.
\end{itemize}

Fig.~\ref{fig:defense-comparison} summarizes the results of \sysname versus other defenses. 
We find that existing defenses often bias towards either privacy (e.g., DP-SGD) or utility (e.g., MemGuard). 
In contrast, \sysname is able to provide strong membership privacy for both members and non-members, and preserve model accuracy. 
\sysname reduces the attack TPR @0.1\% FPR by 94\% and the attack TNR @0.1\% FNR by 97\% respectively, with only 0.46\% accuracy loss on average. This represents a much better privacy-utility trade off than other defenses.

%% file: background.tex
\label{sec:background}

\subsection{Machine Learning Primer}
\label{sec:ml-primer}
This work focuses on supervised training for classification problem.
A ML model can be expressed as a function $F_{\theta} : X \rightarrow Y$, where $X \in \mathbb{R}^d$ denotes the input space and $Y \in \mathbb{R}^k$ the output space, and $F$ is parameterized by weights $\theta$.  
During training, the network is given a training set $(x, y)\in D_{tr}$ where $y$ is the ground truth label.
$y$ is commonly expressed in the one-hot encoding format, where the ground-truth class is indicated with 1 and 0 elsewhere. 
The training objective is to minimize the prediction loss on the training set:
\begin{equation}
\label{eq:training}
\mins_{\theta} \frac{1}{|D_{tr}|} \sum_{x \in D_{tr}} \mathcal{L}(F_{\theta}(x), {y}),
\end{equation}

\noindent where $|D_{tr}|$ denotes the size of the training set, and $\mathcal{L}$ the prediction loss such as cross-entropy loss. 
The model's output $F_{\theta}(x)$ indicates the probability of $x$ belonging to each class with $\sum_{j=0}^{k-1}F_{\theta}(x)_j =1$ that sums up to 1. 

To prevent the model from overfitting on the training set, a separate validation set different from $D_{tr}$ is commonly used to serve as an unbiased proxy of the testing set.
One can use the accuracy on the validation set to assess how good the model will be when evaluated on test data and prevent overfitting.

Hereafter, we refer to $F$ as the trained model $F_{\theta}$, $F(x)$ as the {output score} of $F$ on $x$, 
and $D_{te}$ as the test set.

\subsection{Threat Model}
\emph{Attacker}.
Following prior work~\cite{jia2019memguard,tang2021mitigating,nasr2018machine}, we assume a black-box adversary who can query the target ML model with any input and observe the prediction output. 
The adversary's goal is to infer the membership of the training samples $(x, {y}) \in D_{tr}$ for a given model $F$.
Like previous defenses~\cite{nasr2018machine,tang2021mitigating,shejwalkar2019membership}, we assume a strong adversary with the knowledge of half of the training members and an equal number of non-members.  
Further, we assume the adversary has full knowledge of the defense technique and can therefore train shadow models in the same way as the target model is trained, which facilitates a strong adversary in evaluating the defenses. 

\emph{Defender}.
We assume the defender has a private set $D_{tr}$ and his/her goal is to train a model that can both achieve high classification accuracy and protect against MIAs. 
{We do not assume the defender has access to any additional data.}

%% file: attack_defense.tex
\label{sec:attack_defense}

\subsection{Membership Inference Attacks}
\label{sec:mia}
The attack model $h(x, {y}, F(x)) \rightarrow [0, 1]$ outputs the membership probability. 
We refer to $D_{tr}^A, D_{te}^A$ as the set of members and non-members that are known to the adversary.
The adversary's goal is to find a $h$ that can best distinguish between $D_{tr}^A$ and $D_{te}^A$.  The empirical gain of the attack can be measured as: 
 
\begin{equation}
\label{eq:mia-gain}
\sum_{(x,y) \in D_{tr}^A} \frac{h(x, {y}, F(x))}{|D_{tr}^A|} +
\sum_{(x,y) \in D_{te}^A} \frac{1-h(x, {y}, F(x))}{|D_{te}^A|} 
\end{equation}

We categorize existing MIAs into \emph{score-based} and \emph{label-only} attacks as follows.

\subsubsection*{Score-based MIAs}
This class of attacks either trains an inference model to infer membership~\cite{nasr2018comprehensive,shokri2017membership} or computes custom metrics such as  prediction loss~\cite{yeom2018privacy} to derive a threshold for distinction. %There are four kinds of attacks. 

\emph{NN-based attack}~\cite{nasr2018comprehensive,shokri2017membership} trains an neural network (NN) $A$, to distinguish the target model's prediction on members and non-members: $A: F(x) \rightarrow [0, 1], x\in [D_{tr}^A, D_{te}^A]$.
By querying the target model with $D_{tr}^A, D_{te}^A$, the resulting output $(F(D_{tr}^A),1)$, $(F(D_{te}^A),0)$ forms the training set for $A$.
In addition to output scores, other features like the ground-truth labels and prediction loss can also be used to train the inference model. 

\emph{Loss-based attack}~\cite{yeom2018privacy} is based on the observation that the prediction loss on training samples is often lower than that on testing samples, as the loss on training samples are explicitly minimized during training.
Specifically, the adversary can query the target model with $D_{tr}^A$, and obtain the average loss on $D_{tr}^A$ as the threshold $\tau = - \frac{1}{|D_{tr}^A|} \sum_{(x,y) \in D_{tr}^A} \mathcal{L}(F_{\theta}(x), {y})$.
Any sample with loss lower than $\tau$ is considered as a member.

\emph{Entropy-based attack}~\cite{shokri2017membership,yeom2018privacy} leverages that the output score of a training sample should be close to the one-hot encoded label, and hence its prediction entropy should be close to 0, which is lower than that on testing samples.
Prediction entropy of a sample can be computed as $-\sum_{j} F(x)_j\text{log}(F(x)_j)$, where $j$ is the class index. 

\emph{Modified-entropy-based attack}~\cite{song2021systematic} is an enhanced version of the entropy-based attack by computing the following metric: $-(1-F(x)_y)\text{log}(F(x)_y) - \sum_{j\neq y} F(x)_j\text{log}(1-F(x)_j)$.
This attack improves by taking into account class-dependent thresholds, as well as the ground truth label $y$, which is shown to achieve higher attack effectiveness.

\emph{Confidence-based attack}~\cite{yeom2018privacy,song2021systematic} exploits the observation that the prediction confidence on training samples $F(x)_y$ is often higher than that on testing samples. 
The attack threshold can be derived similar to the entropy-based attacks, and samples predicted with high confidence are deemed as members.

\emph{Likelihood Ratio Attack (LiRA)}~\cite{carlini2022membership} is a state-of-art attack that can successfully infer membership when calibrated at low false positive. 
In LiRA, the adversary trains N shadow models, half of which are trained with target sample (called IN models) and the remaining half are trained without the target sample (called OUT models). 
It then fits two Gaussian distributions to approximate the output distributions by the IN and OUT models (a logit scaling step on the logit values is taken to ensure the outputs follow a Gaussian). 
Finally, LiRA conducts a parametric likelihood-ratio test to conduct membership inference (e.g., a sample is deemed as a member if its output is estimated to come from the IN models with high probability).

\subsubsection*{Label-only MIAs} 
These attacks exploit training members' \emph{higher} degree of robustness to different perturbations (like adversarial perturbations, random noise), and develop different proxies to distinguish the degree of robustness by members and non-members.

\emph{Prediction-correctness attack}~\cite{yeom2018privacy} is the baseline label-only attack that simply determines any samples that are correctly classified as members.
This attack is effective when the training accuracy is higher than the testing accuracy. 

\emph{Boundary attack}~\cite{choquette2021label,li2020membership} is based on the observation that it is easier to perturb a testing sample to change the prediction label than a training sample. 
This is because testing samples are often {closer} to the decision boundary and therefore more susceptible to perturbations. 
Using common attacks such as CW2 attack~\cite{carlini2017towards}, the adversary measures the magnitude of perturbation needed to perturb $x \in [D_{tr}^A, D_{te}^A]$, based on which $\tau$ can be derived.
A sample is deemed as a member if the amount of perturbation needed to change the prediction label is higher than $\tau$ (i.e., more difficult to be perturbed).

The adversary can also inject random noise to the samples (instead of adversarial perturbations), which is more efficient and useful in the cases where constructing the adversarial sample is difficult (e.g., for inputs with binary features)~\cite{choquette2021label}.

\emph{Augmentation attack}~\cite{choquette2021label} makes use of the samples' robustness to data augmentation and the idea is that training samples are often more resilient to data augmentation than testing samples.
For instance, if an image was used to train a model, it should still be classified correctly when it is slightly translated. 
For each input $x$, the adversary first generates multiple augmented versions of $x$, and computes how many of them are correctly classified. 
Based on the classification outcome, the adversary trains an attack inference model to predict whether or not $x$ is a member.

\subsection{Defenses against MIAs}
\label{sec:defense-overview}

This section presents an overview of representative defenses against MIAs (a comprehensive survey of existing defenses is in Section~\ref{sec:related_work}). 

\emph{Adversarial regularization (AdvReg)}~\cite{nasr2018machine} trains the model to both achieve good model performance and protection against a shadow MIA adversary.
During training, the defender first trains an attack inference model that tries to maximize the MIA gain, after which the protected model is trained to minimize the MIA gain and maximize the classification accuracy. This is instantiated as a min-max game in \cite{nasr2018machine}.
 
\emph{Distillation for membership privacy (DMP)}~\cite{shejwalkar2019membership}. 
Shejwalkar et al. propose DMP to defend against MIAs based on  knowledge distillation.
The idea is to distill the knowledge from an undefended model (trained on a private dataset) into a new public model using a new reference set.  
Privacy protection is enabled by thwarting the access of the public model to the private dataset as the public model is trained on a {separate} reference set. 
{
Such a reference set can be curated by assuming the availability of a public dataset or by using synthetic data. We consider the latter since we do not assume access to external data. 
{This is because in many domains such as healthcare, the training data is private/proprietary, and thus such a public dataset may not be available. 
We hence consider a more realistic scenario in which the defender has no access to external data (similar to \cite{tang2021mitigating}).
}
}

\emph{SELf ENsemble Architecture (SELENA)}~\cite{tang2021mitigating}. 
SELENA  also uses knowledge distillation. 
Its key idea is to partition the private dataset into different subsets and train a sub model on each of the subset ({another technique with similar idea is proposed in \cite{chourasia2021knowledge}}).
For each sub model, there exists a subset of private dataset that was {not} used in its training, i.e.,  ``reference set'' for that sub model.
Each sub model assigns the output scores on its ``reference set'', which constitutes the knowledge to the distilled.
The knowledge from the ensemble of sub models is finally distilled into a new public model.

\emph{Early stopping}~\cite{song2021systematic,caruana2001overfitting}. 
As the training proceeds, the model tends to overfit the training data and become susceptible to MIAs. 
Early stopping is a general solution in reducing overfitting~\cite{caruana2001overfitting} by training models with fewer epochs.
Song et al.~\cite{song2021systematic} find that this is useful in mitigating MIAs and we follow to include it as as a benchmark defense mechanism.

\emph{Differential privacy (DP) based defenses}~\cite{abadi2016deep}. 
DP-based defenses leverage the formal framework of differential privacy to achieve rigorous privacy guarantee.
This is done via injecting noise to the learning objective during training such as DP-SGD that adds noise to the gradients~\cite{abadi2016deep}.
However, DP-based defenses often produce models with considerable accuracy drop, resulting in a poor privacy-utility tradeoff. 

\emph{MemGuard}~\cite{jia2019memguard}. 
Jia et al. propose to defend against MIAs via obfuscating the prediction scores. 
The idea is to fool the MIA adversary by constructing a noise vector to be added to the input (analogous to constructing adversarial samples), and make the outputs on members and non-members indistinguishable by the adversary.

\emph{Label Smoothing}~\cite{szegedy2016rethinking}. LS is a common regularization technique to improve model accuracy by using soft labels. LS replaces the one-hot label with a mixture of the one-hot label and uniform distribution using a smoothing intensity parameter. E.g., for a smoothing intensity of 0.3, the soft label becomes 80\% cat, 10\% dog, 10\% frog; and a smoothing intensity of 0.6 yields 60\% cat, 20\% dog, 20\% frog. 
LS trains with different smoothing intensities to produce model with high accuracy.  

Both LS and \sysname use soft labels in their training, but they are two techniques built with different principles that require {different} soft labels. 
LS is used to improve model performance, which necessitates training with \emph{low}-entropy soft labels. 
Unlike LS, \sysname consists of \emph{high}-entropy soft labels, an entropy-based regularizer and a novel testing-time defense (details in the next section), which is to improve membership privacy while preserving model accuracy. 
This consequently results in the different privacy implications by the two techniques: LS improves model performance but the resulting model still suffers from \emph{high} MIA risk~\cite{kaya2021does}, while \sysname consistently contributes to very \emph{low} MIA risk.  
We refer to detailed comparison in Section~\ref{sec:comparison-ls}.

%% file: methodology.tex
\label{sec:methodology}
{The main insight behind \sysname in mitigating diverse MIAs is to identify a common exploitation thread among different MIAs. \sysname is designed to overcome this exploitation so that it can defend against different MIAs regardless of their specific approaches. 
We first explain how existing MIAs can be unified via a common thread in Section~\ref{sec:mia-analysis}, and then discuss how we build \sysname to overcome this exploitation. 
}

\subsection{{Overconfident Prediction Leads to Membership Leakage}}
\label{sec:mia-analysis}
While existing MIAs employ diverse approaches to infer membership, 
we {unify} them by viewing them all as exploiting {the model's overconfidence in predicting training samples}. 
We explain below how different attacks can be viewed as different forms to quantify whether a model is overly confident in predicting a specific sample, in order to infer its membership.

Score-based MIAs leverage the prediction scores to infer membership through different proxies. 
The model's overconfident prediction on training samples can be exposed through high confidence scores~\cite{yeom2018privacy}, low prediction entropy~\cite{shokri2017membership,song2021systematic}, low prediction loss~\cite{yeom2018privacy}, or using a neural network~\cite{shokri2017membership,nasr2018comprehensive}. 
For boundary and augmentation attacks, samples predicted with high confidence can be viewed as exhibiting high robustness against adversarial perturbations and data augmentation.
Training samples can therefore be identified by the adversary based on whether they are more resilient to adversarial perturbation~\cite{choquette2021label,li2020membership} or data augmentation~\cite{choquette2021label}.

\emph{What leads to the model's overconfidence in predicting training samples?}  
As mentioned before, common training algorithms make use of the one-hot hard labels to minimize the prediction loss.
Minimizing the training objective function (\ref{eq:training}) is equivalent to encouraging the model to produce outputs that are consistent with the labels, i.e., 100\% for the ground-truth class and 0\% for any other classes. 

While training with hard labels has achieved success in a broad class of classification problems, we find that it undesirably contributes to the model's overconfidence in predicting training samples, which eventually leads to membership leakage. 
{For example, on Purchase100, the difference between the average prediction confidence on training and testing samples is $>$25\%, which means the model is much more confident in predicting training samples. 
Such differential behavior can be identified by the adversary to obtain $>$14\% attack TPR @0.1\% FPR. 
This indicates training with one-hot hard labels undesirably enables the adversary to identify a large fraction of training samples with very low false positives (and similarly identifying testing samples with low false negatives). }
This inspires our design principle of enforcing less confident prediction to mitigate MIAs, based on which we introduce a novel training and testing defense that can achieve both strong membership privacy and high model accuracy.

\subsection{Overview}
\label{sec:overview}
\begin{figure}[t]
  \centering
  \includegraphics[width=3.2in, height=1.7in]{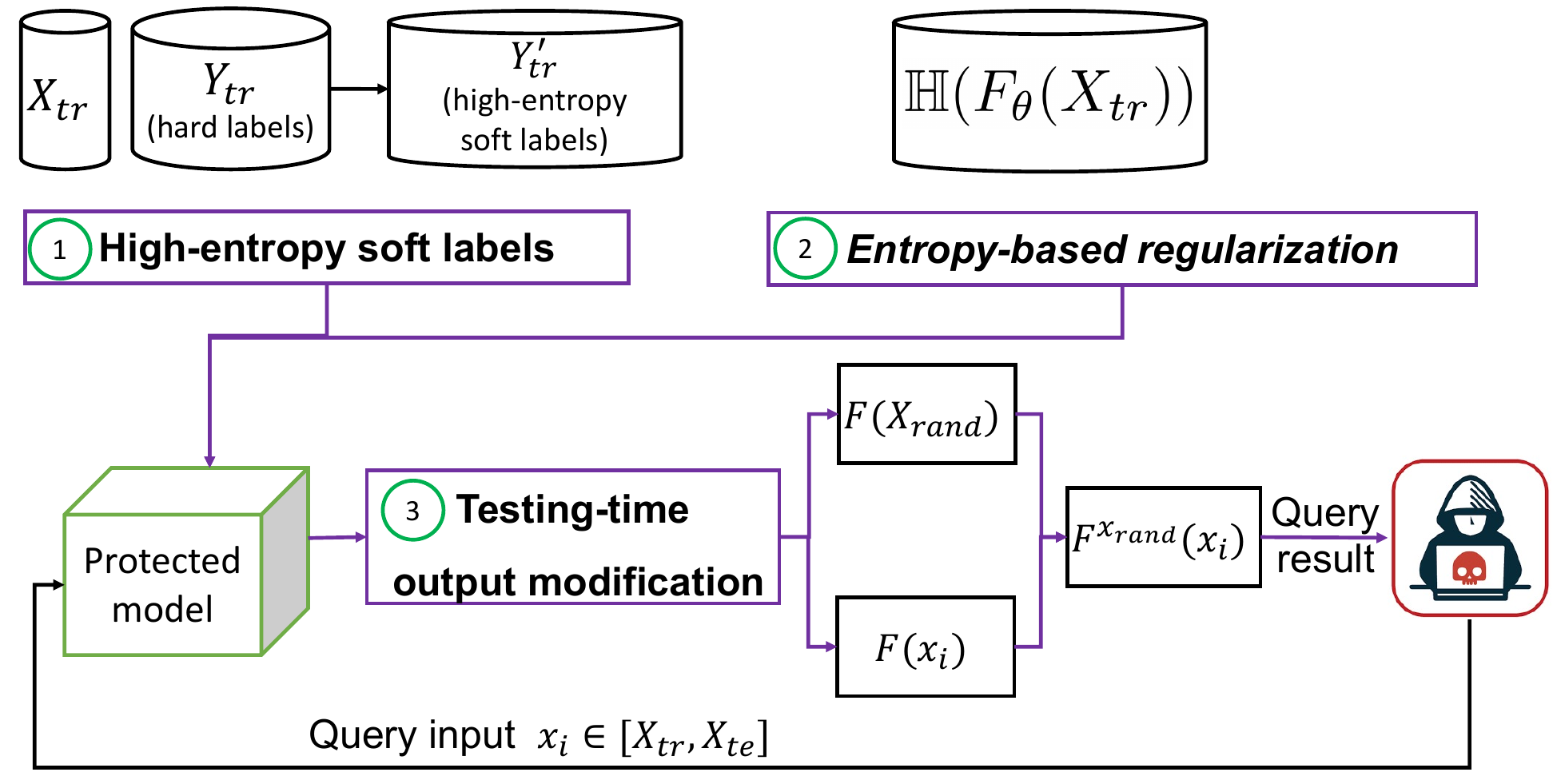}
  \caption{Overview of our training-  and testing-time defense.}
  \label{fig:method} 
  \vspace{-4mm}
\end{figure}

Fig.~\ref{fig:method} shows an overview of \sysname. It has two parts.

\textbf{Training-time defense}.
Inspired by the observation in Section~\ref{sec:mia-analysis}, our main idea is to \emph{train the model to produce less confident prediction even on training samples}, thereby enforcing the model to behave similarly on training and testing samples. 
We achieve this by two innovations: (1) replacing the hard labels with \emph{high-entropy soft labels}; and (2) introducing an \emph{entropy-based regularizer}. 

The first step is to generate soft labels with high entropy from the hard labels. 
These high-entropy soft labels explicitly induce the model to produce less confident output during training by assigning a much lower probability for the ground-truth class.  
For instance, a hard label of [0, 1] can be changed into a soft label of [0.4, 0.6], which guides the model to predict the ground-truth class with 60\% probability (instead of 100\%).
The probability of each class is determined by an \emph{entropy threshold} parameter, and a higher threshold generates a soft label with higher entropy (e.g., [0.5, 0.5] has the highest entropy) - details in the next section. 
The ground-truth class remains the same so that the model can learn correctly, e.g., a dog image is still trained to be predicted as a dog.

Second, we introduce an entropy-based regularizer to penalize the model for predicting any output with low entropy.
Prediction entropy measures the prediction uncertainty, and can be used to regularize the confidence level of the prediction, e.g., low entropy indicates high-confidence output, and can be mitigated by the proposed regularizer to become low-confidence output.

The high-entropy soft labels encourages the model to produce outputs consistent with the labels, while the regularization term allows the model to produce any low-confidence outputs, {even if} the outputs do not closely match the labels.
Both components are important for \sysname to mitigate overconfident prediction and achieve strong membership privacy.

\textbf{How \sysname's training-time defense mitigates membership leakage from different sources?}
There are two sources leading to membership leakage,  and we discuss below how \sysname can reduce leakage from both sources.

\emph{Output scores.}
With the high-entropy soft labels and entropy-based regularizer, \sysname forces the model to produce output scores on training samples with higher entropy (i.e., lower confidence), which resemble the output scores on testing samples. 
E.g., on Purchase100, the average prediction entropy on members and non-members are 0.389 and 0.576 on the undefended model, which are 4.485 and 4.490 on the \sysname model. 
\sysname therefore reduces the entropy difference by 31x (from 0.187 to 0.006) and effectively enforces the output scores on members and non-members to be indistinguishable (more details in Appendix~\ref{sec:entropy}). 
Some score-based MIAs leverage both output scores \emph{and} label information (e.g., \cite{song2021systematic,nasr2018comprehensive}) and we explain next how \sysname prevents membership leakage from the labels.

\emph{Prediction labels.}
\sysname's training-time defense mitigates privacy leakage from the prediction labels by pushing the training samples {closer} to the decision boundary, so that training samples lie \emph{similarly close} to the decision boundary as the testing samples. 
We next use the boundary and augmentation attacks to explain (both attacks exploit label information in different manners to infer membership).

Boundary attack exploits the training samples' higher adversarial robustness than testing samples. 
{Without \sysname, the adversary can discern that the training samples require more perturbations than the testing samples. 
With \sysname however, training samples are predicted with lower confidence, and therefore it takes a similar amount of perturbation to perturb training and testing samples. 
For instance, on CIFAR100, the average amount of perturbation to perturb the training samples on the undefended model is 0.342, and 0.226 on the testing samples. 
With \sysname, the perturbation on the training samples become 0.289 and 0.234 on the testing samples, which effectively reduces the perturbation difference between training and testing samples by $>$53\%.}
This means the members and non-members become indistinguishable from the perspective of their adversarial robustness.

Augmentation attack exploits the training samples' higher resistance to data augmentation, i.e., the augmented variants of training samples are \emph{more likely} to be classified correctly. 
Performing data augmentation on the original samples can be viewed as drawing neighboring variants around the original samples in the sample space. 
Since the training samples are closer to the decision boundary under \sysname, their augmented variants are more likely to cross the decision boundary, and hence be classified \emph{incorrectly}, which is similar to how testing samples would behave. 
{We also evaluate the model's performance on the inputs added with random augmentations. We find \sysname mainly reduces the performance on the augmented training samples (from 64.38\% to 55.12\%), and the performance on the augmented testing samples remain similar before and after \sysname (46.12\% and 46.36\%). This reduces the accuracy difference between members and non-members from 18.26\% to 8.76\% (a 52\% reduction), and enables them to exhibit similar resistance to data augmentation. }

\sysname's training-time framework is able to reduce the model's overconfident prediction on training samples \emph{without} compromising the model's performance, i.e., strong membership privacy and prediction accuracy.  
Nevertheless, membership privacy can be further improved such as by pushing the training samples closer to the decision boundary, but at the cost of accuracy, which is  undesirable. 
In light of this, we introduce a testing-time output modification defense that can attain higher membership privacy \emph{without} degrading accuracy.

\textbf{Testing-time defense}. 
Our idea is to modify all the output scores to become low-confidence scores, hence making the output scores from members and non-members less distinguishable.
The key observation that underpins the testing-time defense is that  {\em randomly-generated samples are often predicted with low confidence, and the low-confidence output scores can be used for output modification.}
Specifically, we first {uniformly} generate random samples, which are highly unlikely to be part of the training set due to the high dimensionality of the input space (e.g., the entire Texas100 dataset contains only $67,330$ samples while the input space has $2^{6170}$ samples).
As these random samples are unlikely to be members of the training set, they are often predicted by the model with low confidence.
We then replace all the entries in each output score with those from random samples, where the replacement is to keep the predicted labels unchanged (all top-\emph{k} labels) and modify the output scores only.  
{In essence, \sysname returns only the ordering of the confidence scores and the ordering is represented by the random output scores arranged in a specific order.}

The random samples \emph{do not} have any prerequisites (e.g., they do not need to come from a specific distribution, nor do they need to produce a  specific prediction label), as long as they are valid inputs (e.g., pixel values are in [0, 255]).

In \sysname, the high-confidence outputs on members and low-confidence outputs on non-members, all become low-confidence outputs after being modified.
This significantly increases the difficulty for the adversary to identify differential behaviors on members and non-members.

In Section~\ref{sec:ablation-study}, we perform detailed ablation study to show that all three defense components in \sysname are crucial in achieving strong membership privacy and preserving high model accuracy. 
We next explain \sysname in details.

\subsection{Training-time Defense}
\label{sec:defense-training}

\emph{Generating high-entropy soft labels}. 
The first step is to generate high-entropy soft labels for training, where the class probabilities in the soft labels are controlled by an \emph{entropy threshold} parameter, denoted as $\gamma$.
The entropy of a soft label $y'$ can be calculated as:
\begin{equation}
\mathbb{H}(y') = - \sum_{j=0}^{k-1} y'_j*\text{log}(y'_j)
\end{equation}

%where $y_j$ is the probability score of class $j$. 
A soft label with uniform probability on each dimension has the highest entropy, based on which we choose a smaller entropy threshold. 
For a $k$-class classification problem, our goal is to find a $y'$ given $\gamma$ such that, 
\begin{equation}
\mathbb{H}(y') \geq \gamma \mathbb{H}(y), y=\{{\frac{1}{k}, ... \frac{1}{k}}\}^k, \gamma \in [0,1],
\end{equation}

\noindent  where $y'$ has the highest probability on its ground-truth class, and the probabilities on the remaining dimension are the same.
For a hard label $y$ whose ground-truth class is $j_{truth}$ ($k$ classes in total), the resulting soft label becomes:
\begin{equation} 
\begin{split}
\label{eq:ls}
\forall y'_j\in y', y'_j={\begin{cases}
        p   &\quad   \text{if}~~ j=j_{truth} \\
        (1-p)/(k-1) &\quad  \text{if}~~ j\neq j_{truth}
        \end{cases}}
\end{split}
\end{equation}
 
\noindent $p$ is the probability on the ground-truth class, and a larger $\gamma$ indicates higher prediction entropy, which leads to a smaller $p$ (i.e., smaller probability on the ground-truth class).

\emph{Entropy-based regularization}.
In addition, we introduce an entropy-based regularizer that measures the prediction entropy during training, and penalizes predictions that have low entropy, as such predictions indicate high-confidence output and may be exploited by the adversary. 

Finally, the overall training objective can be formulated as:

\begin{equation}
\begin{split}
\label{eq:kl-loss}
\mathcal{L_{\text{KL}}}(F_\theta(x), y) = \sum_{j=0}^{k-1} y_j\text{log}(\frac{y_j}{F_\theta(x)_j}),
\end{split}
\end{equation}

\begin{equation}
\begin{split}
\label{eq:objective-func}
\mins_{\theta} [\mathcal{L_{\text{KL}}}( (F_\theta(X_{tr}), Y^{'}_{tr}), \theta) - \alpha \mathbb{H}(F_\theta(X_{tr}))],
\end{split}
\end{equation}

\noindent where $Y^{'}_{tr}$ is the high-entropy soft labels, $L_{\text{KL}}$ the Kullback-Leibler divergence loss, $\alpha$ is to control the strength of regularization. 
Our goal is to train the model to mitigate the overconfident prediction on training samples while maintaining high prediction accuracy. 
We achieve this by using a large $\gamma$ to train the model with soft labels in high entropy, and a $\alpha$ to regularize the prediction entropy. 
Section~\ref{sec:setup} explains how to select the parameters $\gamma, \alpha$ in \sysname ($p$ in Equation~\ref{eq:ls} is determined by $\gamma$).

\subsection{Testing-time Defense}
\label{sec:defense-testing}
The testing-time defense uniformly modifies the runtime outputs to achieve stronger privacy without jeopardizing accuracy.
We first generate uniform random samples $x_{rand}$, e.g., for Purchase100 with binary features, each feature is assigned with 0 or 1 with equal probability.  
For each runtime input $x \in [D_{tr}, D_{te}]$, all the entries in $F(x)$ (that indicate the probability for each class) are replaced by those in $F(x_{rand})$, the resulting output is denoted as $F^{x_{rand}}(x)$.
The replacement is to only modify the entries in $F(x)$ while ensuring $F(x)$ and $F^{x_{rand}}(x)$ give the same prediction labels. 
For example, let $x \in [D_{tr}, D_{te}], F(x) = [0.85, 0.05, 0.1]$, and $x' \in X_{rand}, F(x') = [0.2, 0.3, 0.5]$, then the final output produced by the model becomes: $F(x_i) = [0.5, 0.2, 0.3]$. 
This enforces the model to produce low-confidence outputs on both members and non-members, and reduces privacy leakage.

\textbf{Overall Algorithm.} 
Algorithm~\ref{alg:overall} gives the overall algorithm of \sysname. 
$\gamma$ and $\alpha$ are the two parameters in \sysname to regulate the confidence level of the model's prediction, e.g., a high entropy threshold or strong regularization can enforce the model to become less confident in prediction.  
Line 2 generates a template of high-entropy soft labels of $y'$, which is then used to generate soft labels for each of the hard labels.
The condition %of  $\forall (Y_{tr}[i], Y'_{tr}[i]) \in(Y_{tr}, Y'_{tr}), \text{argmax}(Y_{tr}[i])\equiv\text{argmax}(Y'_{tr}[i])$ 
in Line 3 ensures that the ground-truth labels remains unchanged so that the model can learn the correct labels. % from the samples.

At test time, each output is replaced by those from a random sample. 
The condition of $\text{argsort}(F^{x_{rand}}(x))=\text{argsort}(F(x))$ in line 13 is to ensure both $F^{x_{rand}}(x)$ and $F(x)$ give the same labels (all top-\emph{k} labels and not just the top-1 label).
Line 11 and Line 12 are independent of each other, and hence can be executed independently to facilitate faster runtime inference (overhead evaluation in Appendix~\ref{sec:overhead}).

\begin{algorithm}[t] 
\footnotesize 
\begin{flushleft}
\hspace*{\algorithmicindent} \textbf{Input:} $(X_{tr}, Y_{tr}) \in D_{tr}$: Training set; \\
\hspace*{1.3cm} $\gamma$: Entropy threshold;   \\
\hspace*{1.3cm} $\alpha$: Strength of regularization; \\  
\hspace*{1.3cm} $F$: an initialized ML model;  
\end{flushleft}  
  \begin{algorithmic}[1]
    \Function{Training}{$(X_{tr}, Y_{tr})$, $\gamma$, $\alpha$, $F$}
    \State Generate high-entropy soft labels $y'$ given $\gamma$
    \State \begin{varwidth}[t]{\linewidth}
    Generate $Y'_{tr}$ from $Y_{tr}$ using $y'$, where $\forall (Y_{tr}[i], Y'_{tr}[i]) \in(Y_{tr}, Y'_{tr}), \text{argmax}(Y_{tr}[i])=\text{argmax}(Y'_{tr}[i])$
    \end{varwidth}
    \For{number of training epochs}
    	\State \begin{varwidth}[t]{\linewidth}Minimize (\ref{eq:objective-func}) using Stochastic Gradient Descent\end{varwidth}
    \EndFor
    \State \textbf{return} $F$
    \EndFunction \\

    \Function{Testing}{$F$, $x$}
    \State Generate $F(x)$
    \State Generate random uniform sample $x_{rand}$ and $F(x_{rand})$
    \State \begin{varwidth}[t]{\linewidth}Generate $F^{x_{rand}}(x)$ by replacing $F(x)$ with $F(x_{rand})$, where $\text{argsort}(F^{x_{rand}}(x))=\text{argsort}(F(x))$ /* top-\emph{k} labels unchanged */\end{varwidth}  
    \State \textbf{return} $F^{x_{rand}}(x)$
    \EndFunction

  \end{algorithmic}
    \caption{Training and testing phase of \sysname}
    \label{alg:overall}
\end{algorithm}
\vspace{-2mm}

%% file: evaluation.tex
\label{sec:eval}

\subsection{Experimental Setup}
\label{sec:setup}
\textbf{Datasets.}
We consider five common benchmark datasets, and we describe them below. 

\textbf{Purchase100}~\cite{shokri2017membership} includes 197,324 shopping records of customers, each with 600 binary features indicating whether a specific item is purchased. 
The goal is to predict the customer's shopping habits (100 different classes in total). 

\textbf{Texas100}~\cite{shokri2017membership} contains 67,330 hospital discharge records, each containing 6,170 binary features indicating whether the patient has a particular symptom or not. 
The data is divided into 100 classes, and the goal is to predict the treatment given the patient's symptoms.

\textbf{Location30}~\cite{shokri2017membership} contains the location ``check-in'' records of different individuals. It has 5,010 data records with 446 binary features, each of which corresponds to a certain loation type and indicates whether the individual has visited that particular location. The goal is to predict the user's geosocial type (30 classes in total).

\textbf{CIFAR100}~\cite{krizhevsky2009learning} is an image classification dataset that has 60,000 images in 100 object classes. Each image has a size of 32$\times$32$\times$3.

\textbf{CIFAR10}~\cite{krizhevsky2009learning} is similar to CIFAR100 that also contains 60,000 images but with 10 different object classes.

We follow \cite{shejwalkar2019membership} to use the fully-connected (FC) networks on Purchase100, Texas100 and Location30, and a DenseNet-12~\cite{huang2017densely} on CIFAR100 and CIFAR10 (Appendix~\ref{sec:diff-arch-eval} conducts evaluation  on more network architectures, including ResNet-18~\cite{he2016deep}, MobileNet~\cite{howard2017mobilenets} and ShuffleNet~\cite{zhang2018shufflenet}). 
Purchase100 is trained with 20,000 samples, Texas100 with 15,000 samples, Location30 with 1,500 samples, CIFAR100 and CIFAR10 are with 25,000 samples. Section~\ref{sec:more-train-size} reports additional experiments on more training sizes (from 2,500 to 50,000).

\textbf{Attacks.} 
We consider all nine attacks as in Section~\ref{sec:mia}. 
For NN-based attack, we use the black-box NSH attack from Nasr et al.~\cite{nasr2018comprehensive}, which uses the model loss, logit values from the target model, and the ground-truth label to train an attack inference model. 
We consider the loss-based attack from Yeom et al.~\cite{yeom2018privacy} and confidence-, entropy- and modified-entropy-based attacks as in Song et al.~\cite{song2021systematic}. 
For LiRA~\cite{carlini2022membership}, we train 128 shadow models for each defense (64 IN and OUT models each), where each shadow model is trained following the \emph{same} procedure as the targeted defense (as per our threat model).  
E.g., for \sysname, this means the shadow model is trained with the same high-entropy soft labels and the entropy-based regularization as the defense model, and the shadow model also performs the same output modification as \sysname does.

We consider the boundary and augmentation attacks from Choquette et al.~\cite{choquette2021label}. 
For the boundary attack on the two image datasets, we use the CW2 attack~\cite{carlini2017towards} to generate adversarial samples and derive the perturbation magnitude threshold to distinguish members and non-members.
Likewise, for the other three non-image datasets that contain binary features, we compute the sample's robustness to random noise instead of adversarial perturbation. 
For each sample $x$, we generate hundreds of noisy variants of $x$, and the number of correctly classified noisy variants of $x$ is used to determine a threshold that best distinguishes between members and non-members. 
For augmentation attack, we consider image translation as the augmentation method, and we similarly consider different degrees of translation to find the best attack.

\textbf{\sysname configuration.}
$\gamma, \alpha$ are the two parameters in configuring \sysname (for generating high-entropy soft labels and controlling the strength of regularization respectively). 
We perform grid search to select the parameters ($\gamma \in [0.5, 0.99], \alpha \in [0.0001, 0.5]$), and select the one with small train-validation gap and high validation accuracy. 
We also conduct evaluation to study how \sysname's performance varies under different parameters (please see Appendix~\ref{sec:tweak-hyperparameter}).

For the testing-time defense, we generate random samples (e.g., random pixels in [0, 255]) and perform output modification as in Section~\ref{sec:defense-testing}. There are no any other requirements. 

Our code is available at \url{https://github.com/DependableSystemsLab/MIA_defense_HAMP}.

\textbf{Related defenses.}
We consider seven major defenses: AdvReg~\cite{nasr2018machine}, MemGuard~\cite{jia2019memguard}, DMP~\cite{shejwalkar2019membership}, SELENA~\cite{tang2021mitigating}, Early stopping~\cite{song2021systematic,caruana2001overfitting}, Label Smoothing (LS)~\cite{szegedy2016rethinking} and DP-SGD~\cite{abadi2016deep}. 
We follow the original work to set up the defenses unless otherwise stated (more details in Appendix~\ref{sec:defense-setup}).

\textbf{Evaluation metrics}. 
An ideal privacy defense should provide strong protection for both members and non-members, for which we follow the best practice~\cite{carlini2022membership} to consider (1) \emph{attack true positive rate} (TPR) evaluated at 0.1\% false positive rate (FPR), which evaluates the protection for members, and (2) \emph{attack true negative rate} (TNR) at 0.1\% false negative rate (FNR), which quantifies the protection for non-members.

\begin{figure*}[!t]
    \centering
    \begin{subfigure}[b]{\textwidth}

		\minipage{0.31\textwidth}
		  \includegraphics[width=0.99\textwidth, height=1.7in]{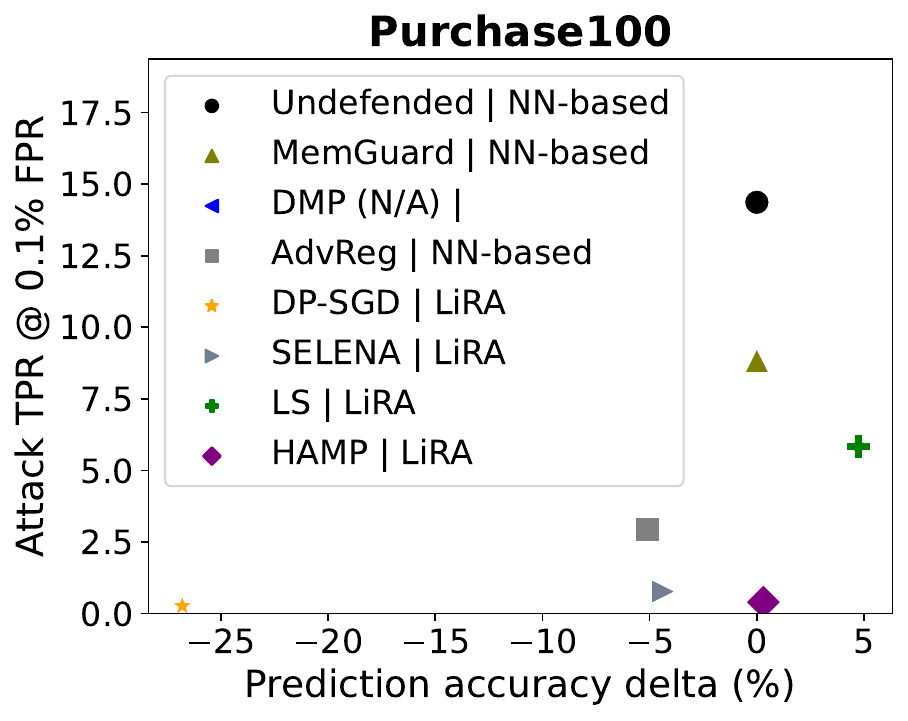}
		  %\subcaption{CIFAR100 (15k)}
		  \label{fig:c15}
		 \endminipage\hfill
		\minipage{0.31\textwidth}
		  \includegraphics[width=0.99\textwidth, height=1.7in]{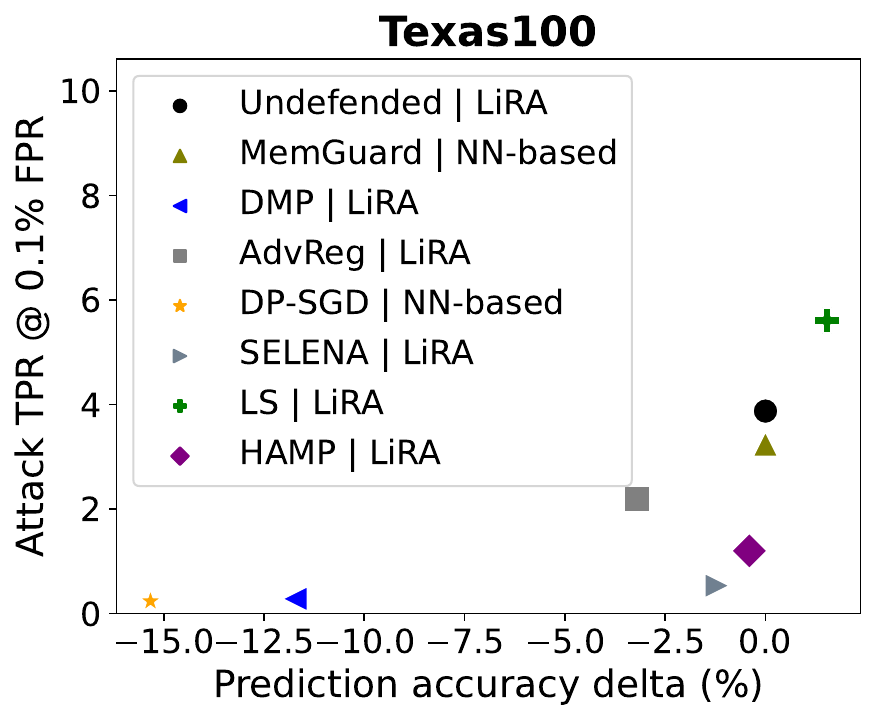}
		  %\subcaption{CIFAR100 (20k)}
		  \label{fig:c20}
		\endminipage\hfill
		\minipage{0.31\textwidth}%
		  \includegraphics[width=0.99\textwidth, height=1.7in]{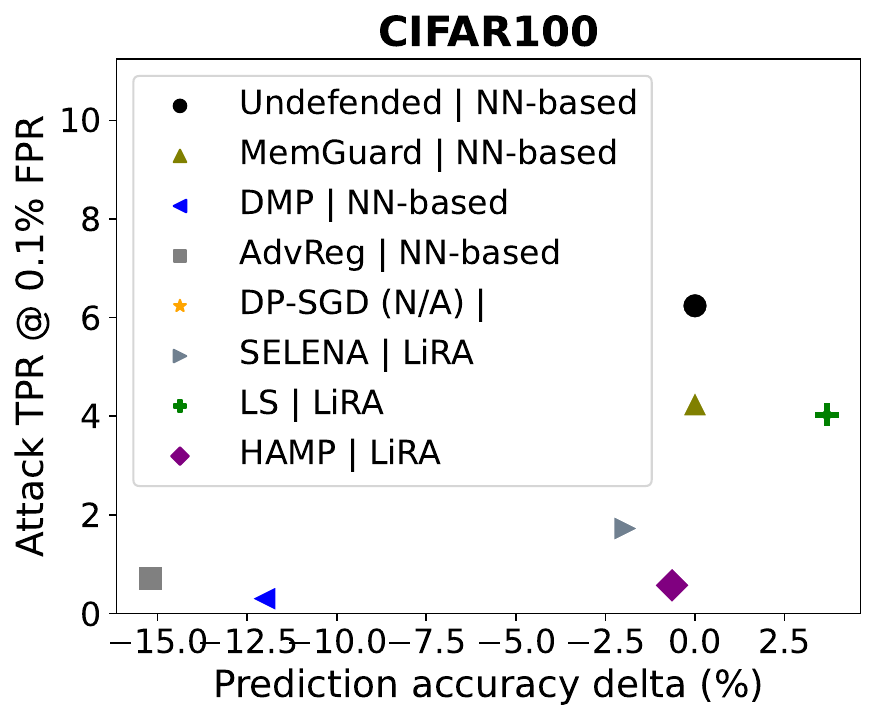}
		  %\subcaption{CIFAR100 (25k)} 
		  \label{fig:c25}
		\endminipage\hfill 

    \end{subfigure} 

    \begin{subfigure}[b]{\textwidth}
		\minipage{0.31\textwidth}
		  \includegraphics[width=0.99\textwidth, height=1.7in]{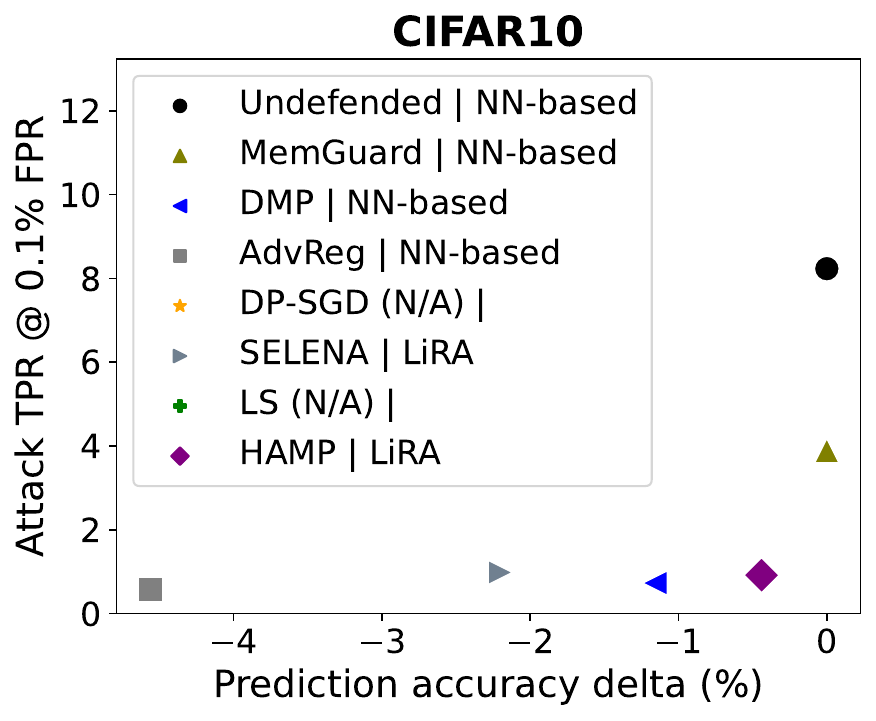}
		  %\subcaption{CIFAR10 (15k)}
		  \label{fig:cifar15}
		 \endminipage\hfill
		\minipage{0.31\textwidth}
		  \includegraphics[width=0.99\textwidth, height=1.7in]{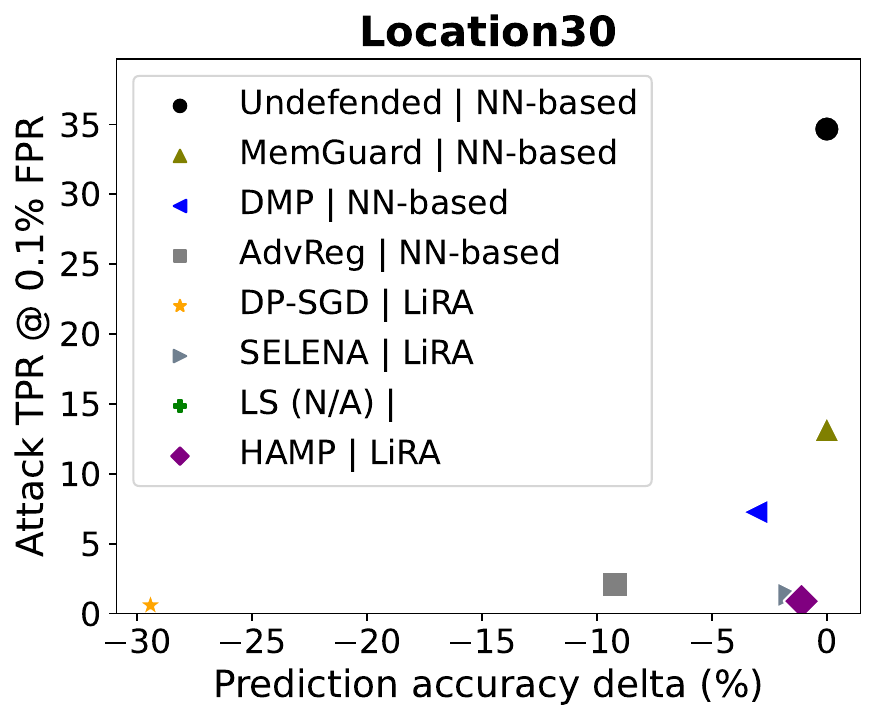}
		  %\subcaption{CIFAR10 (20k)}
		  \label{fig:cifar20}
		\endminipage\hfill
		\minipage{0.31\textwidth}%
		  \includegraphics[width=0.99\textwidth, height=1.7in]{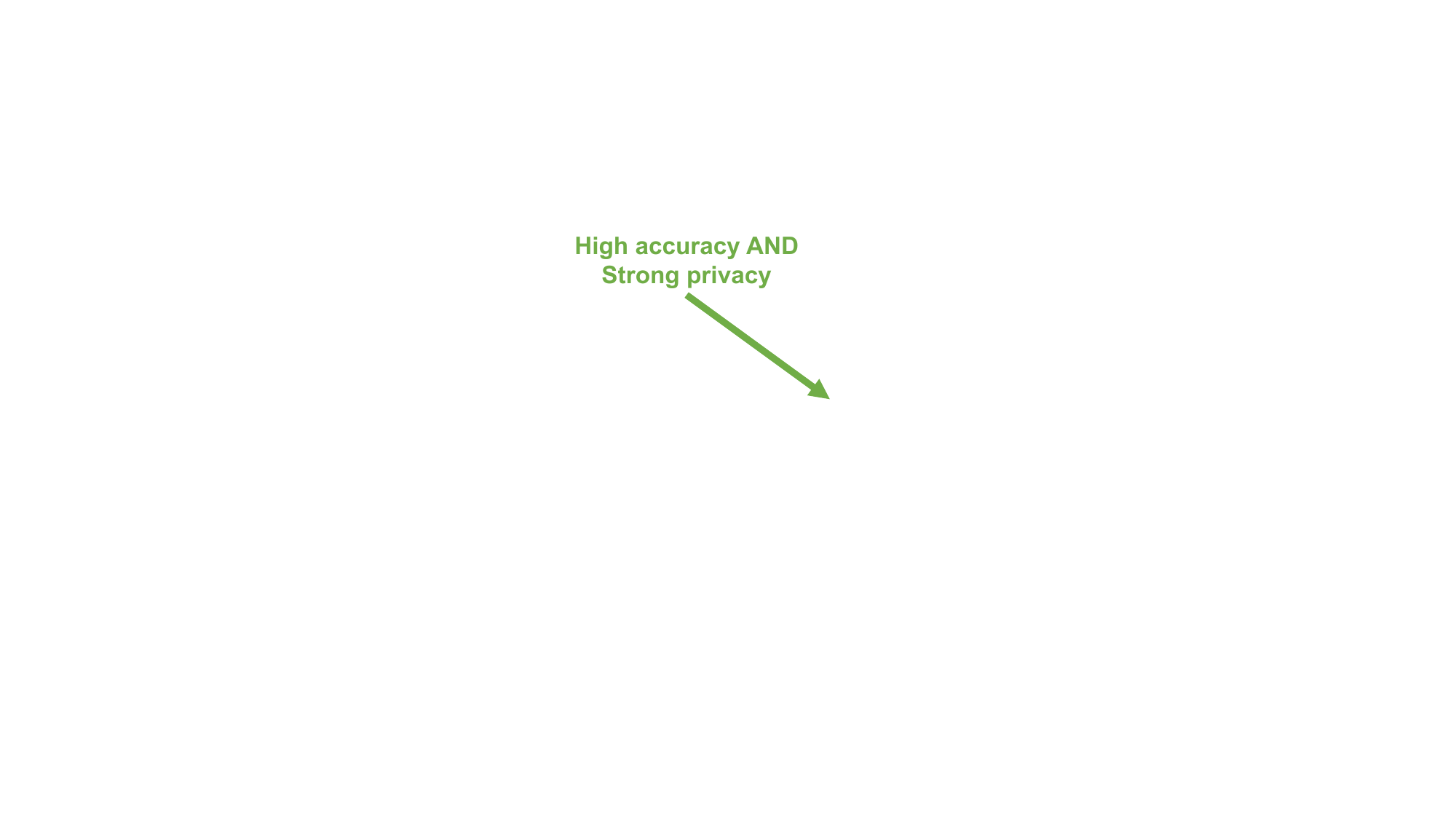}
		  %\subcaption{CIFAR10 (25k)} 
		  \label{fig:c25}
		\endminipage\hfill 

    \end{subfigure}  

    \par\noindent\rule{\textwidth}{0.5pt} 

    \begin{subfigure}[b]{\textwidth}

		\minipage{0.31\textwidth}
		  \includegraphics[width=0.99\textwidth, height=1.7in]{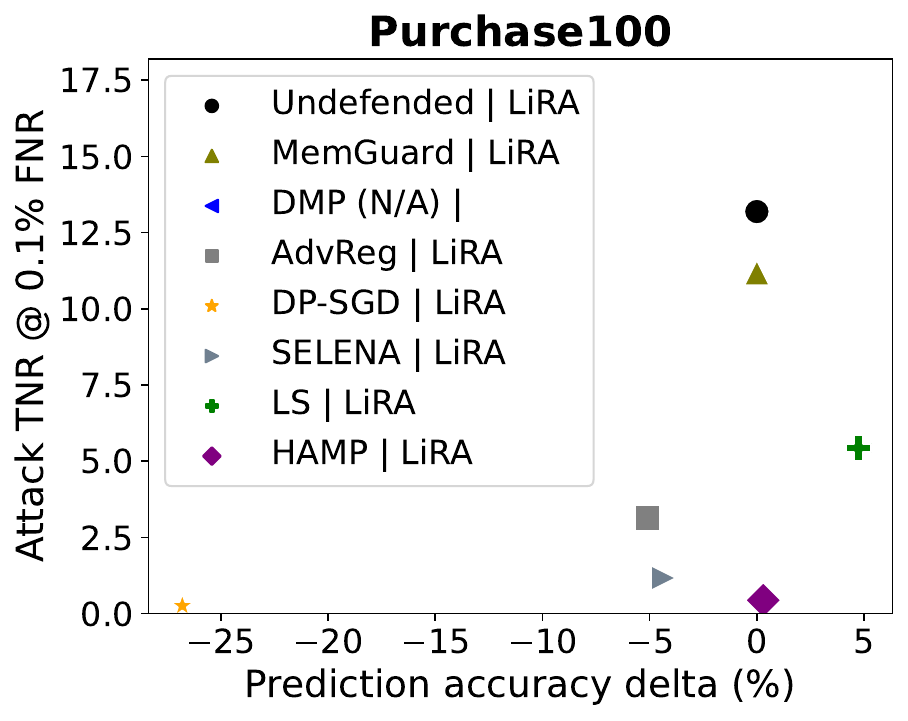}
		  %\subcaption{CIFAR100 (15k)}
		  \label{fig:c15}
		 \endminipage\hfill
		\minipage{0.31\textwidth}
		  \includegraphics[width=0.99\textwidth, height=1.7in]{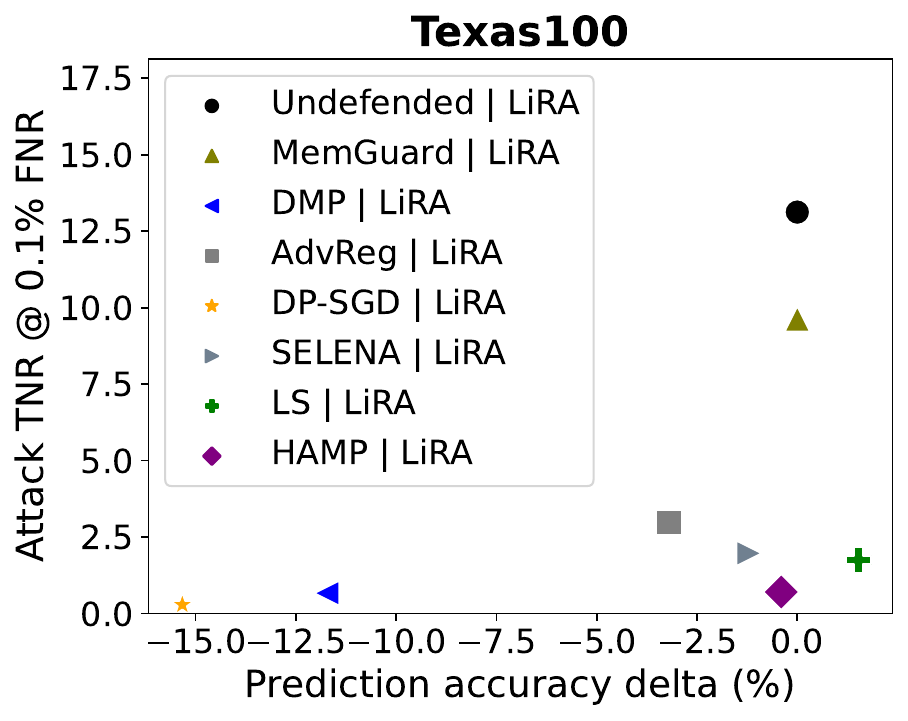}
		  %\subcaption{CIFAR100 (20k)}
		  \label{fig:c20}
		\endminipage\hfill
		\minipage{0.31\textwidth}%
		  \includegraphics[width=0.99\textwidth, height=1.7in]{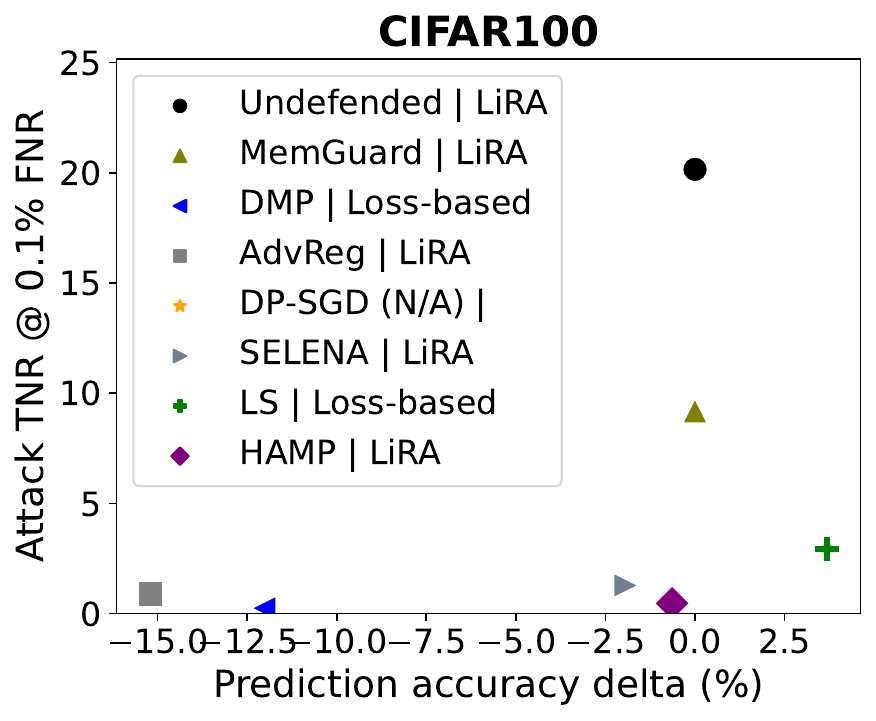}
		  %\subcaption{CIFAR100 (25k)} 
		  \label{fig:c25}
		\endminipage\hfill 

    \end{subfigure} 

    \begin{subfigure}[b]{\textwidth}

		\minipage{0.31\textwidth}
		  \includegraphics[width=0.99\textwidth, height=1.7in]{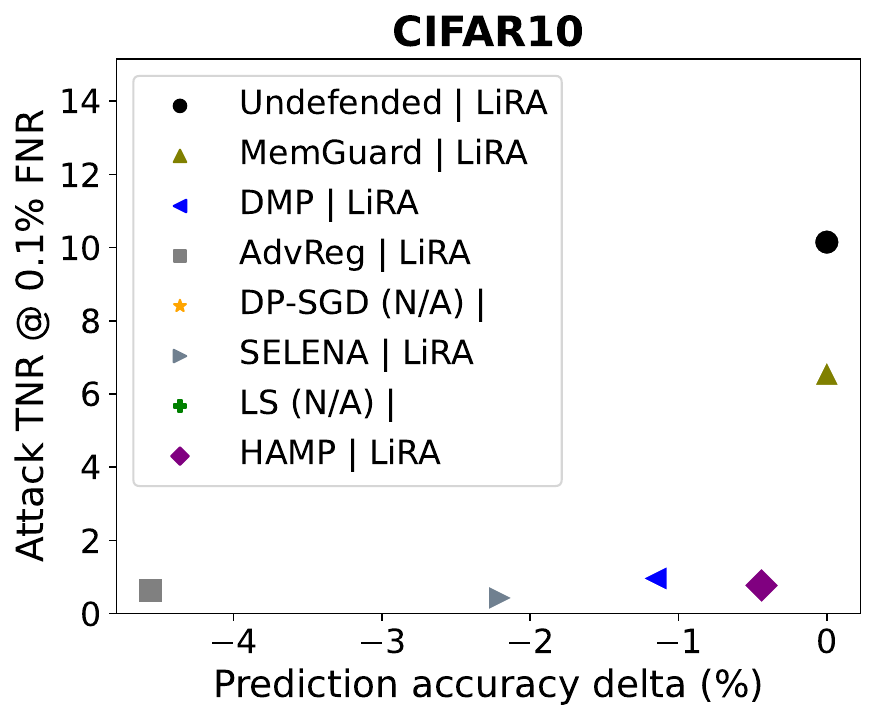}
		  %\subcaption{CIFAR10 (15k)}
		  \label{fig:cifar15}
		 \endminipage\hfill
		\minipage{0.31\textwidth}
		  \includegraphics[width=0.99\textwidth, height=1.7in]{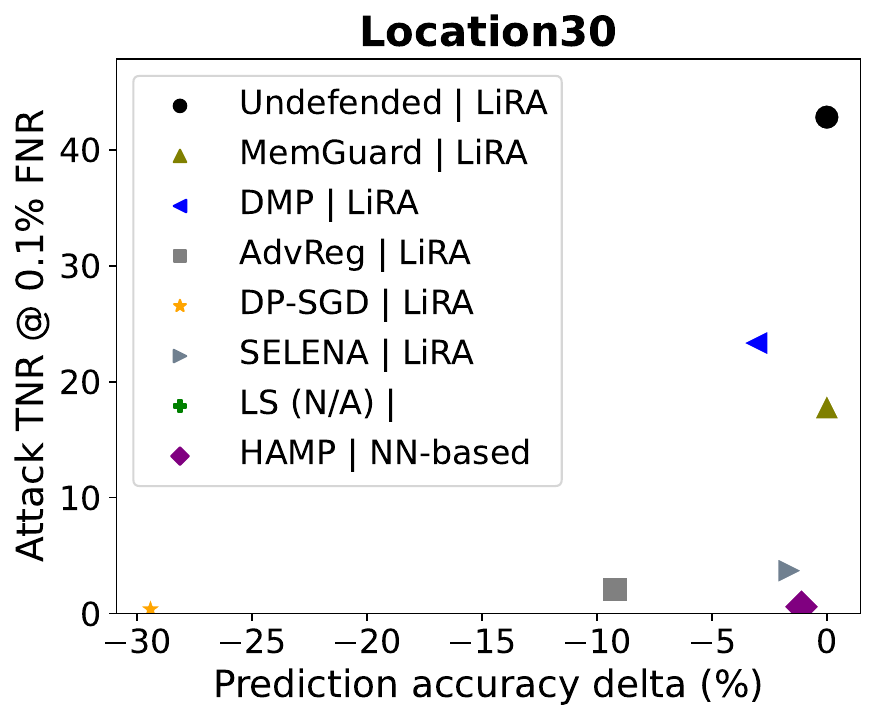}
		  %\subcaption{CIFAR10 (20k)}
		  \label{fig:cifar20}
		\endminipage\hfill
		\minipage{0.31\textwidth}%
		  \includegraphics[width=0.99\textwidth, height=1.7in]{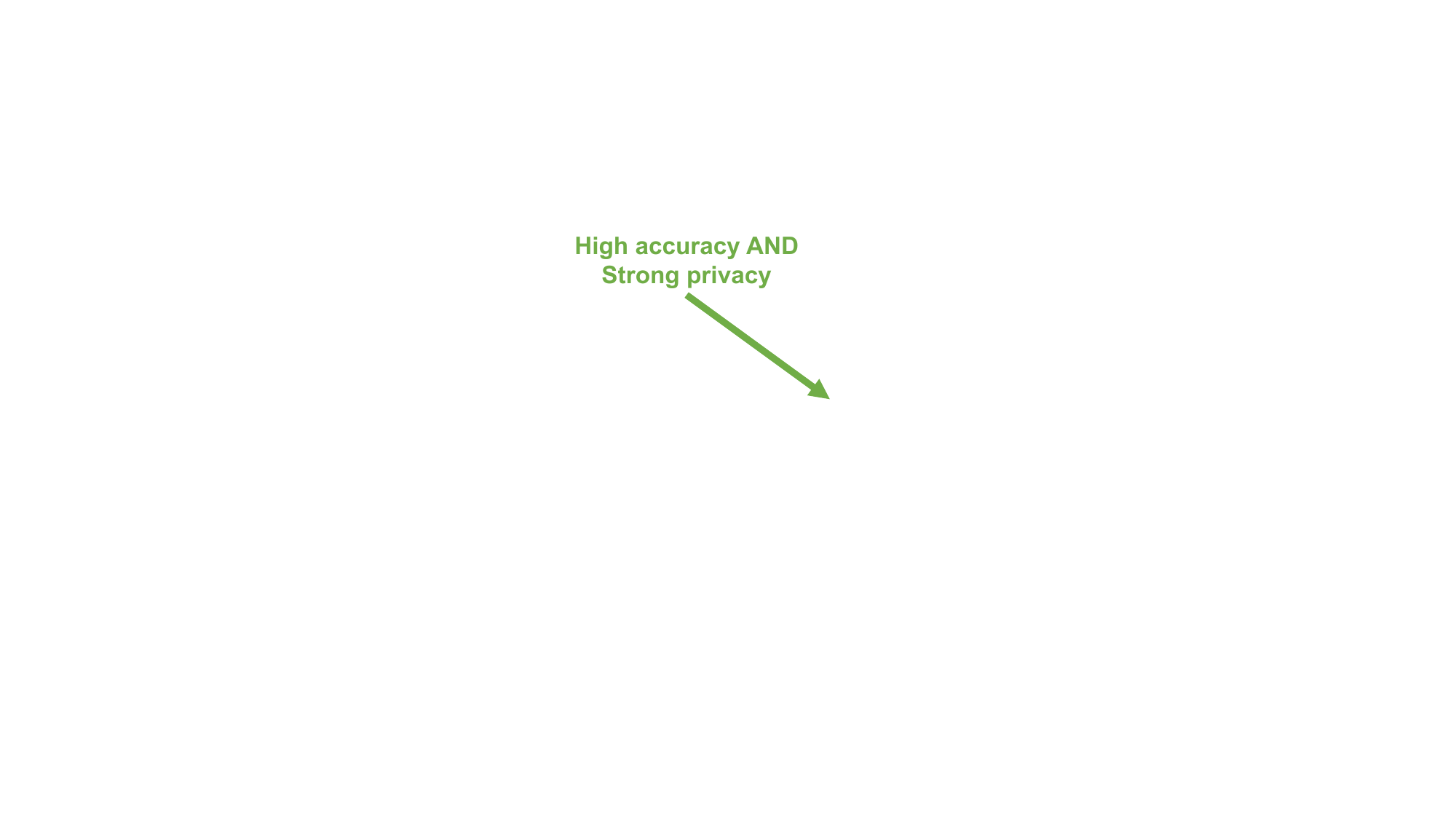}
		  %\subcaption{CIFAR10 (25k)} 
		  \label{fig:c25}
		\endminipage\hfill 

    \end{subfigure} 

    \caption{\textbf{Attack TPR @ 0.1\% FPR} (first two rows) and \textbf{Attack TNR @ 0.1\% FNR} (last two rows) on different datasets. The legend indicates the attack that yields the highest attack TPR/TNR. 
    Negative {prediction accuracy delta} means accuracy drop compared with the undefended models. 
    DP-SGD is reported at $\epsilon=4$, and it is not evaluated on CIFAR100 and CIFAR10 due to its significant accuracy drop (similar case as DMP on Purchase100). 
    LS is not evaluated on CIFAR10 and Location30 as LS did not lead to accuracy improvement. 
    \emph{Overall, \sysname offers strong privacy protection for both members and non-members, while preserving high model accuracy, thereby yielding a superior privacy-utility trade off over other defenses.  }
    }
    \label{fig:tpr-01}
      \vspace{-4mm}
\end{figure*}

\textbf{Result organization.} 
Table~\ref{tab:acy} reports the model accuracy for every defense. 
Fig.~\ref{fig:tpr-01} compares each defense in terms of their membership privacy and model utility. 
Each defense is evaluated with multiple attacks, and we report the ones that achieve the highest attack TPR or TNR (detailed results for each attack are in Appendix~\ref{sec:all-atk-res}). 
Fig.~\ref{fig:auc} presents the average attack AUC (area under curve) by each defense, and the full ROC curves are in Appendix~\ref{sec:roc}. 
We leave the comparison with early stopping in Appendix~\ref{sec:comparison-early-stop} due to space constraint. 
Section~\ref{sec:ablation-study} presents an ablation study, and Appendix~\ref{sec:overhead} reports training and inference overhead evaluation.  
We next discuss the results by comparing \sysname with other defenses. % (including the undefended models).  

\begin{table}[t]
\caption{Model accuracy for each defense. Accuracy delta measures the accuracy difference with the undefended model. 
 }
\label{tab:acy}
\centering  
\footnotesize
\renewcommand{\arraystretch}{1.1}
\begin{tabular*}{\columnwidth}{llllc}
\hline
 {Dataset} & Defense & Training acy & Testing Acy & Acy delta \\
\hline
\hline
\multirow{7}{5em}{Purchase100  } 	& Undefended & 99.36 & 80.85 & 0.00 \\
									& MemGuard & 99.36 & 80.85 & 0.00 \\
									& AdvReg & 93.97 & 75.75 & -5.10 \\
									& DPSGD & 61.06 & 54.05 & -26.80 \\
									& LS & 99.54 & 85.60 & +4.75 \\
									& SELENA & 85.19 & 76.50 & -4.35 \\ 
									& HAMP & 91.12 & 81.15 & +0.30 \\
\hline
\multirow{8}{5em}{CIFAR100  }		& Undefended & 86.21 & 59.56 & 0.00 \\
									& MemGuard & 86.21 & 59.56 & 0.00 \\
									& AdvReg & 55.78 & 44.36 & -15.20 \\ 
									& DMP & 53.37 & 47.52 & -12.04 \\
									& LS & 88.80 & 63.24 & +3.68 \\
									& SELENA & 62.15 & 57.64 & -1.92 \\ 
									& HAMP & 68.44 & 58.92 &  -0.64  \\
\hline 
\multirow{8}{5em}{Location30  }		& Undefended & 99.56 & 57.40 & 0.00 \\
									& MemGuard & 99.56 & 57.40 & 0.00 \\
									& AdvReg & 69.70 & 48.20 & -9.20 \\
									& DPSGD & 36.37 & 28.00 & -29.40 \\
									& DMP & 92.81 & 54.30 & -3.10 \\
									%& LS & 99.93 & 55.90 & -1.50 \\
									& SELENA & 67.41 & 55.80 & -1.60 \\ 
									& HAMP & 78.22 & 56.30 &  -1.10  \\
\hline
\multirow{8}{5em}{CIFAR10  }		& Undefended & 98.72 & 86.72 & 0.00 \\
									& MemGuard & 98.72 & 86.72 & 0.00 \\
									& AdvReg & 86.73 & 82.16 & -4.56 \\ 
									& DMP & 91.08 & 85.56 & -1.16 \\
									%& LS & 94.68 & 86.44 & -0.28 \\
									& SELENA & 86.86 & 84.52 & -2.20 \\ 
									& HAMP & 95.88 & 86.28 &    -0.44 \\
\hline
\multirow{8}{5em}{Texas100  }		& Undefended & 76.79 & 54.80 & 0.00 \\
									& MemGuard & 76.79 & 54.80 & 0.00 \\
									& AdvReg & 62.76 & 51.60 & -3.20 \\
									& DPSGD & 43.08 & 39.47 & -15.33 \\
									& DMP & 46.92 & 43.07 & -11.73 \\
									& LS & 75.52 & 56.33 & +1.53 \\
									& SELENA & 58.58 & 53.60 & -1.20 \\ 
									& HAMP & 68.56 & 54.40 &  -0.40  \\
\hline
\hline
\multirow{4}{5em}{\emph{Average accuracy delta} } 	& Undefended & 0.00 & MemGuard & 0.00 \\
											&AdvReg & -7.45 & DPSGD & -23.84 \\
											& LS & +2.42 &DMP & -7.01  \\
											&SELENA & -2.25 &\sysname & -0.46\\

\hline

\end{tabular*}  
\vspace{-4mm}
\end{table}

\begin{figure}[t]
    \centering
    \includegraphics[width=2.in, height=1.6in]{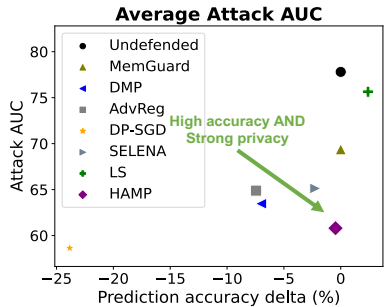}
    \caption{Average attack AUC by each defense (detailed results for each dataset can be found in Appendix~\ref{sec:auc}).} 
    \label{fig:auc} 
  \vspace{-6mm}
\end{figure}

\subsection{Comparison with Undefended Models}
\label{sec:comparison-undefened} 
\textbf{\sysname significantly reduces the MIA risk against both members and non-members. } 
Compared with the undefended models, \sysname achieves significantly lower attack TPR and TNR. 
The average attack TPR on the undefended model is 13.48\%, which is reduced to 0.8\% by \sysname (a 94.1\% reduction). 
Similarly, \sysname reduces the attack TNR by 97\%, from 19.89\% to 0.59\%. 
This effectively thwarts the adversary in inferring members or non-members from the target model.

In addition, we find that NN-based attack yields the highest attack TPR on the undefended models in many cases (as in Fig.~\ref{fig:tpr-01}), and we explain the reason in Appendix~\ref{sec:explain-nn-based}. 

\textbf{\sysname achieves strong membership privacy while preserving high model accuracy.} 
Across the five diverse datasets, \sysname is able to consistently produce models with comparable accuracy as the undefended models. 
\sysname has an accuracy drop of at most 1.1\% (on Location30), and the average accuracy drop by \sysname is only 0.46\%.

\subsection{{Comparison with MemGuard~\cite{jia2019memguard}}}
\label{sec:comparison-memguard}
\textbf{Both MemGuard and \sysname are capable of preserving model accuracy}. 
MemGuard does not incur any accuracy drop since it is a post-processing technique, and does not change the prediction label. 
Likewise, \sysname only incurs a minor accuracy drop of 0.46\%. 

\textbf{\sysname achieves considerably stronger membership privacy than MemGuard.} 
MemGuard offers very limited privacy protection because MemGuard only modifies the output scores without changing the prediction labels, which cannot prevent privacy leakage from the label information. 
On the contrary, \sysname consists of a training-time defense that can mitigate membership leakage from both output scores and label information (explained in Section~\ref{sec:overview}), and achieves much stronger membership privacy than MemGuard. 
The average attack TPR on MemGuard is 6.7\%, which is 8.4x relative to that of \sysname. 
Similarly, the attack TNR by MemGuard is 10.9\%, which is 18.3x relative to that of \sysname.

\subsection{Comparison with AdvReg~\cite{nasr2018machine}}
\label{sec:comparison-advreg} 
\textbf{\sysname outperforms AdvReg with higher model accuracy and stronger membership privacy}. 
In terms of accuracy, \sysname consistently achieves higher accuracy than AdvReg. 
AdvReg incurs an average 7.45\% accuracy drop, while \sysname incurs only 0.46\% (94\% lower than AdvReg). 

In terms of privacy, \sysname outperforms AdvReg with both much lower attack TPR and TNR. 
The attack TPR is 1.70\% with AdvReg and 0.8\% with  \sysname, which translate to a 87\% and 94\% reduction from those of the undefended models. 
Similarly, AdvReg reduces the attack TNR by 90\% while \sysname reduces it by 97\%, which is much higher.

\subsection{{Comparison with DMP~\cite{shejwalkar2019membership}}}
\label{sec:comparison-dmp} 
DMP~\cite{shejwalkar2019membership} uses generative adversarial networks (GANs) trained on the private dataset to produce synthetic data as the reference set for knowledge distilation. 
We follow Shejwalker et al. \cite{shejwalkar2019membership} to train the two image datasets on DC-GAN~\cite{radford2015unsupervised}. 
The defender can generate unlimited data from the GAN, and hence he/she can create a reference set that is larger than the original training set. 
Therefore, we use 150K synthetic samples to train the model with higher accuracy (we do not consider more synthetic images as the improvement is negligible).

For the three datasets with binary features, we use CTGAN~\cite{xu2019modeling} for modeling tabular data. 
We use 100K synthetic samples for Texas100, 10k for Location30. We do not consider Purchase100 as it incurs significant accuracy drop (over 30\%).
To validate that synthetic samples are useful for the domain task, we compare the performance of the models trained with GAN-generated synthetic data and those with {random} data (i.e., all features are randomly selected as 0 or 1 with equal probability) using Texas100. 
We find that models trained with random data only achieve accuracy from 5.8\% to 14.8\%; while those with GAN-generated data achieve over 40\% accuracy.

\textbf{\sysname outperforms DMP by being able to consistently achieve strong privacy protection with high model accuracy across different datasets}. 
%On average, DMP yields an attack TPR of 2.13\% and an attack TNR of 6.13\% vs. 0.89\% and 0.63\% by \sysname. 
In terms of membership privacy, we find that  DMP is  able to achieve strong results in many (but not all) cases, and it achieves an average attack TPR of 0.44\%  and TNR of 0.38\% on Texas100, CIFAR100 and CIFAR10, where \sysname achieves 0.9\% TPR and 0.65\% TNR (DMP is slightly better). 
However, DMP's performance does not generalize across datasets. 
For instance, on Location30, DMP suffers from a much higher attack TPR of 7.26\% and TNR of 23.33\%. 
This is because the model is trained with limited data (1,500), and the GAN is \emph{not} able to generate diverse data that are different from the original training data. As a result, the teacher model assigns high confidence to the synthetic data, from which the student model learns to predict the training members with high confidence that eventually leads to high MIA risk. 
To validate this, we compare the difference between the prediction confidence on members and non-members by the DMP models. On Location30, the average difference is $>$30\%, and only $<$5\% on the other datasets, which is why DMP exhibits poor privacy protection on Location30. 
On the same dataset, \sysname yields a low TPR of 0.89\% and TNR of 0.59\%, and this trend is consistent across datasets.

In terms of accuracy, DMP suffers from different degrees of accuracy loss that are much higher than \sysname's. 
DMP incurs $>$30\% accuracy loss on Purchase100 (as mentioned earlier),  $\sim$12\% accuracy drop on Texas100 and CIFAR100, 3.1\% on Location30, and 1.2\% on CIFAR10 (smaller accuracy loss as CIFAR10 has 10 classes only). 
In contrast, \sysname incurs average accuracy drop of $<$0.5\% (at most 1.1\%), which is significantly better than DMP.

\subsection{{Comparison with SELENA~\cite{tang2021mitigating}}}
\label{sec:comparison-selena}
\textbf{Both SELENA and \sysname achieve similarly strong privacy protection}. 
On average, \sysname has a slightly better membership privacy than SELENA, but neither technique has consistently better membership privacy overall (Fig.~\ref{fig:tpr-01}). 
The attack TPR of SELENA is $0.53\%\sim1.72\%$, with an average of 1.1\%, and that of \sysname is $0.4\%\sim1.2\%$, with an average of 0.8\%. 
They are able to reduce the attack TPR by 92\% (SELENA) and by 94\% (\sysname). 
In addition, the attack TNR of SELENA is $0.42\%\sim3.7\%$, with an average of 1.7\%, and that of \sysname is $0.44\%\sim0.77\%$, with an average of 0.6\%. 
This translates to a TNR reduction of 91\% (SELENA) and 97\% (\sysname), respectively.  

\textbf{While providing comparable privacy benefits, \sysname outperforms SELENA by having lower accuracy loss, hence providing a better privacy-utility trade off}.  
The largest accuracy drop by SELENA is 4.4\% and that by \sysname is only 1.1\%. 
On average, SELENA incurs a 2.25\% accuracy drop, while  \sysname incurs a much smaller drop of 0.46\%.  
Moreover, our additional experiment in Section~\ref{sec:more-train-size} shows that \sysname continues to outperform SELENA with much lower accuracy drop when evaluated on a variety of different training sizes (2.2\%$\sim$5.2\% by SELENA and 0.04\%$\sim$0.98\% by \sysname).

\subsection{{Comparison with Label Smoothing (LS)~\cite{szegedy2016rethinking}}}
\label{sec:comparison-ls} 
\textbf{Though LS is able to improve model accuracy, the model trained with LS still suffers from \emph{high} MIA risk. 
In contrast, the model trained with \sysname can maintain high model accuracy and exhibit very \emph{low} MIA risk}. 
For LS, we follow prior work by Kaya  et al. ~\cite{kaya2021does} to train with different smoothing intensities from 0.01 to 0.995, and select the model with the highest accuracy (we omit CIFAR10 and Location30 as LS did not lead to accuracy improvement).  
We first discuss the qualitative difference between LS and \sysname, and then quantitatively compare their privacy risk.

While LS and \sysname use soft labels in their training, they are built with different purposes that require different soft labels. 
LS is used as a regularization technique to improve model accuracy, which necessitates training with \emph{low}-entropy soft labels, and is able to increase the accuracy by 2.4\% on average. 
However, the resulting model still suffers from high MIA risk, as LS causes the model to overfit on the smooth labels and exhibit discernible behavior on the training samples~\cite{kaya2021does}. 
In contrast, \sysname is built to improve membership privacy, which consists of \emph{high}-entropy soft labels, an entropy-based regularizer and a novel testing-timd defense to force the model to make less confident predictions, and to behave similarly on the training and testing samples. 

To quantitatively compare the different soft labels used by both techniques,  we measure the soft label entropy in LS and \sysname, and find that the label entropy in \sysname is considerably higher than that in LS, and is 4x$\sim$50x relative to that in LS (average 9x). 
This contributes to the {low} membership privacy risk by \sysname, unlike LS. 

The average attack TPR on the LS models is 5.1\%, 7.1x relative to that by \sysname (on the same datasets). 
The attack TNR on LS is 6.3x relative to that by \sysname (we observe a similar trend even when we train LS with other smoothing intensities that have comparable accuracy improvement - see  Appendix~\ref{sec:ls-diff-smooth-intensities}). 
Moreover, our results reveal that LS may {amplify} the MIA risk and render the model \emph{more vulnerable} than the undefended model. 
On Texas100, LS increases the attack TPR from 3.87\% (on the undefended model) to 5.61\%, which increases the MIA risk against training members by 45\%. 
This suggests that LS may constitute a hidden privacy risk for the practitioners (a similar finding was identified recently by Kaya et al.~\cite{kaya2021does}). 
On the contrary, \sysname consistently leads to low MIA risk and outperforms LS with significantly better membership privacy.

\subsection{Comparison with DP-SGD~\cite{abadi2016deep}}
\label{sec:comparion-dpsgd} 
We use the canonical implementation of DP-SGD using Pytorch Opacus~\cite{opacus}. 
We first consider a fixed privacy budget $\epsilon=4$ as per Tang et al. \cite{tang2021mitigating}, and then evaluate DP-SGD with different values of $\epsilon$. 

\subsubsection{DP-SGD with fixed $\epsilon=4$.} 
In this setting, the average attack TPR of the DP-SGD models is 0.36\% and 0.3\%, both of which are the lowest among all the defenses we evaluated. In comparison, \sysname yields 0.8\% attack TPR and 0.6\% TNR, which are slightly higher than DP-SGD. 
However, DP-SGD suffers from considerable accuracy loss, with an average loss of 23.84\%, while \sysname a significantly smaller loss of 0.46\%.

\begin{figure}[t]
    \centering
		  \includegraphics[width=3.5in, height=1.5in]{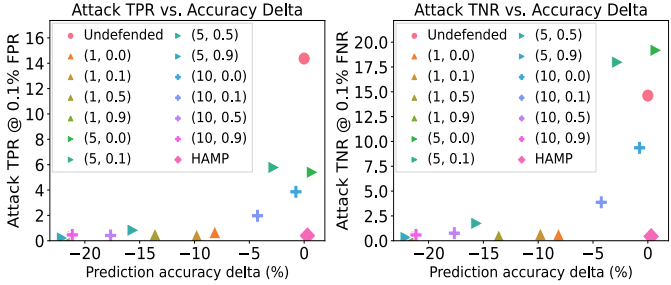}
    \caption{Results on DP-SGD under different clipping norms $\in [1, 5, 10]$, and noise\_multipliers $\in [0.0, 0.1, 0.5, 0.9]$. }
    \label{fig:dp-more-epsilons}  
  \vspace{-4mm}
\end{figure}

\subsubsection{DP-SGD with different $\epsilon$.} 
We next evaluate DP-SGD by considering different noise\_multipliers and clipping norms. 
We consider Purchase100, on which we used a noise\_multiplier of 1.7 and a clipping norm of 1, for $\epsilon=4$ in the earlier evaluation. 
We select different noise\_multiplier values of 0.0 (no noise injected), 0.1 ($\epsilon=12069.1$), 0.5 ($\epsilon=62.5$) and 0.9 ($\epsilon=10.9$); and clipping norm values of 1, 5 and 10, totalling 12 different configurations. 
We report the results in Fig.~\ref{fig:dp-more-epsilons}.

Reducing the amount of injected noise and using a larger clipping norm allows DP-SGD to provide empirical privacy protection (but with a very large provable bound of  $\epsilon$), and reduce the amount of accuracy loss. 
For instance, by using a clipping norm of 10  {\em without} injecting any noise, DP-SGD is able to reduce the accuracy loss to be $<$1\%, which can also reduce the attack TPR by 73\% (from 14.37\% to 3.86\%), and the attack TNR by 36\% (from 14.62\% to 9.36\%). 
Nevertheless, this performance is still considerably inferior to that of \sysname, which can reduce the attack TPR and TNR by 97.2\% and 96.7\%, respectively. 

Using a tighter clipping norm or injecting more noise can improve the membership privacy even more, but this comes at the cost of accuracy loss (the earlier result has negligible accuracy loss).  
For example, by using a small clipping norm of 1, the attack TPR can be reduced to 0.67\% and attack TNR to 0.62\%.
However, this results in 8.2\% accuracy loss. 
Increasing the noise\_multiplier can further reduce privacy leakage, e.g., using a noise\_multiplier value of 0.5 can reduce the attack TPR to 0.5\% and attack TNR to 0.49\% (and with a large $\epsilon$ of 62.5), which are comparable to the 0.4\% TPR and 0.44\% TNR values by \sysname. 
However, DP-SGD degrades the accuracy by 13.6\%,  while \sysname incurs negligible accuracy drop. 

Therefore, training a model with a small amount of noise or with a tight clipping norm  is also a viable defense against MIAs, though it still incurs much larger accuracy loss than \sysname and results in large provable bounds $\epsilon$.

\begin{figure*}[!t]

  \begin{subfigure}[b]{\textwidth}

  \minipage{0.31\textwidth}
    \includegraphics[width=0.95\textwidth, height=1.7in]{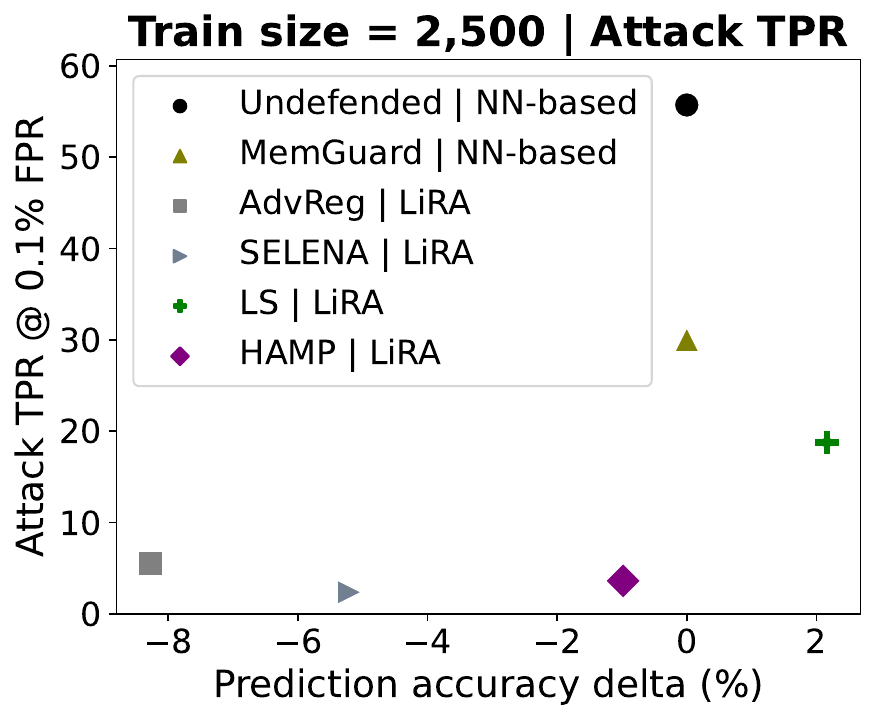}
    %subcaption{Purchase100}% (20k)}
   \endminipage\hfill
  \minipage{0.31\textwidth}
    \includegraphics[width=0.95\textwidth, height=1.7in]{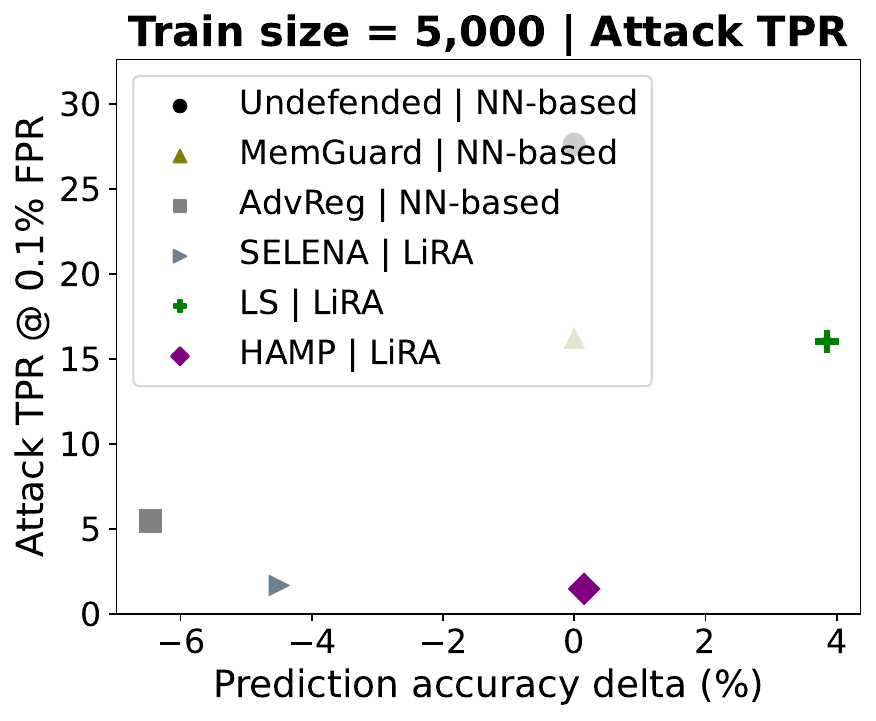}
    %subcaption{Texas100}% (15k)}
  \endminipage\hfill
  \minipage{0.31\textwidth}%
    \includegraphics[width=0.95\textwidth, height=1.7in]{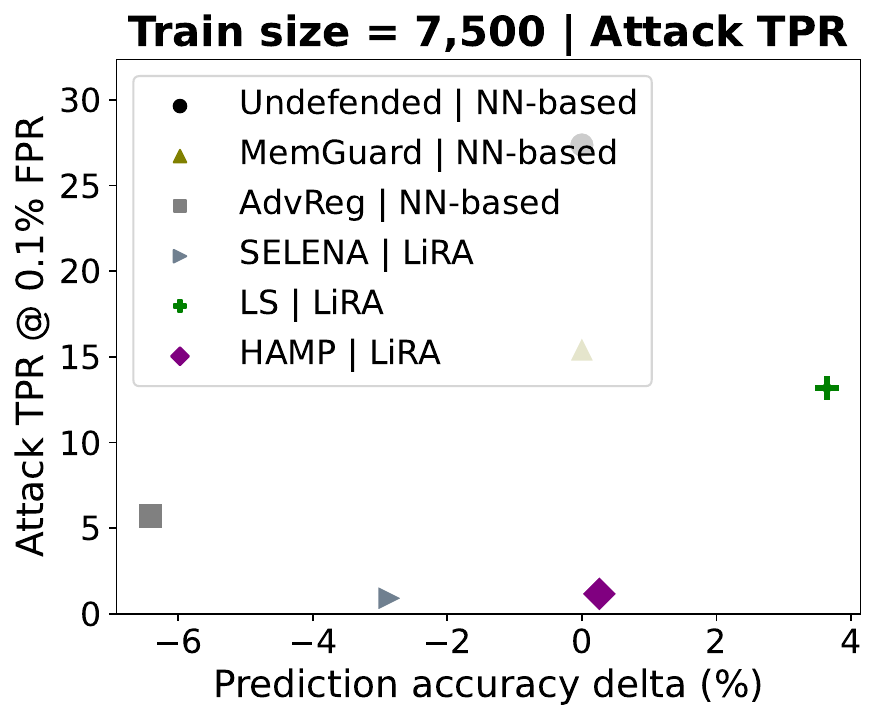}
    %subcaption{CIFAR100}% (20k)} 
    \label{fig:awesome_image3}
  \endminipage\hfill 
  \end{subfigure} 

  \begin{subfigure}[b]{\textwidth}

  \minipage{0.31\textwidth}%
    \includegraphics[width=0.95\textwidth, height=1.7in]{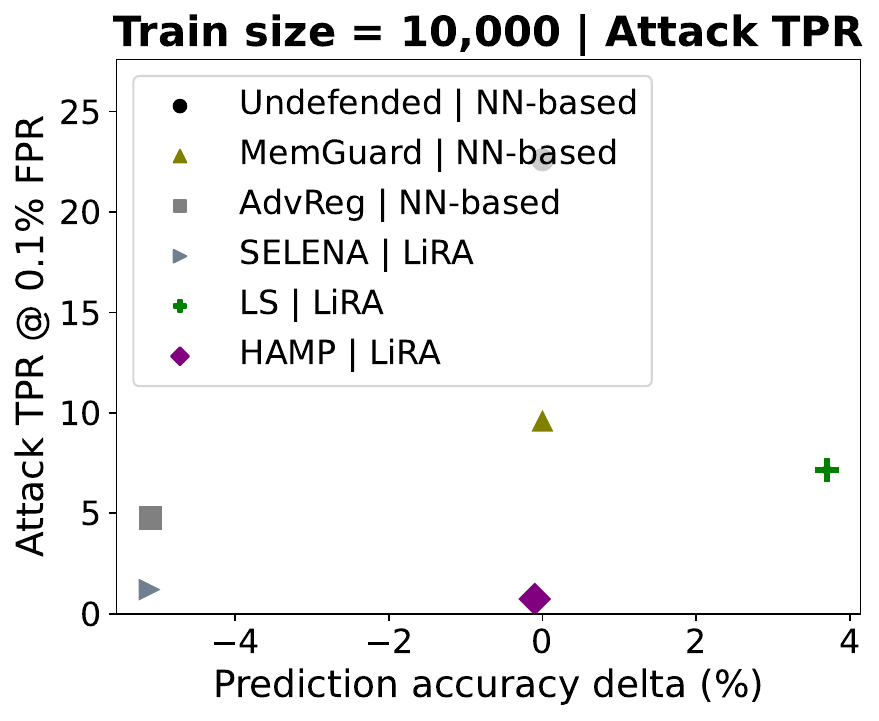}
    %subcaption{CIFAR10}% (20k)} 
    \label{fig:awesome_image4}
  \endminipage\hfill 
  \minipage{0.31\textwidth}%
    \includegraphics[width=0.95\textwidth, height=1.7in]{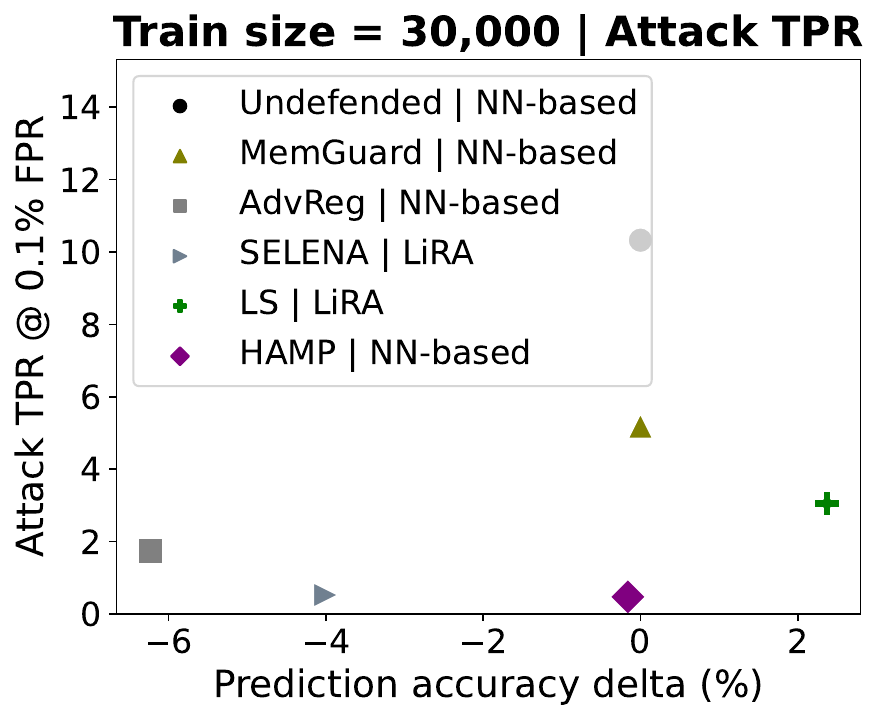}
    %subcaption{CIFAR10}% (20k)} 
    \label{fig:awesome_image4}
  \endminipage\hfill 
  \minipage{0.31\textwidth}%
    \includegraphics[width=0.95\textwidth, height=1.7in]{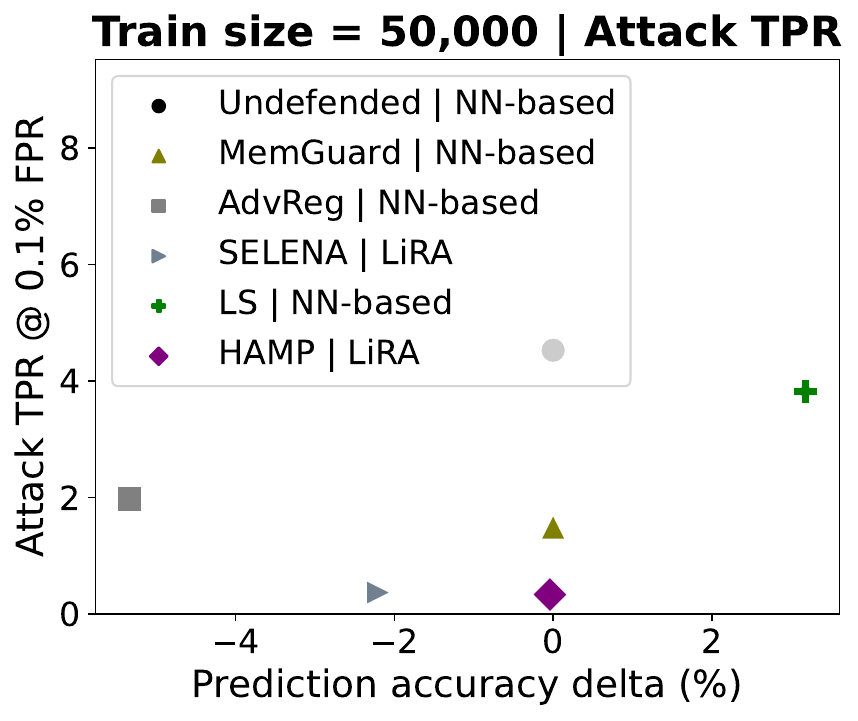}
    %subcaption{Location30}% (20k)} 
    \label{fig:awesome_image4}
  \endminipage\hfill 

  \end{subfigure}

      \par\noindent\rule{\textwidth}{0.5pt}

  \begin{subfigure}[b]{\textwidth}

  \minipage{0.31\textwidth}
    \includegraphics[width=0.95\textwidth, height=1.7in]{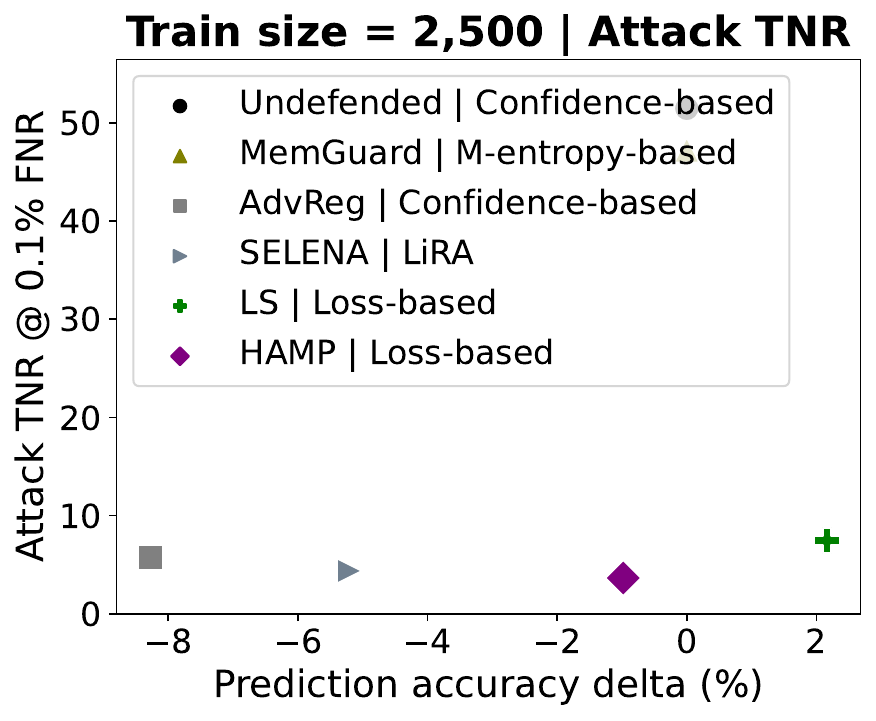}
    %subcaption{Purchase100}% (20k)}
   \endminipage\hfill
  \minipage{0.31\textwidth}
    \includegraphics[width=0.95\textwidth, height=1.7in]{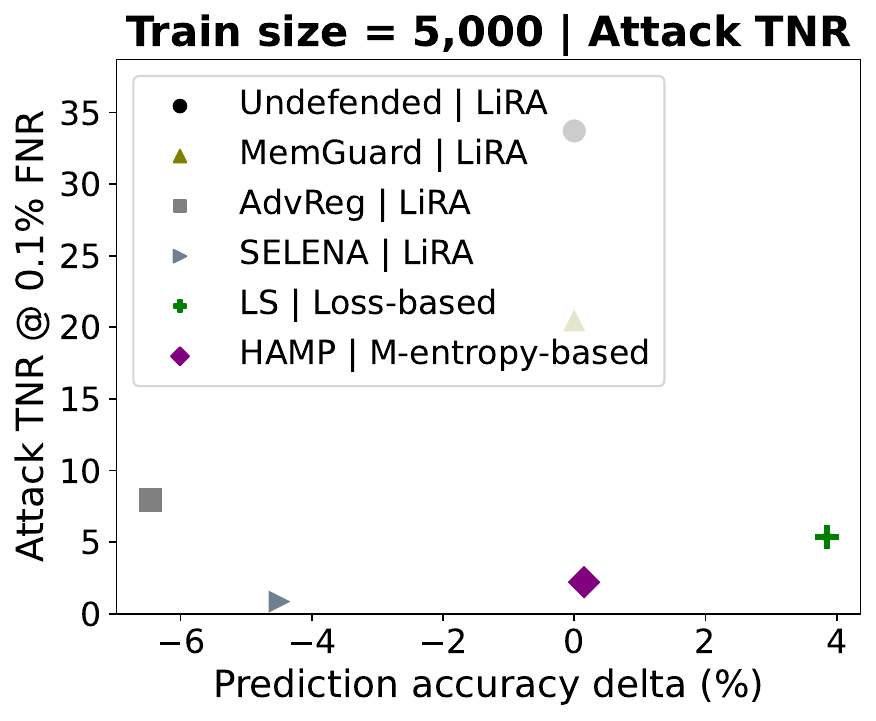}
    %subcaption{Texas100}% (15k)}
  \endminipage\hfill
  \minipage{0.31\textwidth}%
    \includegraphics[width=0.95\textwidth, height=1.7in]{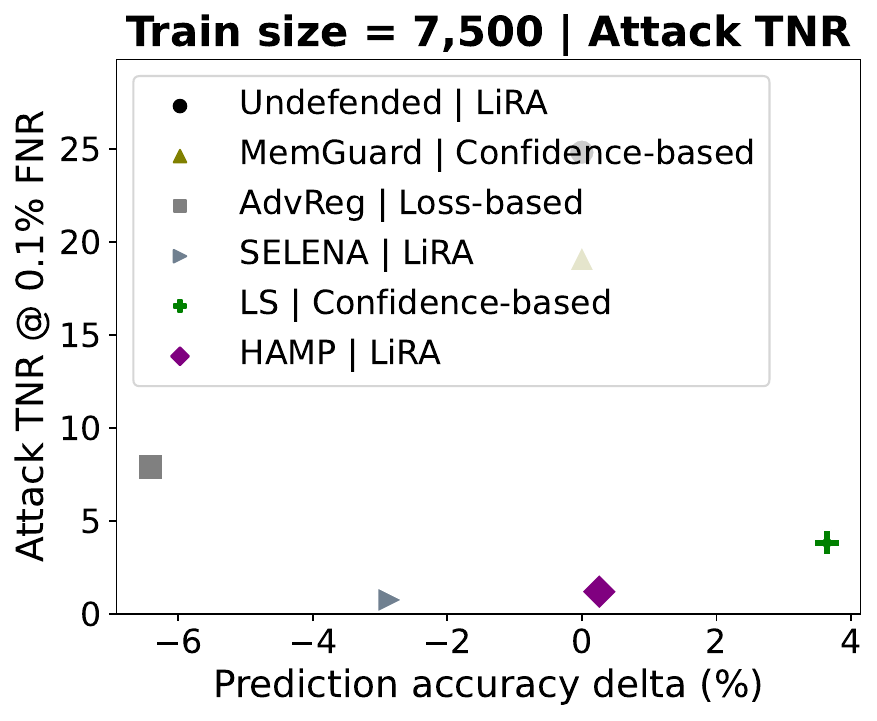}
    %subcaption{CIFAR100}% (20k)} 
    \label{fig:awesome_image3}
  \endminipage\hfill 
  \end{subfigure} 

  \begin{subfigure}[b]{\textwidth}
  \minipage{0.31\textwidth}%
    \includegraphics[width=0.95\textwidth, height=1.7in]{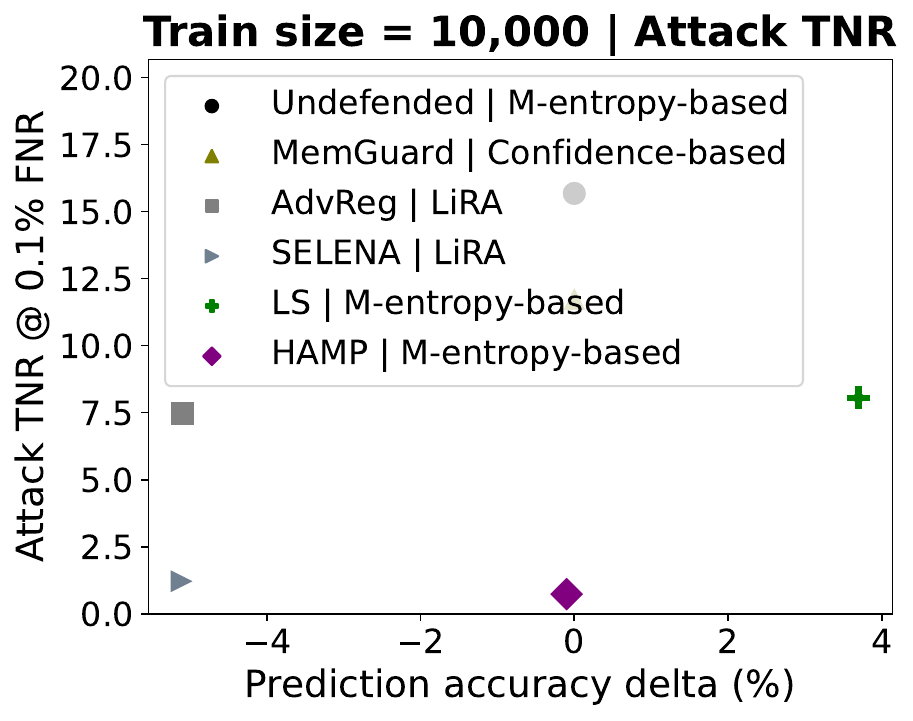}
    %subcaption{CIFAR10}% (20k)} 
    \label{fig:awesome_image4}
  \endminipage\hfill 
  \minipage{0.31\textwidth}%
    \includegraphics[width=0.95\textwidth, height=1.7in]{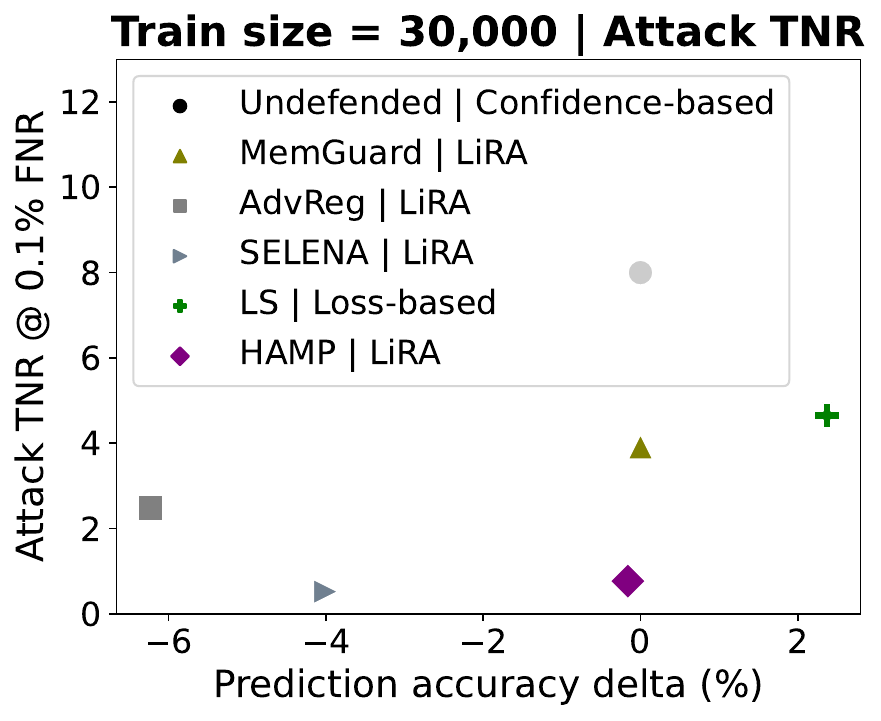}
    %subcaption{CIFAR10}% (20k)} 
    \label{fig:awesome_image4}
  \endminipage\hfill 
  \minipage{0.31\textwidth}%
    \includegraphics[width=0.95\textwidth, height=1.7in]{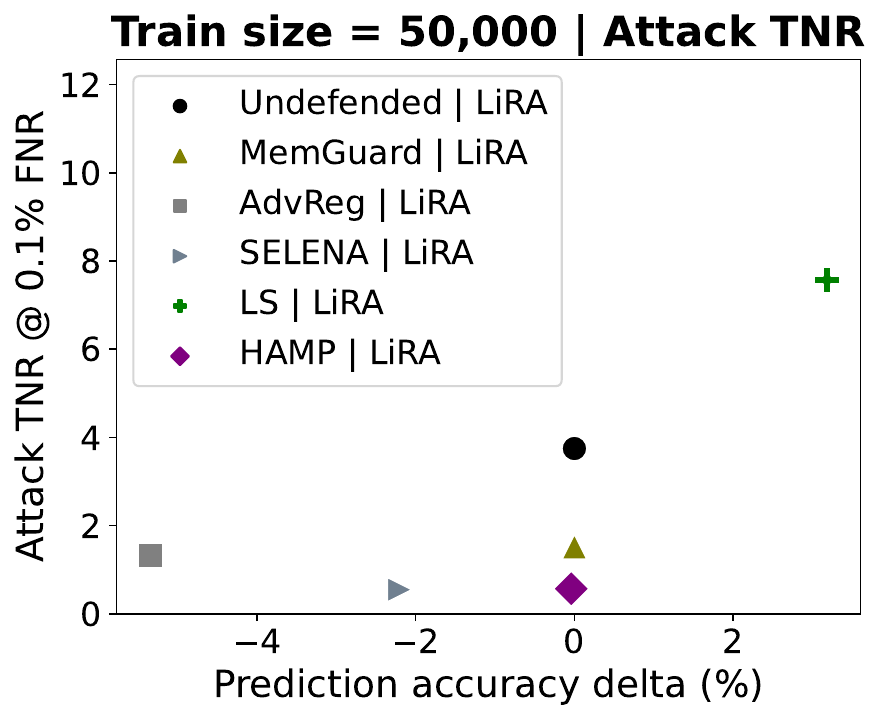}
    %subcaption{Location30}% (20k)} 
    \label{fig:awesome_image4}
  \endminipage\hfill 

  \end{subfigure} 

  \caption{Defense evaluation on models trained with different amounts of training data. The first two rows evaluate attack TPR and the last two rows evaluate attack TNR. 
  \sysname consistently achieves strong privacy protection while preserving model accuracy.}
  \label{fig:more-size-eval} 
  \vspace{-4mm}
\end{figure*}

%% file: limitation.tex
\label{sec:discussion}

\begin{table}[t]
\caption{Ablation study on different components of \sysname: 
\circled{1}: High-entropy soft labels; \circled{2}: Entropy-based regularizer; \circled{3}: Testing-time output modification. 
}
 \renewcommand{\arraystretch}{1.1}
\label{tab:ablation}
\centering  
\footnotesize
\begin{tabular*}{\columnwidth}{c|cc|cc}
\hline

Defense  & Training  & Testing  & Attack TPR & Attack TNR \\
component & accuracy & accuracy & @0.1\% FPR & @0.1\% FNR \\

\hline
\hline
None (undefended) & 99.36 & 80.85 & 14.37 & 14.62 \\
\circled{1} &  94.58 & 81.75 & 4.76 & 4.22 \\
\circled{2} &  98.06 & 81.10 & 3.39 & 4.19 \\
\circled{3} &  99.36 & 80.85 & 8.51 & 5.34 \\
\circled{1} + \circled{2} &  91.12 & 81.15 & 1.86 & 1.07  \\
\circled{1} + \circled{3} &  94.58 & 81.75 & 0.82 & 1.23\\
\circled{2} + \circled{3} &  98.06 & 81.10 & 2.90 & 3.76 \\ 
\hline
\circled{1} + \circled{2} + \circled{3}  &  \multirow{2}{2em}{91.12} & \multirow{2}{2em}{81.15} & \multirow{2}{2em}{0.40} & \multirow{2}{2em}{0.44} \\
(full defense) &  & & & \\
\hline
\end{tabular*}  
\vspace{-2mm}
\end{table} 

\subsection{{Ablation Study}}
\label{sec:ablation-study}
\sysname consists of three components, and we perform a  detailed ablation study  to investigate the effectiveness of each of these components - this includes a total of six configurations. We present the results in Table~\ref{tab:ablation}.

The second to fourth rows in Table~\ref{tab:ablation} shows the results on models using a single component in \sysname. 
For instance, training with high-entropy soft labels alone is able to produce a model with similar accuracy as the undefended model (trained with the one-hot hard labels), and reduce the attack TPR from 14.37\% to 4.76\%, and attack TNR from 14.62\% to 4.22\%. 
This also validates our earlier observation in Section~\ref{sec:mia-analysis} that training with one-hot hard labels could lead to high MIA risk, and the proposed high-entropy soft labels can be used to mitigate the high MIA risk. 
However, this is not enough as the model still suffers from relatively high TPR and TNR. We observe similar trends in the other two settings where we either train with the entropy-based regularizer alone, or directly perform output modification on the undefended model. 

Strengthening the model with more defense components can further reduce the MIA risk while preserving model accuracy. For example, training with high-entropy soft labels and the entropy-based regularizer (fifth row in Table~\ref{tab:ablation}) achieves a low TPR of 1.86\% and a low TNR of 1.07\%. 
We observe a similar trend even if we change to different configurations, as in the sixth and seventh rows in Table~\ref{tab:ablation}, both of which exhibit better privacy protection than models equipped with a single component. 
Furthermore, we find that the resulting model continues to maintain high model accuracy, which means the different defense components in \sysname can be used together to improve membership privacy without jeopardizing model accuracy. 
Finally, the full defense consisting of all three defense components, as in \sysname, exhibits the best privacy protection while maintaining competitive model accuracy.

\subsection{Evaluation on Different Training Sizes}
\label{sec:more-train-size}

This section reports additional experiments where we vary the size of the training set. 
We evaluate six more different sizes on Purchase100, which is the largest dataset in our evaluation and allows us to comprehensively evaluate a wide range of sizes, namely: 2,500, 5,000, 7,500, 10,000, 30,000, 50,000 (up to 20x difference). 
We trained 64 shadow models in the LiRA attack for each defense, with over 2,300 different shadow models in total. 
Fig.~\ref{fig:more-size-eval} shows the results. 

\textbf{We find that even when evaluated under a broad range of training sizes, \sysname consistently achieves superior performance on both privacy protection and model utility. }
The average attack TPR on the undefended model is 24.7\% and attack TNR 22.9\%. 
MemGuard achieves an average attack TPR of 13\% and attack TNR 17.4\%, both of which are significantly higher than the 1.3\% and 1.5\% by \sysname. 
AdvReg incurs an average accuracy loss of 6.3\% while \sysname incurs only 0.2\%. 
\sysname also outperforms AdvReg with better privacy protection: AdvReg reduces the attack TPR by 83\% and attack TNR by 76.1\%, while \sysname reduces them by 94.8\% and 93.4\%, respectively. 
LS improves the accuracy by 3.2\%, but it still suffers from high MIA risk: its attack TPR and TNR are 8x and 4.1x relative to that of \sysname. 
Both SELENA and \sysname have similarly strong membership privacy: the average attack TPR on SELENA is 1.2\%, and 1.3\% on \sysname; the attack TNR are 1.4\% and 1.5\%, respectively. 
Under a similar privacy protection, \sysname still outperforms SELENA with a much lower accuracy drop. 
On average, SELENA degrades the accuracy by 3.97\% (up to 5.2\%), while \sysname degrades accuracy by only 0.15\% (up to 0.98\%).

\subsection{{Evaluation against Data-poisoning-based MIA~\cite{tramer2022truth}}} 
Recent work by Tramer et al.~\cite{tramer2022truth} shows that a more capable adversary can significantly amplify the MIA risk through data poisoning. 
Therefore, we conduct additional evaluation on whether \sysname can protect against such more capable attack. 

The Tramer et al. attack increases the membership leakage against target points, by poisoning the training set to transform the target points into outliers. 
Each target point is replicated $n$ times with a wrong label, and these replicas are added as the poison samples. 
If the target point is a member in the training set, the model will be fooled into believing that the correctly-labeled target point is ``mislabeled'' (due to the presence of other poisoned replicas), which would have a large influence on the model's output and can be identified by the adversary.  

We follow \cite{tramer2022truth} to conduct the evaluation on CIFAR10, and select 250 random target points (containing both members and non-members), each replicated 8 times. 
We train 128 shadow models, which include a total of 32,000 target points. 
Without data poisoning, the adversary achieves 8.23\% attack TPR and 10.15\% attack TNR on the undefended model. 
These are increased to 52.44\% and 24.52\% after data poisoning, respectively. 
Even under such a powerful attack, \sysname is able to reduce the attack TPR from 52.44\% to 0.34\%, and attack TNR from 24.52\% to 0.71\%. Further, \sysname achieves such strong protection with a negligible accuracy drop of 0.6\%.

\subsection{Limitation} 
First, it requires re-training and hence incurs additional training overhead. 
Nevertheless, re-training is commonly required by many existing defenses~\cite{nasr2018machine,shejwalkar2019membership,tang2021mitigating}, and training is a one-time effort prior to deployment. 
Further, our evaluation shows that \sysname incurs only a modest training overhead compared with other defenses (see Appendix~\ref{sec:overhead}). 

The second limitation is that \sysname's testing-time defense incurs an overhead in every inference, which may be undesirable for the computations that have stringent real-time constraints. 
Nevertheless, \sysname incurs a low latency of only 0.04$\sim$0.38\emph{ms} per inference. In comparison, MemGuard, the other defense that also contains post-processing modification, introduces a latency of 335.42$\sim$391.75\emph{ms}. 
{In addition, this process also changes the output scores to be randomized scores, which may affect the usefulness of the output scores. Nevertheless, we try to reduce the impact by ensuring the prediction labels derived from the output scores remain unchanged (all top-k labels), and thus the model accuracy is unaffected. This can still provide meaningful information in the output scores without leaking membership privacy.}

Finally, though \sysname empirically provides superior privacy-utility tradeoff, it does not offer provable guarantees. 
This is a limitation  common to  all practical defenses~\cite{nasr2018machine,shejwalkar2019membership,tang2021mitigating,jia2019memguard}. 
Hence, a more capable adversary may mount stronger attacks, such as the data poisoning attack by Tramer et al.~\cite{tramer2022truth}. 
Our preliminary evaluation shows that \sysname still exhibits strong privacy protection and preserves model accuracy even under the presence of such a  data-poisoning adversary, but we leave further investigation to future work.

%% file: related_work.tex
\label{sec:related_work}

\emph{Membership inference attacks.}
Depending on the adversary capabilities, MIAs can be divided into  black-box~\cite{shokri2017membership,yeom2018privacy,hui2021practical,carlini2022membership,song2021systematic,choquette2021label,ye2021enhanced,li2020membership} and white-box attacks~\cite{leino2020stolen,jayaraman2021revisiting,nasr2018comprehensive}.
The former has access only into the output of the target model while the latter has visibility into information such as the internal model gradients to facilitate membership inference.
Black-box MIA assumes a more realistic adversary,  and hence is hence widely adopted in prior defense studies~\cite{jia2019memguard,tang2021mitigating,nasr2018machine} (and in \sysname).
Such attacks can be mounted by either shadow-training~\cite{shokri2017membership,nasr2018machine,yeom2018privacy} or computing statistical metrics based on the partial knowledge of the private dataset~\cite{song2021systematic,choquette2021label,li2020membership}.
Many of those attacks require full or partial access to the output scores by the model, and may be defeated if the model only reveals the prediction label.
This motivates a new class of attacks called, label-only attacks, which can be launched either  with~\cite{choquette2021label} or without~\cite{li2020membership} partial knowledge of the membership information. 
Carlini et al.~\cite{carlini2022membership} introduce the LiRA attack that can succeed in inferring membership when controlled at low false positive or false negative, through a well-calibrated Gaussian likelihood estimate.

In addition to supervised classification, MIAs have also been explored in other domains, including contrastive learning~\cite{liu2021encodermi}, generative models~\cite{chen2020gan,hayes2019logan}, federated learning~\cite{nasr2018comprehensive}, graph neural networks~\cite{zhang2021inference}, and recommender systems~\cite{zhang2021membership}.

\emph{Defenses against membership inference attacks.} 
These defenses can be divided into provable and practical defenses. 
The former can provide rigorous privacy guarantee, such as DP-SGD~\cite{abadi2016deep}, PATE~\cite{papernot2016semi}. 
Nevertheless, these defenses often incur severe accuracy drop when used with acceptable provable bounds~\cite{rahman2018membership,papernot2021tempered}. 
Another line of practical defenses aim to achieve empirical privacy without severely degrading accuracy. 
Common regularization techniques such as dropout~\cite{srivastava2014dropout}, weight decay~\cite{krogh1992simple} are shown to be able to reduce privacy leakage, but with limited effectiveness~\cite{shokri2017membership,shejwalkar2019membership}.  
Other defenses enforce specific optimization constraint during training to mitigate MIAs~\cite{nasr2018machine,li2021membership}, or perform output obfuscation~\cite{jia2019memguard,yang2020defending}.  
Knowledge distillation is used by different techniques to mitigate MIAs, including PATE~\cite{papernot2016semi}, DMP~\cite{shejwalkar2019membership}, SELENA~\cite{tang2021mitigating} and {KCD}~\cite{chourasia2021knowledge}. 
However, existing defenses are often biased towards either privacy or utility. In contrast, \sysname both achieves strong membership privacy and high accuracy, which offers a much better privacy-utility trade off.

\emph{Other privacy attacks}.
In addition to membership privacy, common ML models are found to leak different private properties~\cite{tramer2016stealing,truong2021data,fredrikson2014privacy,fredrikson2015model,ganju2018property,carlini2021extracting}.
Model extraction attacks can duplicate the functionality of a proprietary model~\cite{tramer2016stealing,truong2021data}.
Model inversion attacks are capable of inferring critical information in the input features such as genomic information~\cite{fredrikson2014privacy,fredrikson2015model}.
Property inference attacks are constructed to infer sensitive properties of the training dataset~\cite{ganju2018property}.

%% file: conclusion.tex
\label{sec:conclusion}

This work introduces \sysname, a defense against Membership Inference Attacks (MIAs) that can achieve both high accuracy and membership privacy. 
\sysname has two innovations: (1) a training framework that consists of high-entropy soft labels and an entropy-based regularizer; and (2) an output modification defense that uniformly modifies the runtime output.
\sysname significantly constrains the model's overconfidence in predicting training samples, and  
forces the model to behave similarly on both members and non-members, thereby thwarting MIAs. 
Our evaluation shows that \sysname outperforms seven leading defenses by offering a better trade off between  utility and membership privacy.

%% file: appendix.tex
\label{sec:appendix}

\subsection{Details of Defense Setup}
\label{sec:defense-setup}

This section provides details of the defense setup in our evaluation.
For each dataset, we use 10\% of the training set as a separate validation set (20\% for Location30 as it has a smaller training size), and select the model with the highest validation accuracy.

\emph{\sysname.} The values of entropy threshold $\gamma$ and $\alpha$ parameter (for controlling the regularizer) are given in Table~\ref{tab:parameter-setup}.
For model training on the two image datasets, in addition to the requirement of yielding high validation accuracy, we empirically set an additional condition that the model needs to gain at least 1\% improvement on validation accuracy in order to be considered the best model. 
This is to prevent the model gaining a marginal improvement on validation accuracy at the cost of significant overfitting on training samples, which could result in a large generalization gap.

\begin{table}[htb]
\caption{Parameter setup in \sysname.}
\label{tab:parameter-setup}
\centering  
\footnotesize
\begin{tabular}{cccc}
\hline
{Dataset} 	& Entropy threshold & Regularization strength\\
\hline
\hline
{Purchase100 } 	 & 0.8 & 0.01 \\
\hline
{Texas100 } & 0.6 & 0.01 \\

\hline
{CIFAR100}		 & 0.5 & 0.005 \\
\hline
{CIFAR10}	 & 0.95 & 0.001 \\
\hline
{Location30}	 & 0.5 & 0.001 \\
\hline
\end{tabular}  
%\vspace{-2mm}
\end{table}

\emph{Adversarial regularization~\cite{nasr2018machine}}: The alpha parameter is for balancing classification accuracy and privacy protection. We set alpha as 3 for Purchase100~\cite{nasr2018machine}, 10 for Texas100~\cite{song2021systematic}, 6 for CIFAR100 and CIFAR10~\cite{nasr2018machine}, and 10 for Location30.

\emph{SELENA~\cite{tang2021mitigating}}: We follow the original authors to set K=25 and L=10, where K is the total number of sub models, and L means for a given training sample, there are L sub models whose training sets do not contain that particular sample.
For these L models, the given training sample can be viewed as an instance in their ``reference set'' for distillation.

\emph{Label Smoothing (LS)~\cite{szegedy2016rethinking}}: We follow  \cite{kaya2021does} to train LS with different smoothing intensities and select the model with the highest accuracy. 
Purchase100 is trained with a smoothing intensity of 0.03, Texas with 0.09 and CIFAR100 with 0.01.

\emph{DP-SGD\cite{abadi2016deep}}: We use PyTorch Opacus~\cite{opacus} to train the DP-SGD model. 
We set microbatch size to be 1.  
Purchase100 is trained with a noise\_multiplier of 1.7, a norm clipping bound of 1.0 and with 200 epochs. 
Texas100 is trained with a noise\_multiplier of 1.44, a norm clipping bound of 1.0 and with 200 epochs. 
Location30 is trained with a noise\_multiplier of 2.91, a norm clipping bound of 3.0 and with 50 epochs.

\begin{figure*}[!t]

	\begin{subfigure}[b]{\textwidth}

	\minipage{0.19\textwidth}
	  \includegraphics[width=1.15\textwidth, height=1.3in]{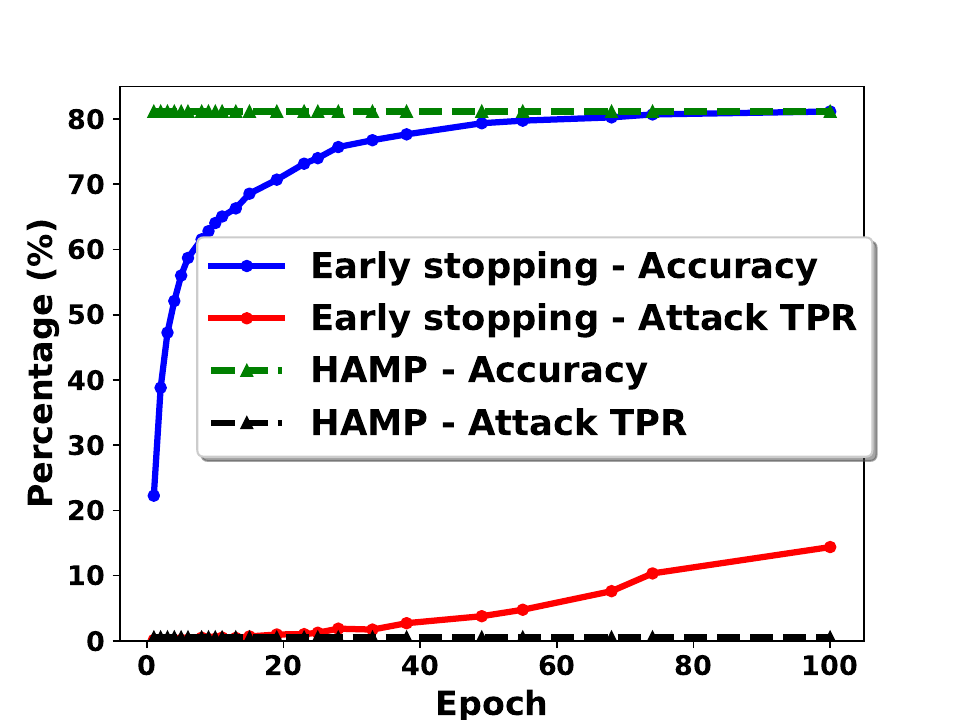}
	  \subcaption{Purchase100}% (20k)}
	 \endminipage\hfill
	\minipage{0.19\textwidth}
	  \includegraphics[width=1.15\textwidth, height=1.3in]{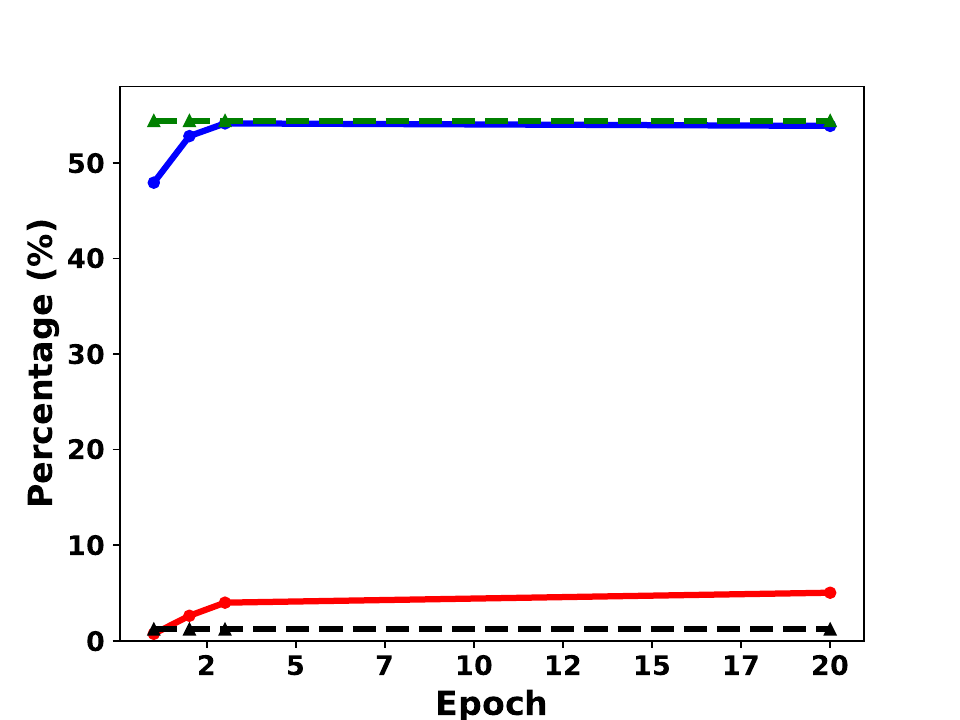}
	  \subcaption{Texas100}% (15k)}
	\endminipage\hfill
	\minipage{0.19\textwidth}%
	  \includegraphics[width=1.15\textwidth, height=1.3in]{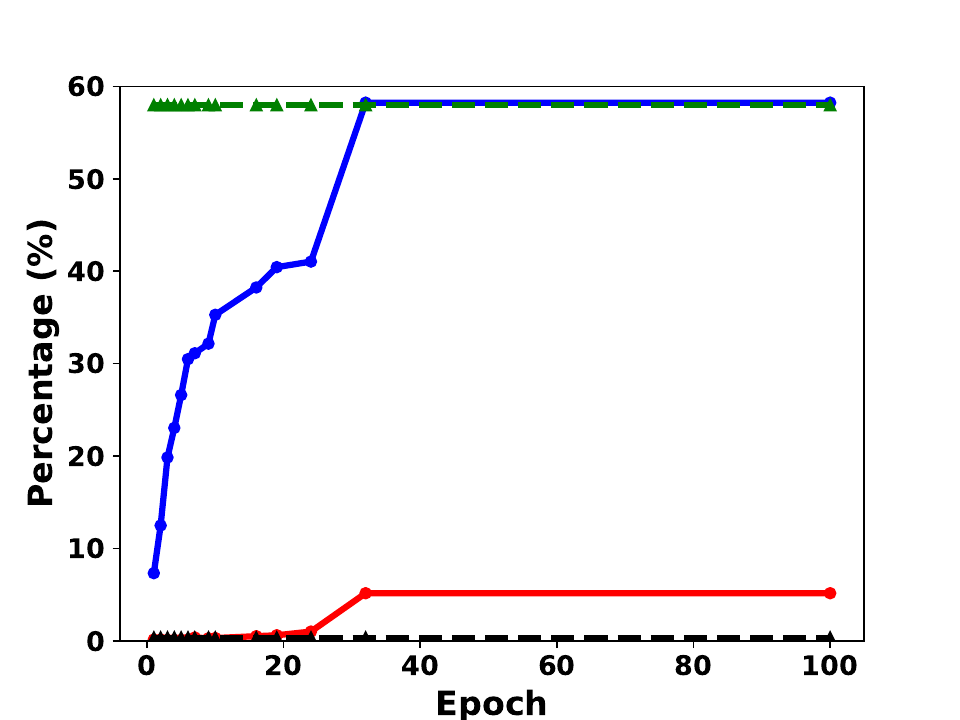}
	  \subcaption{CIFAR100}% (20k)} 
	  \label{fig:awesome_image3}
	\endminipage\hfill 
	\minipage{0.19\textwidth}%
	  \includegraphics[width=1.15\textwidth, height=1.3in]{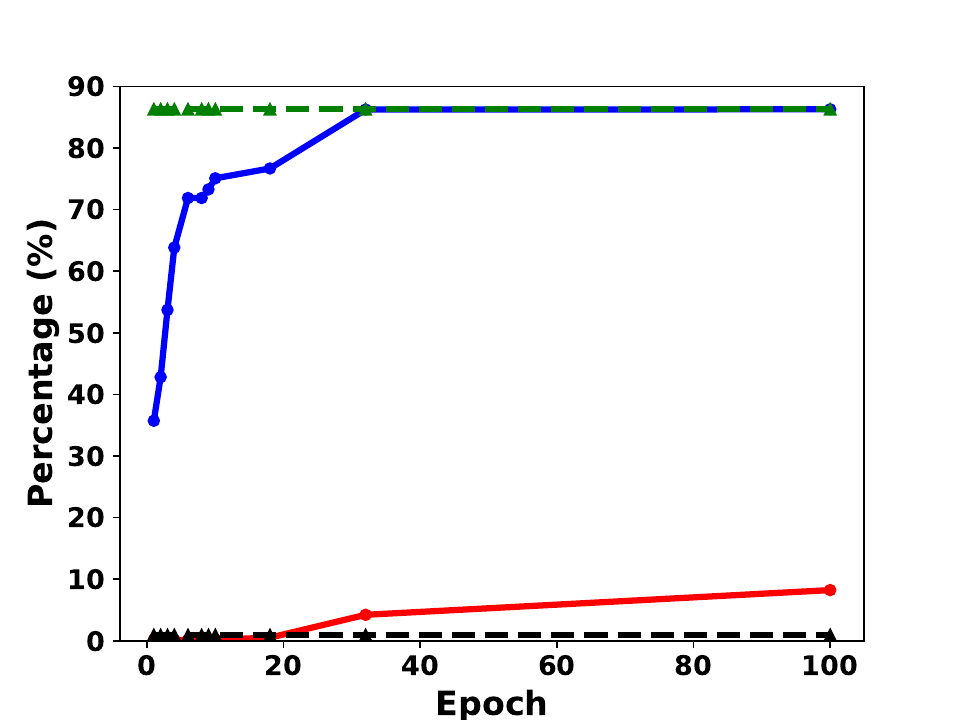}
	  \subcaption{CIFAR10}% (20k)} 
	  \label{fig:awesome_image4}
	\endminipage\hfill 
	\minipage{0.19\textwidth}%
	  \includegraphics[width=1.15\textwidth, height=1.3in]{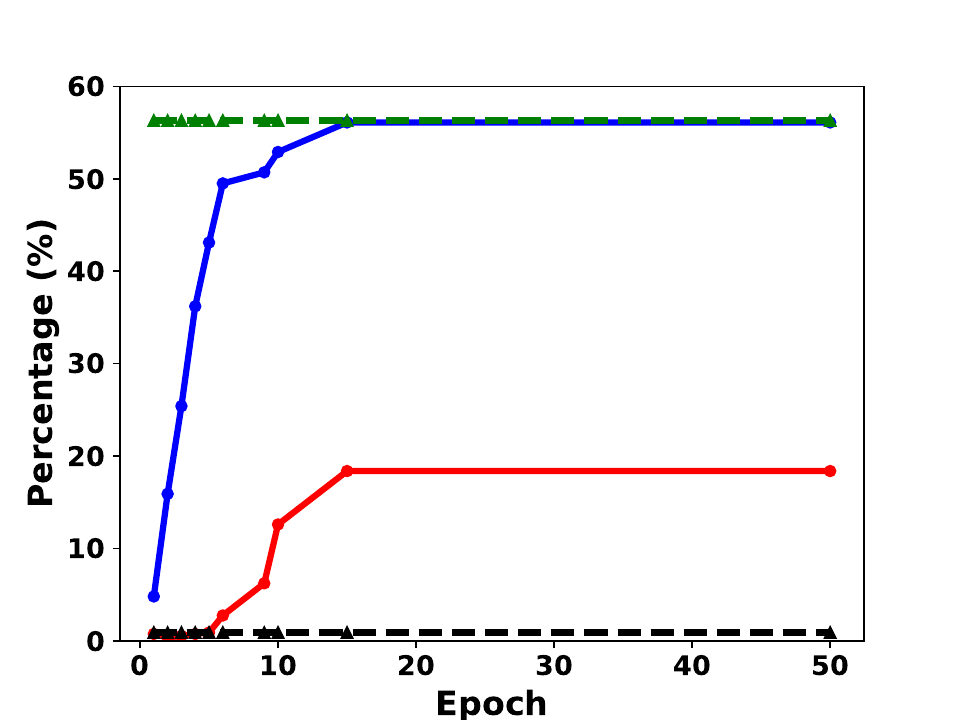}
	  \subcaption{Location30}% (20k)} 
	  \label{fig:awesome_image4}
	\endminipage\hfill 

	\end{subfigure} 

	\begin{subfigure}[b]{\textwidth}

	\minipage{0.19\textwidth}
	  \includegraphics[width=1.15\textwidth, height=1.3in]{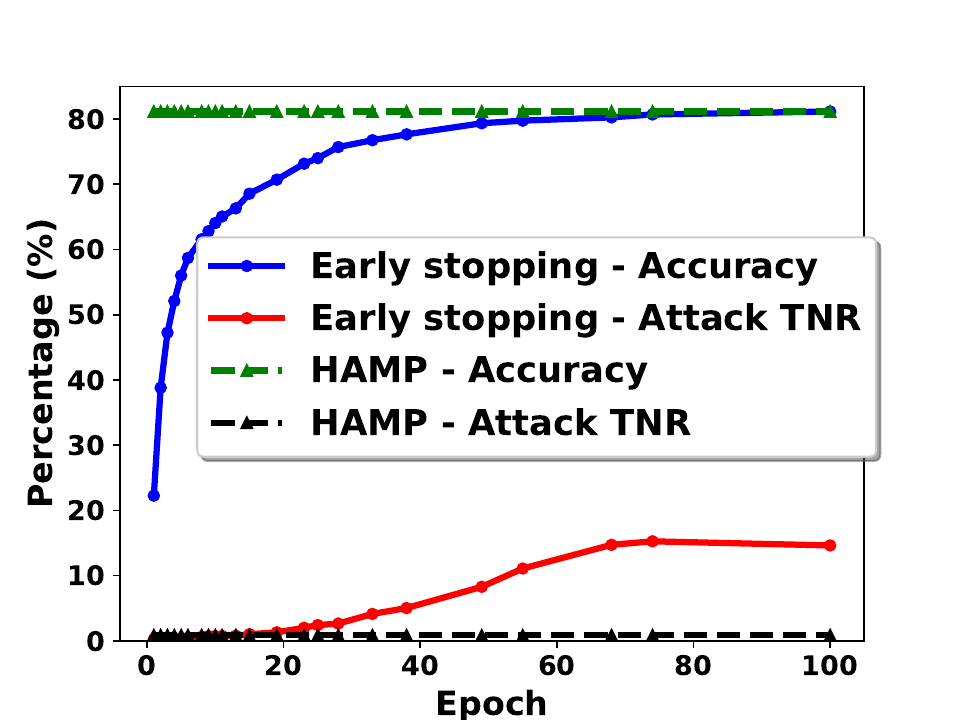}
	  \subcaption{Purchase100}% (20k)}
	 \endminipage\hfill
	\minipage{0.19\textwidth}
	  \includegraphics[width=1.15\textwidth, height=1.3in]{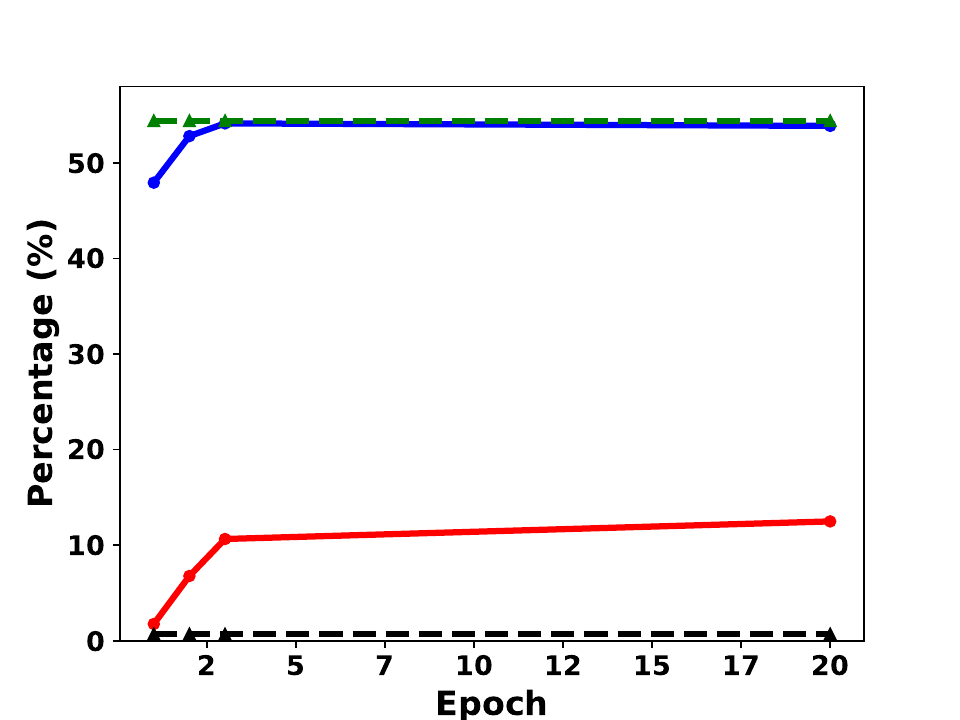}
	  \subcaption{Texas100}% (15k)}
	\endminipage\hfill
	\minipage{0.19\textwidth}%
	  \includegraphics[width=1.15\textwidth, height=1.3in]{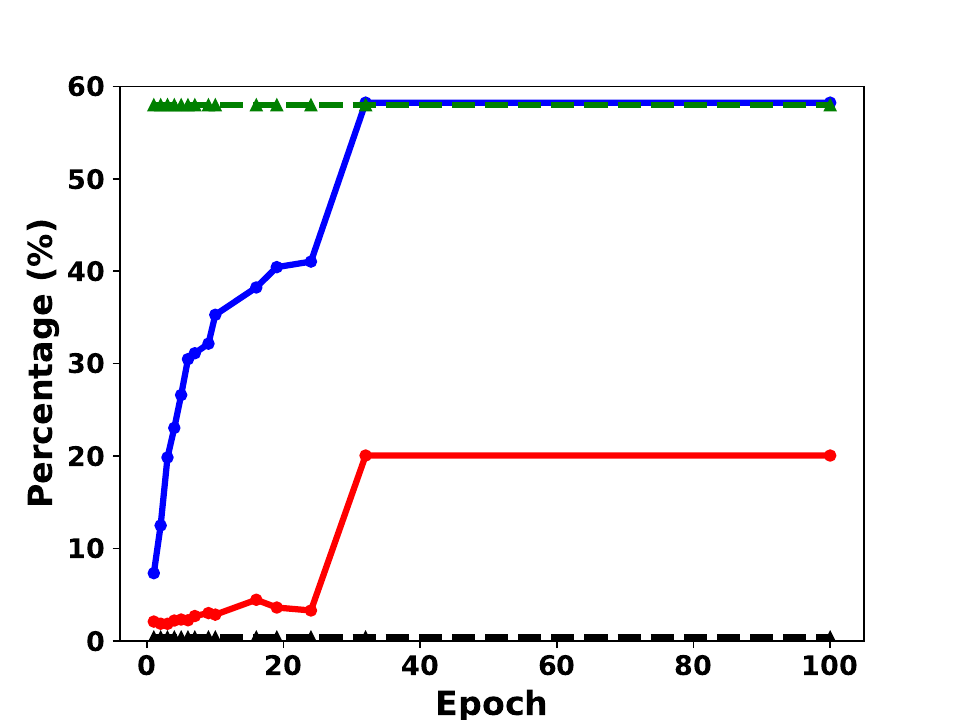}
	  \subcaption{CIFAR100}% (20k)} 
	  \label{fig:awesome_image3}
	\endminipage\hfill 
	\minipage{0.19\textwidth}%
	  \includegraphics[width=1.15\textwidth, height=1.3in]{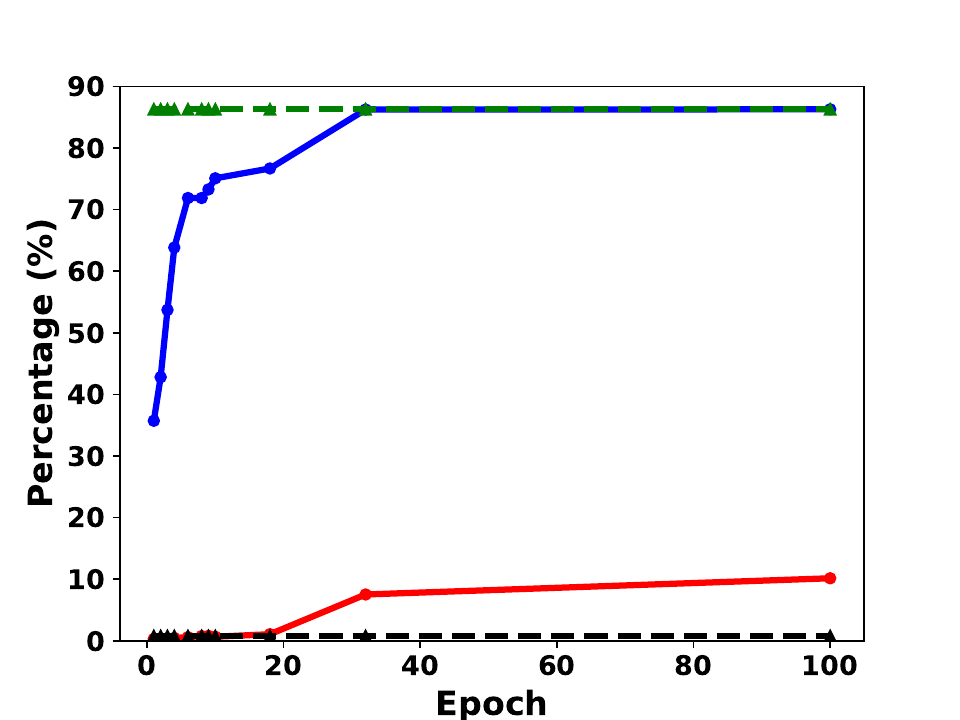}
	  \subcaption{CIFAR10}% (20k)} 
	  \label{fig:awesome_image4}
	\endminipage\hfill 
	\minipage{0.19\textwidth}%
	  \includegraphics[width=1.15\textwidth, height=1.3in]{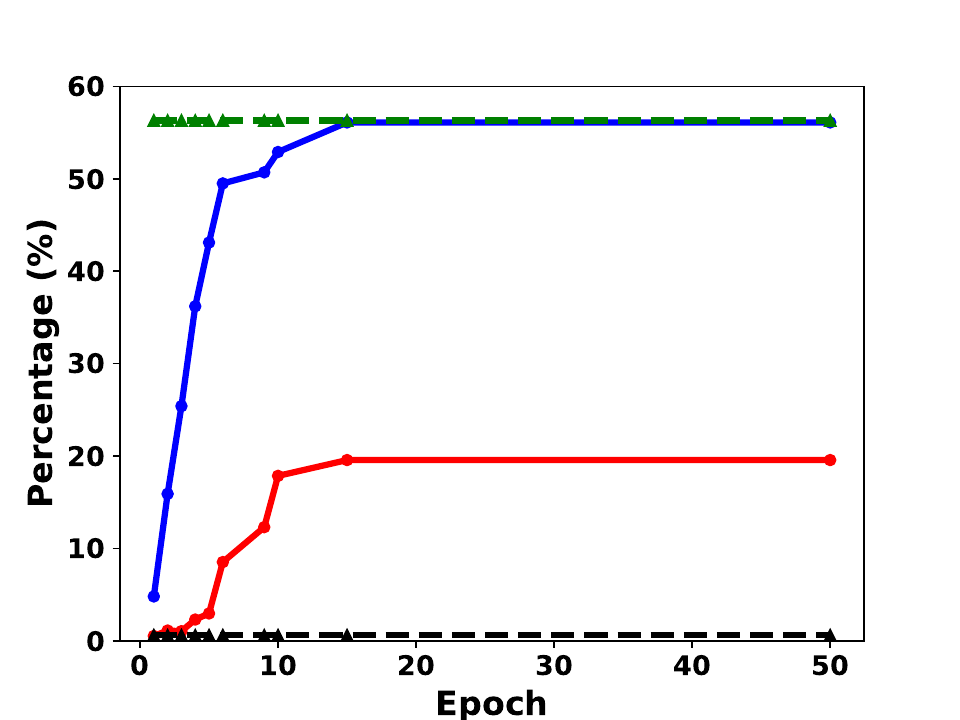}
	  \subcaption{Location30}% (20k)} 
	  \label{fig:awesome_image4}
	\endminipage\hfill 

	\end{subfigure} 

	\caption{Comparison of our defense (\sysname) with early stopping on each dataset. First row compares the attack TPR @0.1\% FPR and second row the attack TNR @0.1\% FNR at different epochs. 
	Dashed lines indicate the results by \sysname, while solid lines are those for early stopping (\sysname is trained till it converges, while early stopping is trained with different epoch sizes before convergence).
	}
	\label{fig:elr-comp} 
	\vspace{-4mm}
\end{figure*}

\subsection{{Measuring Prediction Entropy by \sysname}}
\label{sec:entropy}
As mentioned in Section~\ref{sec:overview}, \sysname reduces privacy leakage from output scores via enforcing the model to predict training samples with higher entropy (i.e., less confident prediction on training samples). 
We validate this by measuring the prediction entropy produced by the models before and after \sysname, and report the results in Table~\ref{tab:entropy}. 

\vspace{-2mm}

\begin{table}[htb]
\caption{Comparing the prediction entropy on the undefended models and \sysname. \sysname significantly increases the prediction entropy on members (i.e., less confident prediction on members) and enforces the model to predict members and non-members with similarly high prediction entropy. 
}
\label{tab:entropy}
\centering 
\footnotesize
\begin{tabular}{ccccc}
\hline

 \multirow{3}{4em}{Model} & \multirow{3}{2em}{Dataset} & \multirow{3}{4em}{Members} & \multirow{3}{4em}{Non-members} & Difference \\
 				&  & & & (the smaller  \\
 				& & & & the better) \\
\hline
\hline
\multirow{4}{4em}{Undefended} & Purchase100 & 0.389 & 0.576 & 0.187 \\
										& Texas100 & 0.505 & 0.771 & 0.266 \\
										& CIFAR100 & 0.685 & 1.020 & 0.337 \\
										& CIFAR10  & 0.102 & 0.226 & 0.125 \\
										& Location30 & 0.224 & 0.567 & 0.343 \\
\hline
\multirow{4}{4em}{\sysname} & Purchase100 & 4.485 & 4.490 & 0.006 \\
											& Texas100 & 4.484 & 4.495 & 0.011 \\
											& CIFAR100 & 4.124 & 4.157 & 0.032 \\
											& CIFAR10 & 2.010 & 2.029 & 0.019 \\
											& Location30 & 2.789 & 2.847 & 0.058 \\
\hline
\end{tabular}  
%\vspace{-2mm}
\end{table}

On the undefended models, the member samples are predicted with much lower entropy than that on non-members, and the entropy difference between members and non-members is 0.125$\sim$0.343. 
Such a large difference indicates the differential behavior on members and non-members that can be distinguished by the MIAs. 

In contrast, the models trained with \sysname predict both members and non-members with much higher prediction entropy (increase by 4.1x$\sim$19.8x), and the average difference between members and non-members is reduced from 0.125$\sim$0.337 (on undefended models) to 0.006$\sim$0.058, which is 6.5x$\sim$32.7x smaller. 
This demonstrates how \sysname enforces the model to behave similarly on members and non-members and therefore reduce privacy leakage.

\subsection{Evaluating Label Smoothing with Different Smoothing Intensities} 
\label{sec:ls-diff-smooth-intensities}
\begin{figure}[!t]
\centering
	\begin{subfigure}[b]{\columnwidth}
\centering
	\minipage{1.\columnwidth}
	  \includegraphics[width=.8\textwidth, height=2.1in]{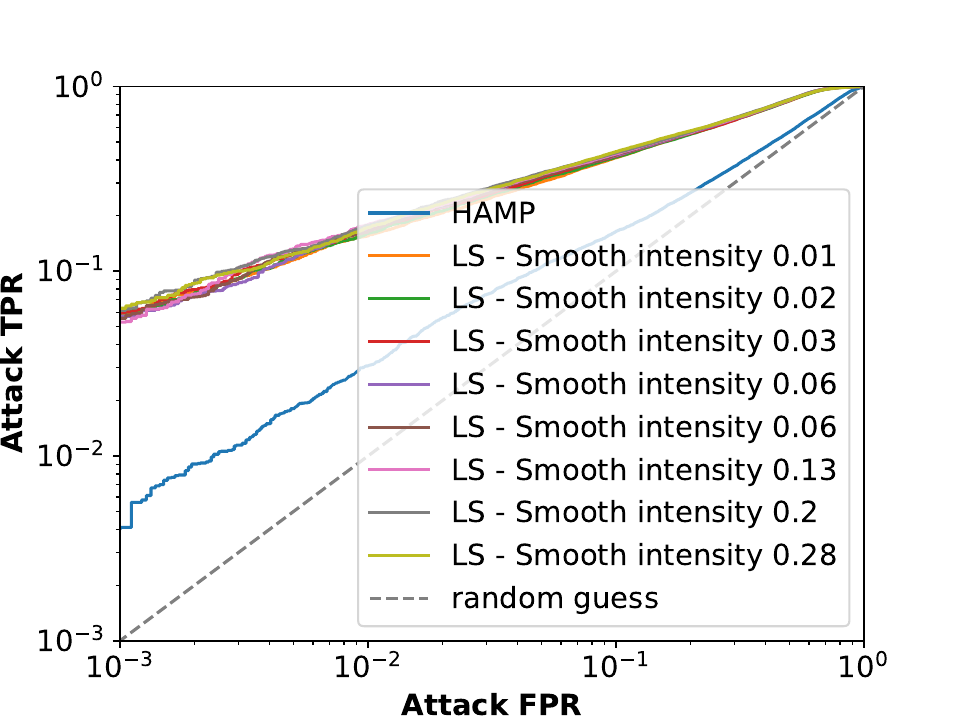} 
	 \endminipage\hfill
	\end{subfigure}

	\begin{subfigure}[b]{\columnwidth}
	\centering
	\minipage{1.\columnwidth}
	  \includegraphics[width=.8\textwidth, height=2.1in]{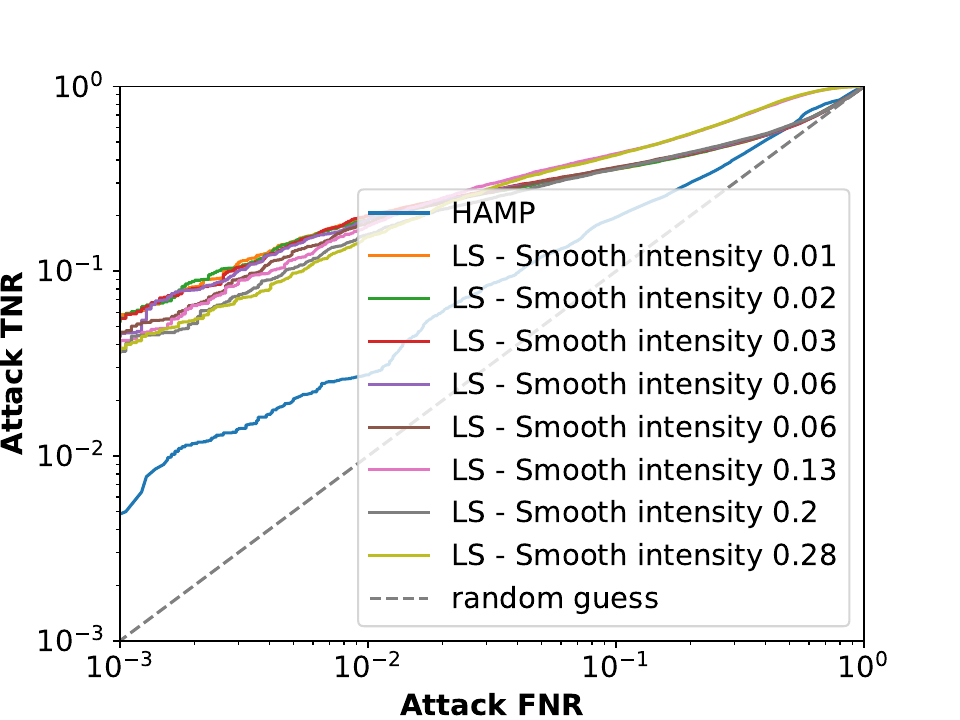} 
	\endminipage\hfill 
	\end{subfigure}

	\caption{Comparing \sysname with Label Smoothing under different smoothing intensities. \sysname consistently achieves significantly better membership privacy than LS. 
	}
	\label{fig:ls-diff-intensities} 
	\vspace{-4mm}
\end{figure}

In Section~\ref{sec:comparison-ls}, we compare \sysname with LS using the smoothing intensity that achieves the highest accuracy, and we found that \sysname achieves significantly lower MIA risk than LS. 
In this section, we evaluate LS with other intensities that achieve similar accuracy improvement. % to validate that similar trend still holds.  
On Purchase100, we select a smoothing intensity of 0.03, which yields the highest accuracy improvement of 4.75\%, and we consider all seven other intensities that achieve comparable accuracy improvement (3.8\%$\sim$4.4\%). 
Fig.~\ref{fig:ls-diff-intensities} presents the results, which show that LS trained with different intensities still exhibit very high MIA risk. 
For example, the attack TPR @ 0.1\% FPR by LS are 13.7x$\sim$15.5x higher than that of \sysname, and the attack TNR are 8.2x$\sim$12.4x higher than that of \sysname.

\subsection{Comparison with Early Stopping}
\label{sec:comparison-early-stop} 
Early stopping produces models trained with fewer epochs to prevent overfitting. 
In our evaluation, we benchmark the classification accuracy and attack TPR/TNR  of the models trained with different epochs before convergence, and compare them with \sysname. 
The results are shown in Fig.~\ref{fig:elr-comp}. 

When the model is trained with a few epochs in early stopping, the model is able to achieve comparable privacy protection as \sysname, but with a large accuracy drop.
For example, on Purchase100, the model trained with 15 epochs yields an attack TPR of 0.67\% and attack TNR of 1.01\%, which are slightly higher than the 0.4\% and 0.44\% by \sysname. 
However, its prediction accuracy is only $68.6\%$, which is much lower than the $81.15\%$ achieved by \sysname.

The model's accuracy improves with more training epochs, but then so does the attack TPR and TNR.  
When the models derived by early stopping  converge, there is a substantial gap between the attack TPR and TNR of \sysname and early stopping  (black dashed line vs. red solid line in Fig.~\ref{fig:elr-comp}). 

To summarize, under a similar MIA risk for members (i.e., similar attack TPR), \sysname achieves an average 12.5\% higher accuracy than early stopping; and 28.6\% higher accuracy than early stopping when under similar attack TNR.

\subsection{Varying the Parameters of \sysname}
\label{sec:tweak-hyperparameter}
This section evaluates the performance of \sysname under different parameters, $\gamma\in (0.1, 0.9), \alpha\in (0.0001, 0.5)$. 
We use Purchase100 and present the results in Table~\ref{tab:vary-parameter}. 

\textbf{Entropy threshold.}
A higher entropy threshold assigns lower probability to the ground-truth class in the labels and enforces the model to become less confident in predicting training samples. 
For instance, for the entropy threshold of 0.9, the probability of the ground-truth class is only 20\%, 
while with a threshold of 0.1, the probability is 94\%. 
Table~\ref{tab:vary-parameter} shows that a higher entropy threshold leads to a model with {lower} classification accuracy and also {lower} MIA risk (on both attack TPR and attack TNR).
The highest entropy threshold, 0.9, produces a model with the lowest test accuracy of 66.7\% and the lowest attack TPR of 0.38\% and attack TNR of 0.26\%

\textbf{Strength of regularization.}
Stronger entropy-based regularization forces the model to produce outputs with higher uncertainty (uncertainty is measured by the prediction entropy), and is useful in preventing the model's overconfidence in predicting training samples. 
The model exhibits strong resistance against MIAs when $\alpha$ is large (e.g., 0.05)

On the other hand, strong regularization results in a model with low classification accuracy.
This is because, when $\alpha$ is large, the overall loss term in objective (\ref{eq:objective-func}) is dominated by the second regularization term, while the first loss term for improving classification accuracy is not optimized sufficiently.

\begin{table}[t]
\caption{Performance of \sysname under different parameters.}
\label{tab:vary-parameter} 
\centering  
\footnotesize
\begin{tabular*}{\columnwidth}{cccccc}
\hline
& \multirow{2}{4em}{Parameter}	& Training & Testing &{Attack TPR} & {Attack TNR} \\
 	&  & Acy  & Acy & @0.1\% FPR &  @0.1\% FNR \\

\hline 
\multirow{9}{3em}{$\gamma$ (Entropy threshold)}	& 0.9 & 73.79 & 66.7 & 0.38 & 0.26  \\
										& {0.8} & 91.12 & 81.15 & 0.4 & 0.44  \\
										& 0.7 & 94.56 & 82.35 &0.53 & 0.68  \\
										& 0.6 & 96.73 & 83.4 & 0.87 & 1.15 \\
										& 0.5 & 97.93 & 83.55 & 0.98 & 1.41 \\
										& 0.4 & 98.49 & 83.5 & 0.9 & 1.97  \\
										& 0.3 & 98.67 & 83.9 & 1.23 & 1.47 \\
										& 0.2 & 98.85 & 84.55 & 1.17 & 2.09 \\
										& 0.1 & 99.06 & 84.45 & 2.02 & 1.91 \\
\hline
\multirow{7}{3em}{$\alpha$ (Regularization strength)} 		
										%& 1.0 & 31.66 & 24.4 & 0.22 & 0.23  \\
										& 0.5 & 31.81 & 29.1 & 0.31 & 0.19  \\
										& 0.1 & 34.27 & 33.25 & 0.15 & 0.2 \\
										& 0.05 & 74.98 & 68.05 & 0.22 & 0.36  \\
										& {0.01} & 91.12 & 81.15 & 0.4 & 0.44 \\
										& 0.005 & 92.53 & 81.45 & 0.44 & 0.46 \\
										&0.001 & 93.8 & 82.1 & 0.56 & 0.53 \\
										& 0.0005 & 94.22 & 81.85 & 0.7 & 0.78 \\
										&0.0001 & 94.69 & 82.6 & 0.81 & 0.91 \\

\hline
\end{tabular*}  
%\vspace{-2mm}
\end{table}

\subsection{{Overhead Evaluation}}
\label{sec:overhead}
\textbf{Training overhead.} 
We compare the training overhead of \sysname with AdvReg, SELENA, LS and DMP.  
We do not compare training overhead with MemGuard as it is a post-processing technique that modifies the prediction vector during inference. Instead, we compare with its  inference overhead. 

For Purchase100, Texas100,  CIFAR100, CIFAR10 and Location30, 
the undefended models and the sub models in SELENA are trained with 100, 20, 100, 100 and 50 epochs; 
For knowledge distillation in DMP and SELENA, we use 200, 100, 200, 200 and 100 epochs. 
LS and \sysname are trained with 200, 100, 200, 200 and 100 epochs.
AdvReg is trained with 50, 20, 200, 200 and 50 epochs, respectively. 
All models converged after training.

The overhead is measured on a single NVidia V100SXM2 GPU with 16 GB memory.
Each measurement is repeated 5 times and we report the average overhead.
The training overhead of each defense is shown in Table~\ref{tab:overhead}.
All defense techniques incur higher training cost compared with the undefended models (as expected), \sysname and LS incur the lowest training overhead among all the defenses (\sysname is slightly higher than LS). 
AdvReg's overhead is 5.4x$\sim$11.4x relative to that of \sysname, and DMP's overhead is 3.2x$\sim$5.6x relative to that of \sysname. 
SELENA's overhead is 4x$\sim$8.8x relative to that of of \sysname. 
Even though the latency of training multiple sub models in SELENA can be hidden by parallel training, its overhead is still 23\%$\sim$66\% higher than that of \sysname.

\begin{table}[t]
\caption{Training overhead comparison. SELENA-parallel means all the sub models are trained in parallel.}
\label{tab:overhead}
\centering 
%\footnotesize  
\scriptsize
%\begin{tabular}{ccccccc}
\begin{tabular*}{\columnwidth}{ccccccc}
\hline
\multirow{2}{3em}{Dataset} 	& \multirow{2}{2em}{None} & \multirow{2}{2.5em}{AdvReg} & SELENA & \multirow{2}{2em}{LS}  &  \multirow{2}{2em}{DMP} & \multirow{2}{2.5em}{\textbf{\sysname}} \\
							& 		&				& sequential/parallel & &  & \\
\hline
\hline
Purchase100 & 177s & 2615s & 3386s / 600s & 376s & N/A & 484s \\ 
\hline
Texas100 & 31s & 1501s & 859s / 263s & 188s & 1000s & 214s \\ 
\hline
CIFAR100 & 0.66h & 29.06h & 15.45h / 4.25h & 2.56h & 8.45h & 2.56h \\ 
\hline
CIFAR10 & 0.66h & 28.88h  & 15.32h / 4.20h & N/A & 8.39h & 2.63h  \\ 
\hline
Location30 & 6.4s & 116.6s & 116.6s / 19.8s & N/A & 74s & 13.2s  \\

\hline
\end{tabular*}   
\end{table}

\textbf{Inference overhead.}
We compare \sysname with MemGuard on their inference overhead (other defenses do not have a post-processing procedure, and hence their inference overheads are the same as the undefended model's). 
For \sysname, the generation of random samples is independent of the runtime inference, so we first generate the random samples and obtain their output scores, and measure only the overhead of performing output modification (i.e., Line 13 in Algorithm~\ref{alg:overall}). 
We measure the inference overhead by performing inference on 500 random member and non-member samples (1,000 samples in total). 

Table~\ref{tab:inference-overhead} shows the average inference overhead per sample. 
The overhead incurred by MemGuard is 25x$\sim$1048x  the overhead incurred by \sysname. This is because MemGuard requires solving a complex optimization to obfuscate the prediciton scores while \sysname only performs output modification on the prediction scores (Line 13 in Algorithm~\ref{alg:overall}), which does not require solving any optimization.

\begin{table}[t]
\caption{Inference overhead comparison with MemGuard.}
\label{tab:inference-overhead}
\centering 
\footnotesize  
%\scriptsize
\begin{tabular}{cccc}

\hline
{Dataset} 	& Undefended & MemGuard & \textbf{\sysname} \\
\hline
\hline
Purchase100 & 0.34\emph{ms} & 354.72\emph{ms} & 0.74\emph{ms} \\
\hline
%Texas100 & 486s & 3270s &  9617s & 932s &594s & 629s \\
Texas100 & 0.31\emph{ms} & 345.61\emph{ms} & 0.69\emph{ms} \\
\hline
CIFAR100  & 16.0\emph{ms} & 407.75\emph{ms} & 16.34\emph{ms} \\
\hline
CIFAR10 & 12.2\emph{ms} & 389.90\emph{ms} & 12.24\emph{ms} \\ 
\hline
Location30 & 0.23\emph{ms} & 335.42\emph{ms} & 0.32\emph{ms}\\ 
\hline
\end{tabular}  
\end{table}

\subsection{Understanding the High Attack Performance by the NN-based Attack~\cite{nasr2018comprehensive}} 
\label{sec:explain-nn-based}
\begin{figure}[htb]
	\centering
	\includegraphics[width=2in, height=1.7in]{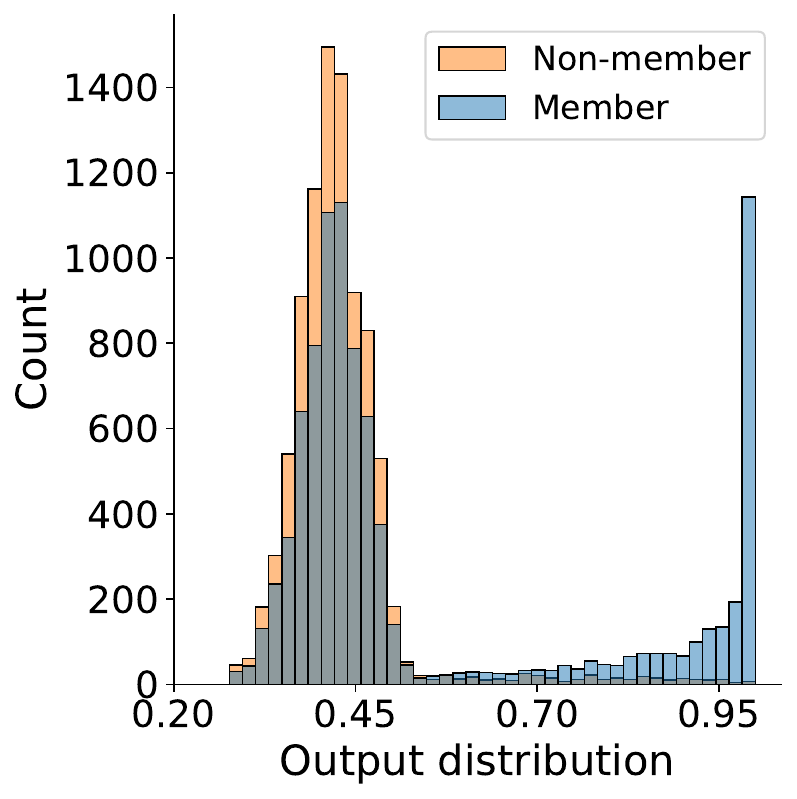}
	\caption{Output distribution by the attack inference model (obtained from the undefended model). By using a large output threshold to infer membership, the NN-based attack~\cite{nasr2018comprehensive} achieves high attack TPR with low FPR.}
	\label{fig:nn-attack} 
  \vspace{-2mm}
\end{figure}

Fig.~\ref{fig:tpr-01} in our earlier evaluation shows that the NN-based attack~\cite{nasr2018comprehensive} achieves the highest TPR with low FPR on the undefended models in many cases. 
We explain the reason. 
The NN attack trains an attack inference model on the known member and non-member samples, which outputs \emph{large values on members} and \emph{small ones on non-members}. 
We first plot in Fig.~\ref{fig:nn-attack} the output distribution by the attack inference model to help understand how different thresholds affect the attack TPR and FPR. 

The default NN attack uses a threshold of 0.5 and predicts any sample with an output $>$0.5 as a member.
As shown in Fig.~\ref{fig:nn-attack}, in order to maintain a low FPR, the attack switches to a larger threshold (as high as over 0.99 in our experiment). 
In this case, low FPR can be achieved because most non-members are predicted with low values (the left region in Fig.~\ref{fig:nn-attack}). 
Likewise, the attack achieves high TPR, because  many members are predicted with large values (in the right most region in Fig.~\ref{fig:nn-attack}), and are correctly recognized as members.

\begin{figure}[!t]
\centering
	\begin{subfigure}[b]{\columnwidth}
\centering
	\minipage{1.\columnwidth}
	  \includegraphics[width=.8\textwidth, height=2.1in]{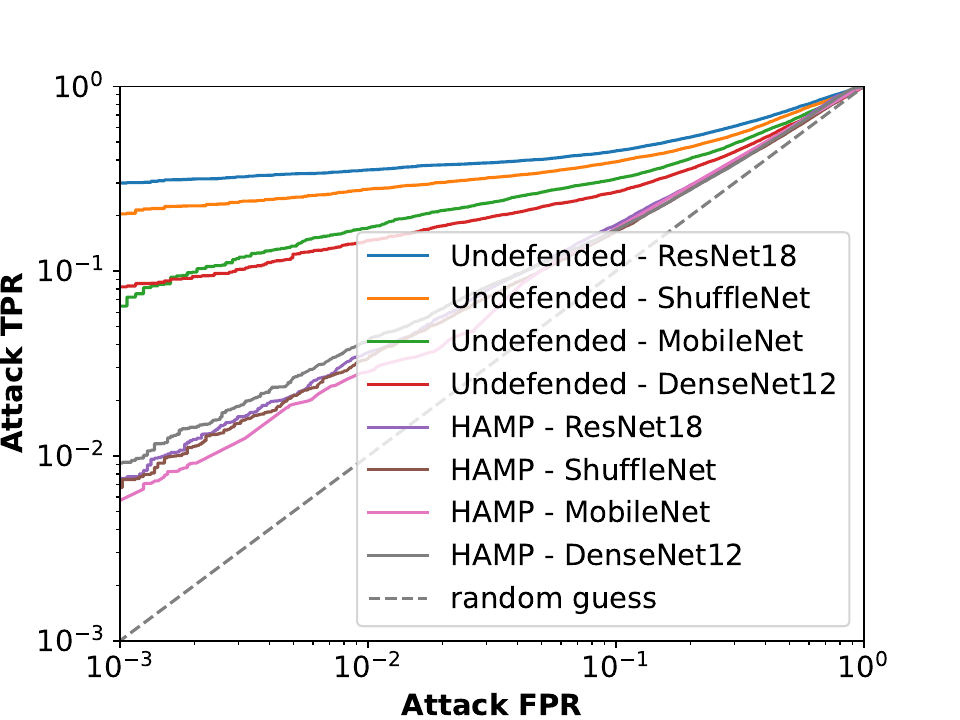} 
	 \endminipage\hfill
	\end{subfigure}

	\begin{subfigure}[b]{\columnwidth}
	\centering
	\minipage{1.\columnwidth}
	  \includegraphics[width=.8\textwidth, height=2.1in]{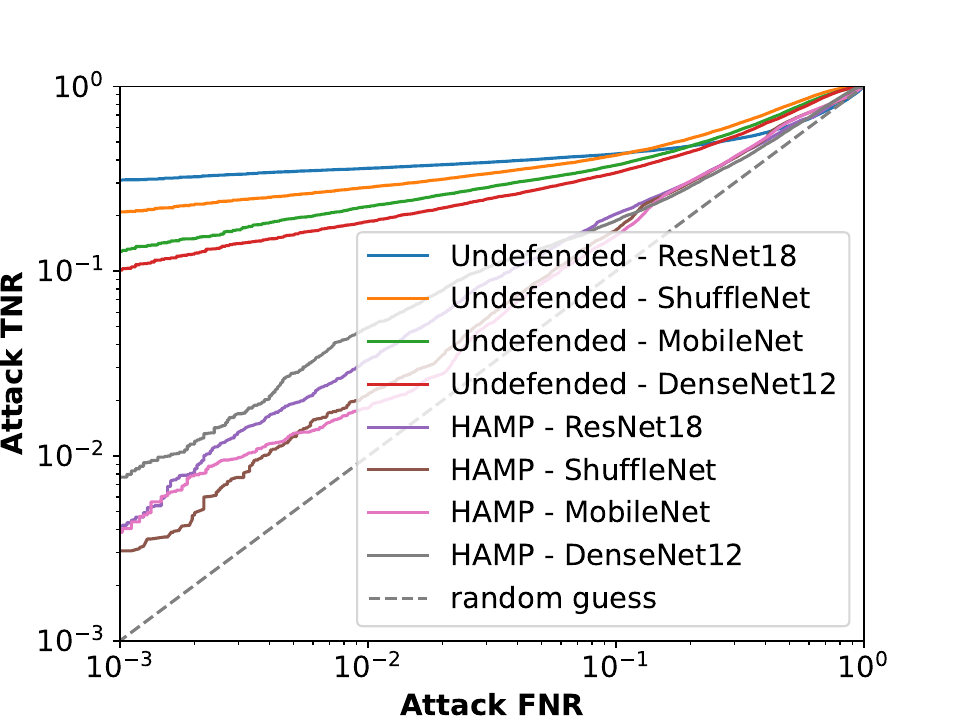} 
	\endminipage\hfill 
	\end{subfigure}

	\caption{Evaluation under different network architectures. While models trained with different architectures exhibit varied degree of MIA risk, \sysname \emph{consistently} contributes to low MIA risk despite the specific architecture. 
	}
	\label{fig:diff-arch} 
	\vspace{-4mm}
\end{figure}

\begin{figure*}[!t]
    \centering
    \begin{subfigure}[b]{\textwidth}

		\minipage{0.31\textwidth}
		  \includegraphics[width=0.99\textwidth, height=1.7in]{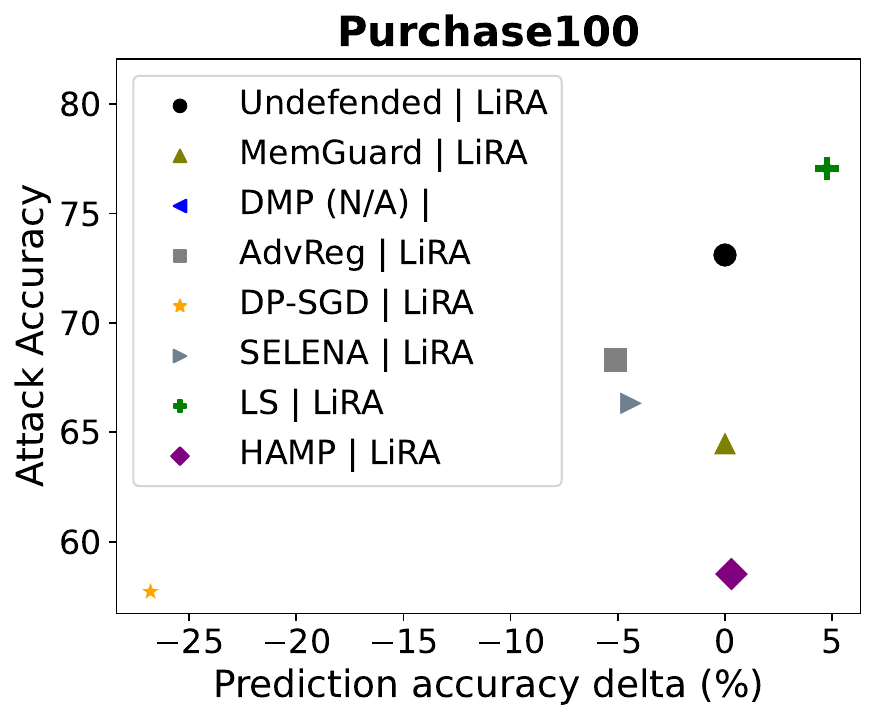}
		  %\subcaption{CIFAR100 (15k)}
		  \label{fig:c15}
		 \endminipage\hfill
		\minipage{0.31\textwidth}
		  \includegraphics[width=0.99\textwidth, height=1.7in]{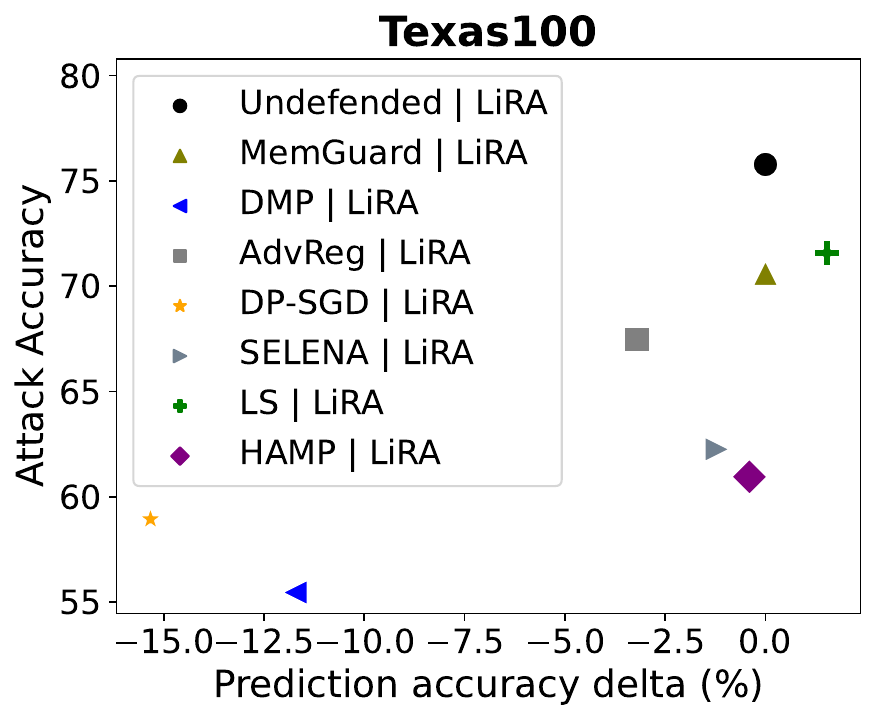}
		  %\subcaption{CIFAR100 (20k)}
		  \label{fig:c20}
		\endminipage\hfill
		\minipage{0.31\textwidth}%
		  \includegraphics[width=0.99\textwidth, height=1.7in]{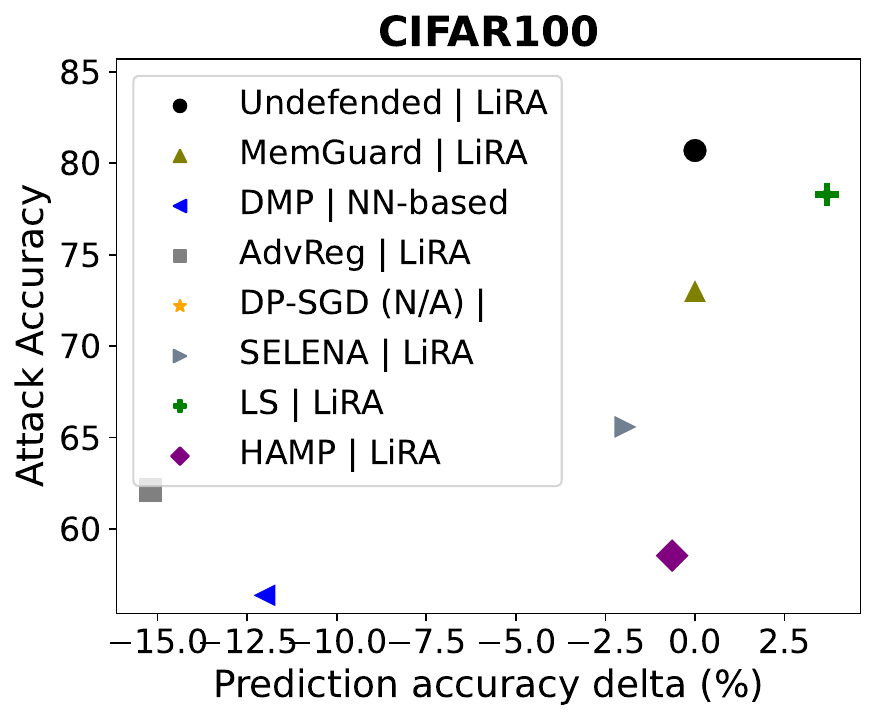}
		  %\subcaption{CIFAR100 (25k)} 
		  \label{fig:c25}
		\endminipage\hfill 

    \end{subfigure} 

    \begin{subfigure}[b]{\textwidth}
		\minipage{0.31\textwidth}
		  \includegraphics[width=0.99\textwidth, height=1.7in]{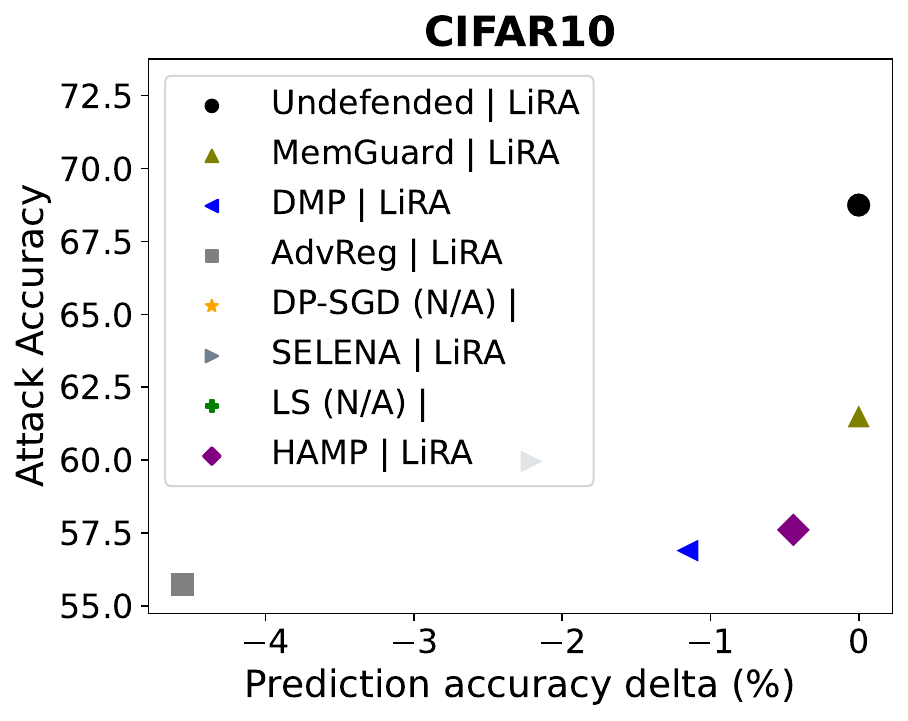}
		  %\subcaption{CIFAR10 (15k)}
		  \label{fig:cifar15}
		 \endminipage\hfill
		\minipage{0.31\textwidth}
		  \includegraphics[width=0.99\textwidth, height=1.7in]{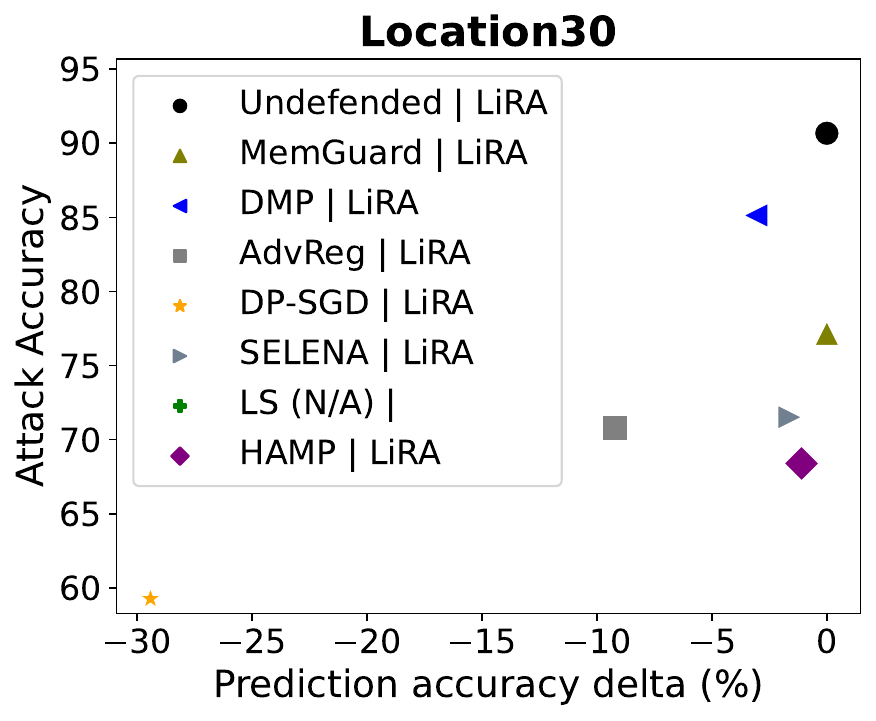}
		  %\subcaption{CIFAR10 (20k)}
		  \label{fig:cifar20}
		\endminipage\hfill
		\minipage{0.31\textwidth}%
		  \includegraphics[width=0.99\textwidth, height=1.7in]{fig/auc-plot/average_res-auc.pdf}
		  %\subcaption{CIFAR10 (25k)} 
		  \label{fig:c25}
		\endminipage\hfill 

    \end{subfigure}

    \caption{Attack AUC comparison. The legend indicates the attack that yields the highest attack AUC.   }
    \label{fig:auc-detailed}
      \vspace{-4mm}
\end{figure*}

\begin{figure}[htb]
	\centering
	\includegraphics[ height=1.8in]{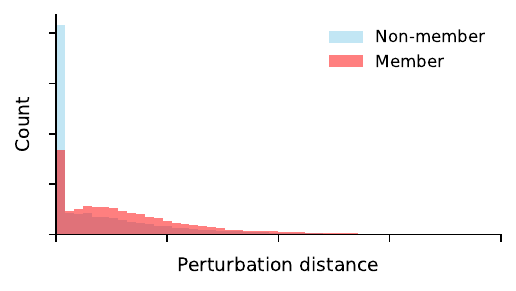}
	\caption{Perturbation distance on the members and non-members (obtained from the undefended model). The poor separation between the members and non-members in terms of the perturbation distance indicates the boundary-based label-only attack can not be calibrated to infer members/non-members with low false positive/negative.}
	\label{fig:boundary-attack} 
  \vspace{-4mm}
\end{figure}

\subsection{Evaluation on Different Network Architectures} 
\label{sec:diff-arch-eval}
This section reports additional evaluation on models trained with different network architectures (using CIFAR10), including DenseNet-12~\cite{huang2017densely}, ResNet-18~\cite{he2016deep}, MobileNet~\cite{howard2017mobilenets}, ShuffleNet~\cite{zhang2018shufflenet}. 
The results are shown in Fig.~\ref{fig:diff-arch}. 

We find that models trained with different architectures exhibit disparate degrees of MIA risk, with the attack TPR @0.1\% FPR being 6.47\%$\sim$30\%, and the attack TNR @0.1\% FNR 10.15\%$\sim$ 31.12\%. 
This gives an average attack TPR of 16.29\% and attack TNR of 18.75\%. 
\sysname is able to consistently reduces the MIA risk, with the attack TPR on \sysname being 0.52\%$\sim$0.92\% and the attack TNR 0.31\%$\sim$0.77\%. 
On average, \sysname reduces the attack TPR by 95.6\% (from 16.29\% to 0.72\%) and the attack TNR by 97.5\% (from 18.75\% to 0.47). 
Further, \sysname achieves such strong privacy protection with only a minor accuracy drop of 0.59\% (at most 1.28\%).

\subsection{Detailed Attack AUC comparison}
\label{sec:auc}
In Section~\ref{sec:setup}, we report the average attack AUC on each defense in Fig.~\ref{fig:auc}, and we provide the detailed results on each dataset in Fig.~\ref{fig:auc-detailed}.

\begin{figure*}[!t]
    \centering
    \begin{subfigure}[b]{\textwidth}
    \minipage{0.33\textwidth}
      \includegraphics[width=1.1\textwidth, height=2.0in]{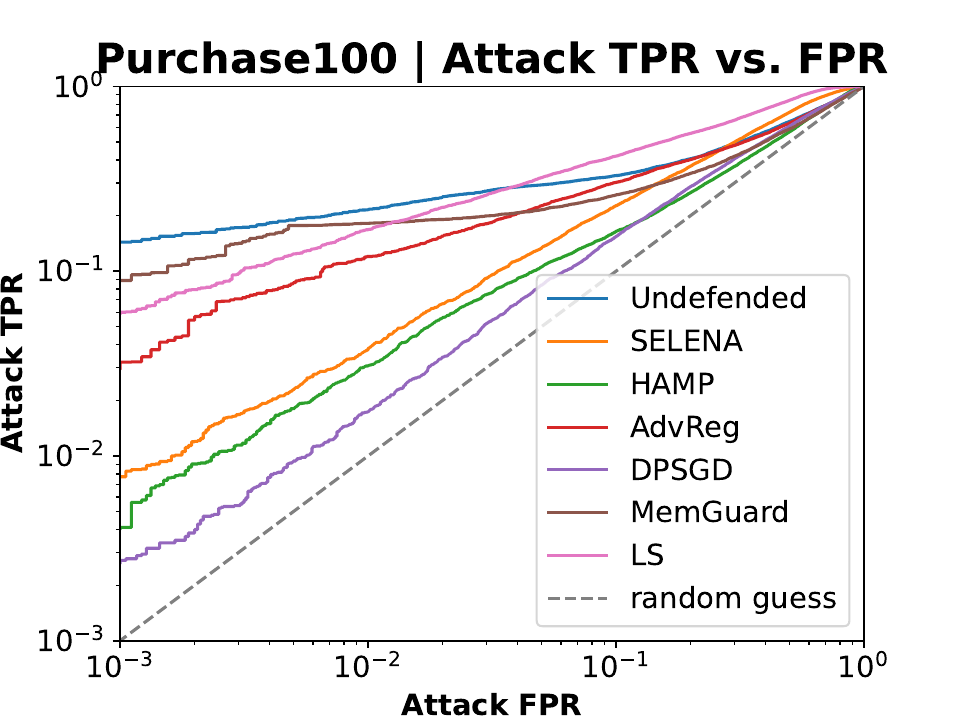}
      %\subcaption{CIFAR100 (15k)}
      \label{fig:c15}
     \endminipage\hfill
    \minipage{0.33\textwidth}
      \includegraphics[width=1.1\textwidth, height=2.0in]{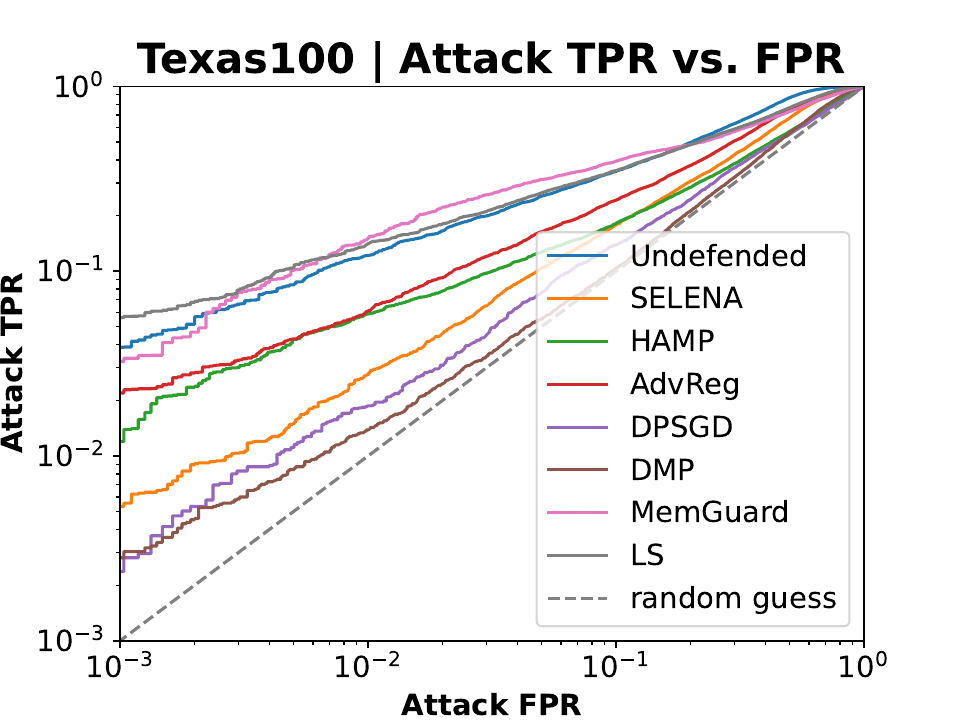}
      %\subcaption{CIFAR100 (20k)}
      \label{fig:c20}
    \endminipage\hfill
    \minipage{0.33\textwidth}%
      \includegraphics[width=1.1\textwidth, height=2.0in]{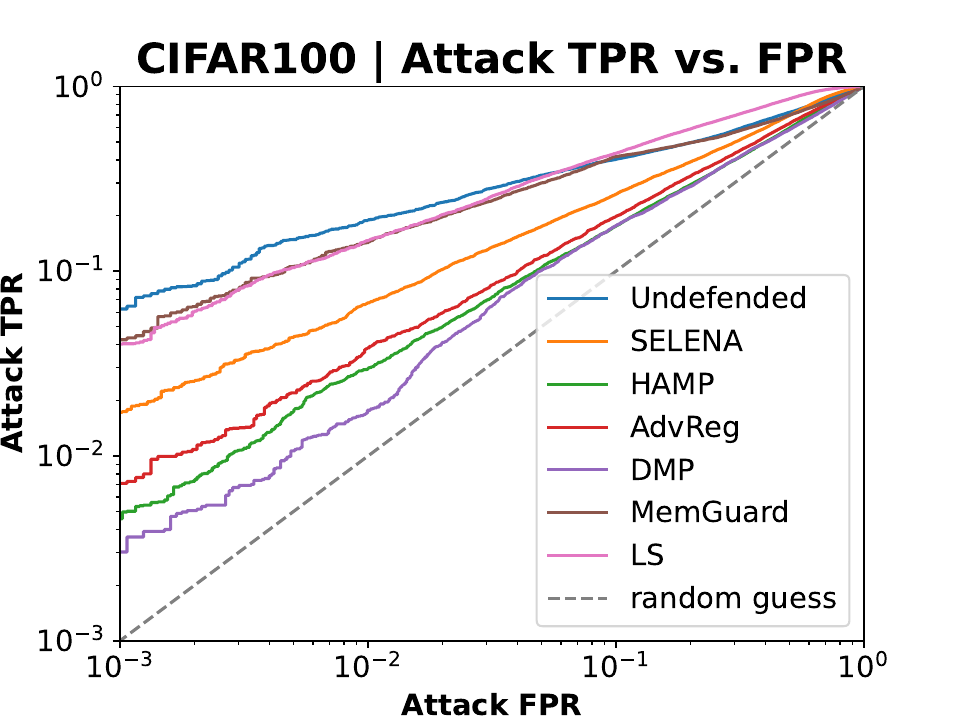}
      %\subcaption{CIFAR100 (25k)} 
      \label{fig:c25}
    \endminipage\hfill 
    \end{subfigure} 

    \begin{subfigure}[b]{\textwidth}
    \minipage{0.33\textwidth}
      \includegraphics[width=1.1\textwidth, height=2.0in]{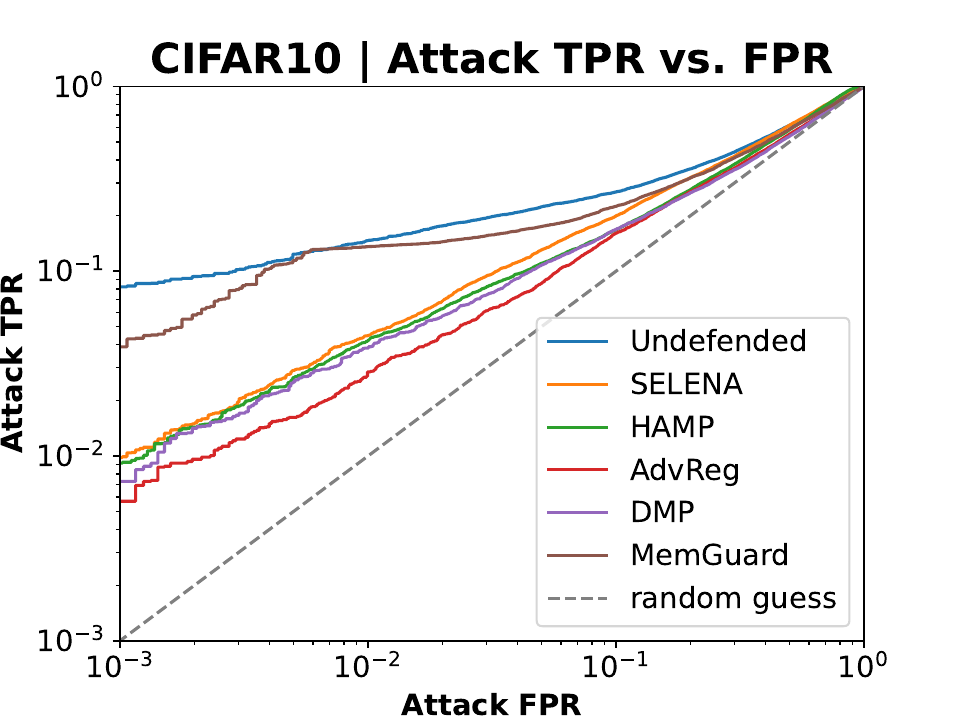}
      %\subcaption{CIFAR10 (15k)}
      \label{fig:cifar15}
     \endminipage\hfill
    \minipage{0.33\textwidth}
      \includegraphics[width=1.1\textwidth, height=2.0in]{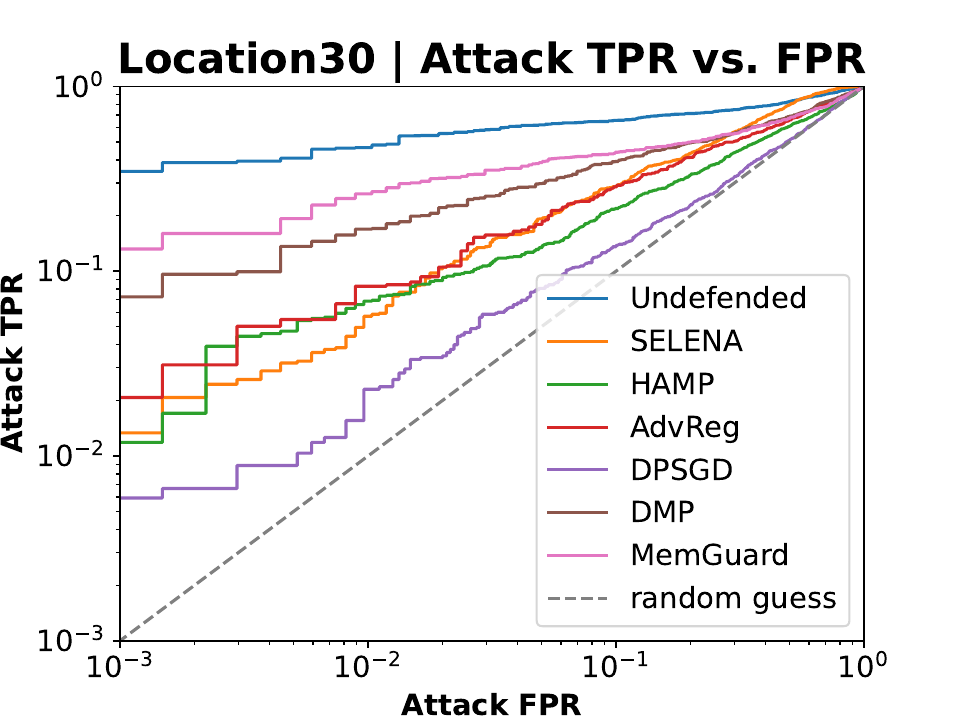}
      %\subcaption{CIFAR10 (20k)}
      \label{fig:cifar20}
    \endminipage\hfill 
        \minipage{0.33\textwidth}
      \includegraphics[width=1.1\textwidth, height=2.0in]{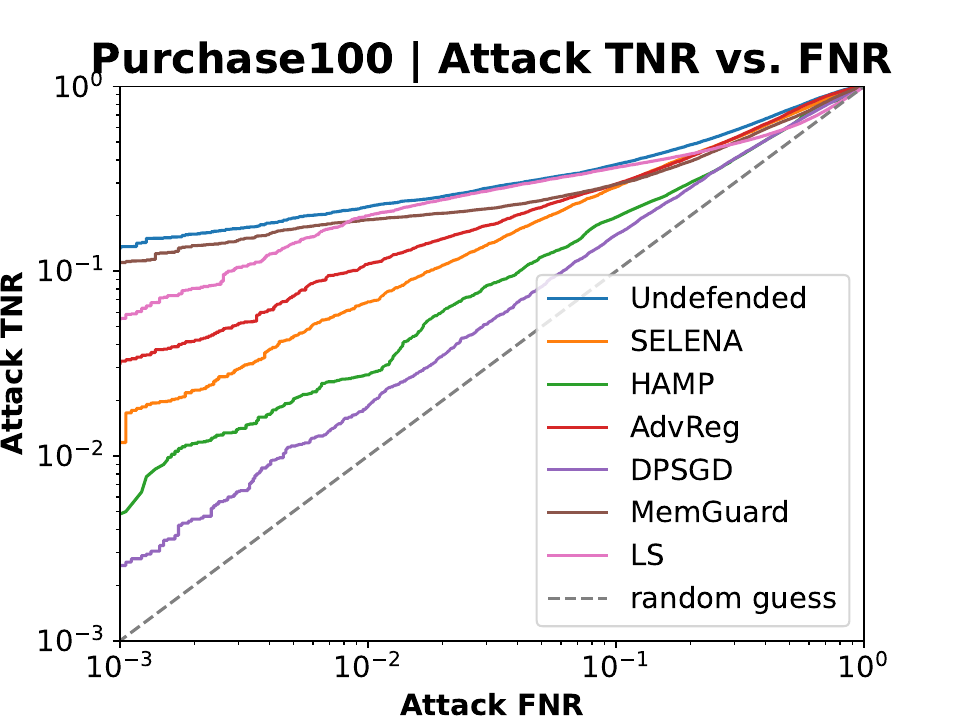}
      %\subcaption{CIFAR100 (15k)}
      \label{fig:c15}
     \endminipage\hfill
    \end{subfigure}

  \begin{subfigure}[b]{\textwidth}

    \minipage{0.33\textwidth}
      \includegraphics[width=1.1\textwidth, height=2.0in]{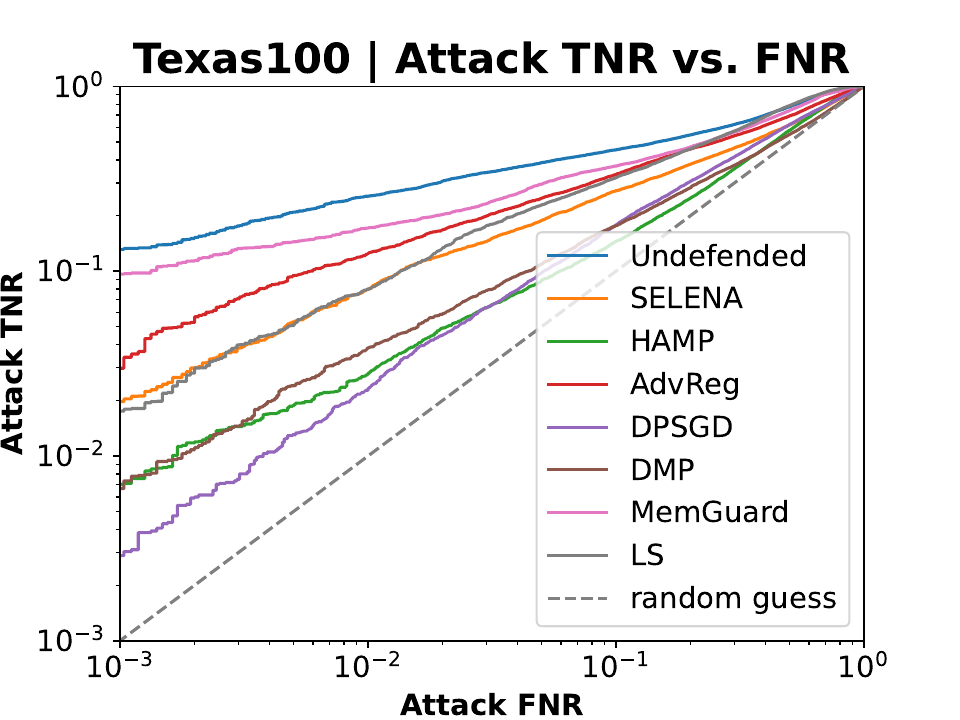}
      %\subcaption{CIFAR100 (20k)}
      \label{fig:c20}
    \endminipage\hfill
    \minipage{0.33\textwidth}%
      \includegraphics[width=1.1\textwidth, height=2.0in]{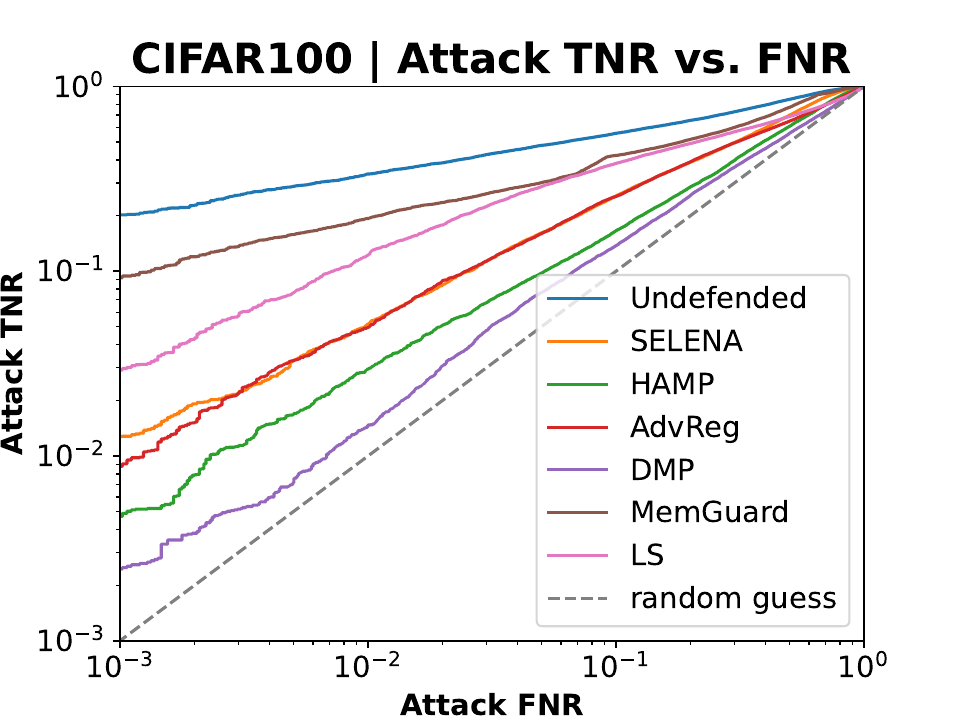}
      %\subcaption{CIFAR100 (25k)} 
      \label{fig:c25}
    \endminipage\hfill 
    \minipage{0.33\textwidth}
      \includegraphics[width=1.1\textwidth, height=2.0in]{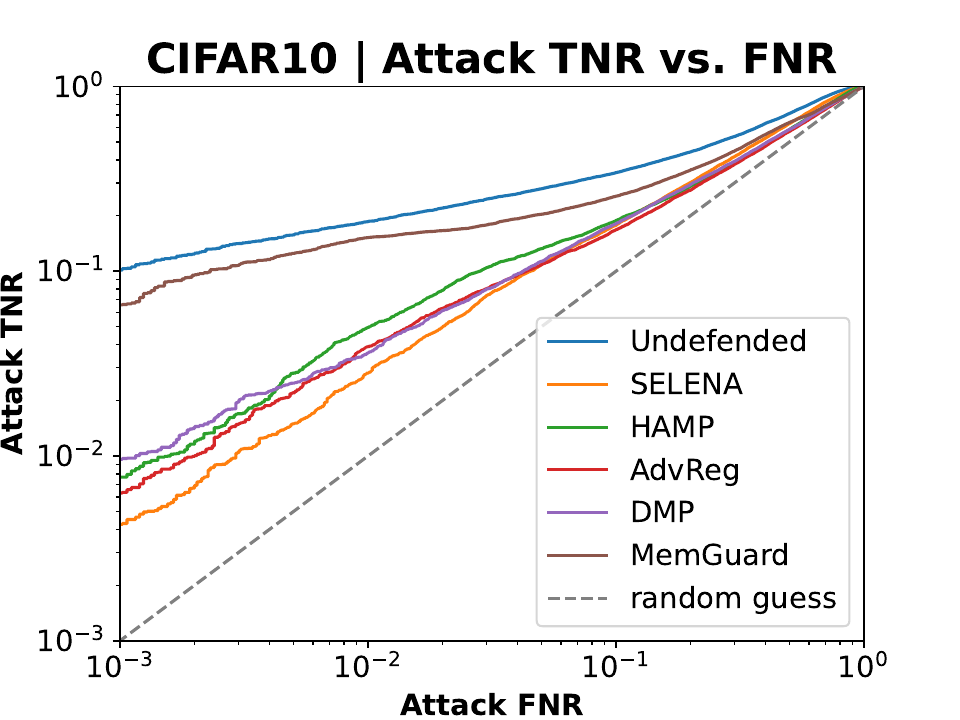}
      %\subcaption{CIFAR10 (15k)}
      \label{fig:cifar15}
     \endminipage\hfill
    \end{subfigure} 

    \begin{subfigure}[b]{\textwidth} 
    \minipage{0.33\textwidth}
      \includegraphics[width=1.1\textwidth, height=2.0in]{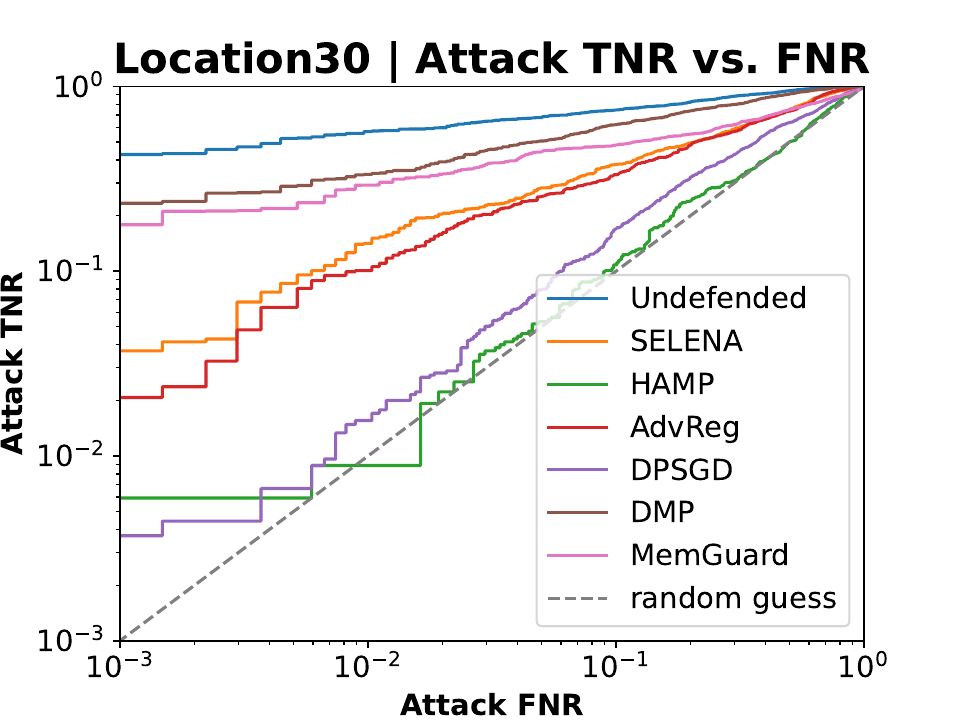}
      %\subcaption{CIFAR10 (20k)}
      \label{fig:cifar20}
    \endminipage\hfill 

    \end{subfigure}  

    \caption{Full ROC curves, showing attack TPR Vs. FPR, and attack TNR Vs. FNR. }
    \label{fig:roc}
\end{figure*}

\subsection{{Full ROC Curves}}
\label{sec:roc}
The full ROC curves from the evaluation in Section~\ref{sec:eval} can be found in Fig.~\ref{fig:roc}.

\subsection{{Detailed Results for Each Attack}}
\label{sec:all-atk-res}
In Section~\ref{sec:eval}, we reported the highest attack results among all evaluated attacks. 
We now provide the detailed results for each attack for completeness (the correctness-based attack is omitted as it does not work when calibrated at a low FPR or FNR), and they can be found in Table~\ref{tab:detailed-atk-res}. 

{
Our results also find that label-only attacks are unsuccessful in inferring members and non-members when controlled at low FPR and FNR regime - this is also a known issue found on many score-based attacks by Carlini et al.~\cite{carlini2022membership}. 
We use the boundary-based attack~\cite{choquette2021label} as an example to illustrate. 

We first plot the perturbation distance on the members and non-members in Fig.~\ref{fig:boundary-attack}. 
As shown, though perturbing the training members requires more perturbations than the testing samples, the distance is not well separated enough to be calibrated for inferring members with low false positive (hence with a low 0.12\% TPR@0.1\% FPR), or inferring non-members with low false negative (hence with a low 0.1\% TNR@0.1\%FNR). 
}

\begin{table*}[t]
  \caption{Detailed attack TPR and TNR for each attack. The highest attack results are highlighted in \textbf{bold}.
   }
  \label{tab:detailed-atk-res}
  \centering  
  \footnotesize
  \begin{tabular*}{\textwidth}{l|ll|llllllll}
\cline{1-11} 

  \multirow{2}{4em}{Dataset} & \multirow{2}{5em}{Defense}  & \multirow{2}{4em}{Metric (\%)} &  \multirow{2}{2em}{NN-based} &  \multirow{2}{2em}{Loss-based} &  \multirow{2}{4em}{Confidence-based} &  \multirow{2}{3em}{Entropy-based} &  \multirow{2}{5em}{M-entropy-based}  &  \multirow{2}{5em}{Augmentation-based} &   \multirow{2}{4em}{Boundary-based} &  \multirow{2}{2em}{LiRA} \\
    & & & & & & & &  \\
\cline{1-11} 
\cline{1-11} 

  \multirow{14}{5em}{Purchase100 }  
              & \multirow{2}{5em}{Undefended} & Attack TPR & \textbf{14.37} & 0.09 & 0.09 & 0.08  & 0.08  & N/A & 0.00 & 1.86  \\
              &                               & Attack TNR & 0.20 & 10.24 & 10.24 & 1.70 & 9.64  &   N/A& 0.00 & \textbf{13.19}  \\
              \cline{2-11} 
              & \multirow{2}{5em}{MemGuard} & Attack TPR 	 & \textbf{8.84} & 0.05 & 0.05 & 0.05 & 0.05 &  N/A & 0.00 & 1.18 \\
              &                               & Attack TNR & 0.10 & 7.23 & 7.23 & 0.16 & 7.32 &  N/A & 0.00 & \textbf{11.19} \\
              \cline{2-11} 
              & \multirow{2}{5em}{LS} & Attack TPR 				 & 3.22 & 0.08 & 0.08 & 0.08 & 0.08 &  N/A & 0.00 & \textbf{5.83} \\
              &                               & Attack TNR & 0.14 & 5.07 & 5.07 & 1.69 & 5.26 &  N/A & 0.00 & \textbf{5.44} \\
              \cline{2-11} 
              & \multirow{2}{5em}{DPSGD} & Attack TPR 		 & 0.04 & 0.11 & 0.11 & 0.14 & 0.12 &  N/A & 0.00 & \textbf{0.26} \\
              &                               & Attack TNR & 0.06 & 0.03 & 0.03 & 0.11 & 0.03 &  N/A & 0.00 & \textbf{0.26} \\
              \cline{2-11} 
              & \multirow{2}{5em}{SELENA} 		& Attack TPR & {0.70} & 0.09 & 0.09 & 0.09 & 0.08 &  N/A & 0.00 & \textbf{0.77} \\
              &                               & Attack TNR & 0.06 & 0.67 & 0.67 & 0.15 & 0.56 &  N/A & 0.00 & \textbf{1.17} \\
              \cline{2-11} 
              & \multirow{2}{5em}{AdvReg} & Attack TPR 		 & \textbf{2.93} & 0.14 & 0.14 & 0.13 & 0.13 &  N/A & 0.00 & {1.39} \\
              &                               & Attack TNR & 0.16 & 2.77 & 2.77 & 0.58 & 1.86 &  N/A & 0.00 & \textbf{3.14} \\
              \cline{2-11} 
              & \multirow{2}{5em}{\sysname} & Attack TPR 	 & 0.39 & 0.09 & 0.09 & 0.11 & 0.09 &  N/A & 0.00 & \textbf{0.4} \\
              &                               & Attack TNR & 0.13 & 0.35 & 0.35 & 0.08 & 0.33 &  N/A & 0.00 & \textbf{0.44} \\
\cline{1-11} 
\cline{1-11} 

  \multirow{16}{5em}{Texas100 }   
              & \multirow{2}{5em}{Undefended} & Attack TPR & 3.67 & 0.17 & 0.17 & 0.16 & 0.17 &  N/A & 0.00 & \textbf{3.87} \\
              &                               & Attack TNR & 0.68 & 2.03 & 2.03 & 0.40 & 1.87 &  N/A & 0.00 & \textbf{13.13} \\
              \cline{2-11} 
              & \multirow{2}{5em}{MemGuard} & Attack TPR 	 & \textbf{3.24} & 0.13 & 0.13 & 0.08 & 0.14 &  N/A & 0.00 & 2.65 \\
              &                               & Attack TNR & 0.56 & 1.93 & 1.93 & 0.23 & 1.88 &  N/A & 0.00 & \textbf{9.64} \\
              \cline{2-11} 
              & \multirow{2}{5em}{LS} & Attack TPR 				 & 1.11 & 0.15 & 0.15 & 0.16 & 0.15 &  N/A & 0.00 & \textbf{5.61} \\
              &                               & Attack TNR & 0.62 & 1.03 & 1.03 & 0.59 & 0.97 &  N/A & 0.00 & \textbf{1.75} \\
              \cline{2-11} 
              & \multirow{2}{5em}{DPSGD} & Attack TPR 		 & \textbf{0.24} & 0.10 & 0.10 & 0.10 & 0.10 &  N/A & 0.00 & 0.13 \\
              &                               & Attack TNR & 0.12 & 0.19 & 0.19 & 0.07 & 0.14 &  N/A & 0.00 & \textbf{0.29} \\
              \cline{2-11} 
              & \multirow{2}{5em}{DMP} & Attack TPR 		   & \textbf{0.24} & 0.05 & 0.05 & 0.04 & 0.09 &  N/A & 0.00 & 0.16 \\
              &                               & Attack TNR & 0.04 & 0.13 & 0.13 & 0.13 & 0.15 &  N/A & 0.00 & \textbf{0.21} \\
              \cline{2-11} 
              & \multirow{2}{5em}{SELENA} & Attack TPR 		 & 0.31 & 0.08 & 0.08 & 0.08 & 0.07 &  N/A & 0.00 & \textbf{0.53} \\
              &                               & Attack TNR & 0.13 & 0.16 & 0.16 & 0.19 & 0.10 &  N/A & 0.00 & \textbf{1.97} \\
              \cline{2-11} 
              & \multirow{2}{5em}{AdvReg} & Attack TPR 		 & 0.07 & 0.16 & 0.16 & 0.17 & 0.13 &  N/A & 0.00 & \textbf{2.19} \\
              &                               & Attack TNR & 0.25 & 0.39 & 0.39 & 0.10 & 0.37 &  N/A & 0.00 & \textbf{2.99} \\
              \cline{2-11} 
              & \multirow{2}{5em}{\sysname} & Attack TPR 	 & 0.31 & 0.12 & 0.12 & 0.07 & 0.12 &  N/A & 0.00 & \textbf{1.20} \\
              &                               & Attack TNR & 0.07 & 0.59 & 0.59 & 0.11 & 0.59 &  N/A & 0.00 & \textbf{0.70} \\
\cline{1-11} 
\cline{1-11} 

  \multirow{14}{5em}{Location30 }   
              & \multirow{2}{5em}{Undefended} & Attack TPR & \textbf{34.67} & 0.15  & 0.15 & 0.07 & 0.15 &  N/A & 0.00 & 16.22 \\
              &                               & Attack TNR & 1.93  & 19.56 & 19.56 & 0.89 & 11.63 &  N/A & 0.00 & \textbf{42.81} \\
              \cline{2-11} 
              & \multirow{2}{5em}{MemGuard} & Attack TPR 	 & \textbf{13.19} & 0.00 & 0.00 & 0.00 & 0.15 &  N/A & 0.00 & 4.00 \\
              &                               & Attack TNR & 0.00 & 16.59 & 16.59 & 1.33 & 16.59 & N/A  & 0.00 & \textbf{17.85} \\
              \cline{2-11} 
              & \multirow{2}{5em}{DPSGD} & Attack TPR 		 & 0.00 & 0.22 & 0.22 & 0.22 & 0.22 & N/A  & 0.00 & \textbf{0.59} \\
              &                               & Attack TNR & 0.15 & 0.22 & 0.22 & 0.00 & 0.22 &  N/A & 0.00 & \textbf{0.37} \\
              \cline{2-11} 
              & \multirow{2}{5em}{DMP} & Attack TPR 			 & \textbf{7.26} & 0.59 & 0.59 & 0.52 & 0.52 &   N/A		& 0.00 & 4.52 \\
              &                               & Attack TNR & 0.00 & 15.48 & 15.48 & 1.19 & 10.74 &  N/A & 0.00 & \textbf{23.33} \\
              \cline{2-11} 
              & \multirow{2}{5em}{SELENA} & Attack TPR 		 & 0.00 & 0.30 & 0.30 & 0.30 & 0.59 &  N/A & 0.00 & \textbf{1.33} \\
              &                               & Attack TNR & 0.59 & 0.67 & 0.67 & 0.67 & 0.81 &  N/A & 0.00 & \textbf{3.70} \\
              \cline{2-11} 
              & \multirow{2}{5em}{AdvReg} & Attack TPR 	   & \textbf{2.07} & 0.15 & 0.15 & 0.15 & 0.15 &  N/A & 0.00 & 1.85 \\
              &                               & Attack TNR & 0.15 & 1.48 & 1.48 & 0.59 & 1.04 &  N/A & 0.00 & \textbf{2.07} \\
              \cline{2-11} 
              & \multirow{2}{5em}{\sysname} & Attack TPR 	 & 0.30 & 0.52 & 0.52 & 0.15 & 0.44 &  N/A & 0.00 & \textbf{1.19} \\
              &                               & Attack TNR & \textbf{0.59} & 0.22 & 0.22 & 0.22 & 0.22 &  N/A & 0.00 & 0.59 \\
\cline{1-11} 
\cline{1-11} 

  \multirow{12}{5em}{CIFAR10 }   
              & \multirow{2}{5em}{Undefended} & Attack TPR & \textbf{8.23} & 0.00 & 0.00 & 0.10 & 0.00 & 0.02 & 0.10 & 2.76 \\
              &                               & Attack TNR & 0.05 & 6.24 & 5.99 & 0.40 & 6.00 & 3.63 & 0.00 & \textbf{10.15} \\
              \cline{2-11} 
              & \multirow{2}{5em}{MemGuard} & Attack TPR 	 & \textbf{3.89} & 0.08 & 0.08 & 0.07 & 0.09 & 0.02 & 0.10 & 1.52 \\
              &                               & Attack TNR & 0.13 & 2.96 & 2.96 & 0.20 & 3.35 & 3.63 & 0.00 & \textbf{6.57} \\
              \cline{2-11} 
              & \multirow{2}{5em}{DMP} & Attack TPR 			 & \textbf{0.73} & 0.06 & 0.00 & 0.10 & 0.00 & 0.10 & 0.10 & 0.11 \\
              &                               & Attack TNR & 0.05 & 0.37 & 0.67 & 0.12 & 0.68 & 0.20 & 0.00 & \textbf{0.72} \\
              \cline{2-11} 
              & \multirow{2}{5em}{SELENA} & Attack TPR 		 & 0.18 & 0.00 & 0.00 & 0.11 & 0.00 & 0.13 & 0.05 & \textbf{0.98} \\
              &                               & Attack TNR & 0.07 & 0.16 & 0.11 & 0.19 & 0.06 & 0.16 & 0.00 & \textbf{0.43} \\
              \cline{2-11} 
              & \multirow{2}{5em}{AdvReg} & Attack TPR 		 & \textbf{0.57} & 0.09 & 0.09 & 0.14 & 0.14 & 0.05 & 0.13 & 0.18 \\
              &                               & Attack TNR & 0.04 & 0.40 & 0.30 & 0.16 & 0.29 & 0.18 & 0.00 & \textbf{0.63} \\
              \cline{2-11} 
              & \multirow{2}{5em}{\sysname} & Attack TPR 	 & 0.39 & 0.00 & 0.11 & 0.00 & 0.00 & 0.08 & 0.12 & \textbf{0.92} \\
              &                               & Attack TNR & 0.17 & 0.17 & 0.26 & 0.00 & 0.51 & 0.38 & 0.00 & \textbf{0.77} \\
\cline{1-11} 
\cline{1-11} 

  \multirow{14}{5em}{CIFAR100 }   
              & \multirow{2}{5em}{Undefended} & Attack TPR & \textbf{6.24} & 0.09 & 0.13 & 0.14  & 0.15  & 0.07 & 0.12 & 3.57  \\
              &                               & Attack TNR & 0.46 & 2.80 & 2.56 & 0.24 & 2.52  & 1.05 & 0.10 & \textbf{20.16}  \\
              \cline{2-11} 
              & \multirow{2}{5em}{MemGuard} & Attack TPR 	 & \textbf{4.26} & 0.15 & 0.15 & 0.12 & 0.10 & 0.07 & 0.12 & 1.86 \\
              &                               & Attack TNR & 0.14 & 2.28 & 2.28 & 0.20 & 2.37 & 1.05 & 0.10 & \textbf{9.21} \\
              \cline{2-11} 
              & \multirow{2}{5em}{LS} & Attack TPR 					& 1.56 & 0.07 & 0.08 & 0.08 & 0.08 & 0.30 & 0.09 & \textbf{4.03} \\
              &                                & Attack TNR & 0.28 & \textbf{2.92} & 2.92 & 0.29 & 2.71 & 0.37 & 0.00 & 2.88 \\
              \cline{2-11} 
              & \multirow{2}{5em}{DMP} & Attack TPR 			 & \textbf{0.30} & 0.11 & 0.20 & 0.20 & 0.21 & 0.21 & 0.09 & 0.21 \\
              &                               & Attack TNR & 0.11 & \textbf{0.24} & 0.20 & 0.17 & 0.23 & 0.12 & 0.00 & 0.21 \\
              \cline{2-11} 
              & \multirow{2}{5em}{SELENA} 		& Attack TPR & 0.15 & 0.06 & 0.06 & 0.07 & 0.06 & 0.19 & 0.08 & \textbf{1.72} \\
              &                               & Attack TNR & 0.15 & 0.21 & 0.19 & 0.11 & 0.20 & 0.06 & 0.00 & \textbf{1.28} \\
              \cline{2-11} 
              & \multirow{2}{5em}{AdvReg} & Attack TPR 		 & \textbf{0.71} & 0.00 & 0.00 & 0.12 & 0.00 & 0.08 & 0.12 & 0.13 \\
              &                               & Attack TNR & 0.31 & 0.60 & 0.57 & 0.08 & {0.65} & 0.12 & 0.00 & \textbf{0.85} \\
              \cline{2-11} 
              & \multirow{2}{5em}{\sysname} & Attack TPR 	 & 0.22 & 0.16 & 0.16 & 0.00 & 0.17 & 0.09 & 0.10 & \textbf{0.46} \\
              &                               & Attack TNR & 0.08 & 0.41 & 0.41 & 0.07 & 0.41 & 0.24 & 0.00 & \textbf{0.47} \\
              \cline{1-11}

	\end{tabular*}  
	%\vspace{-2mm}
\end{table*}